%% file: main.tex
\documentclass[acmsmall]{acmart}
\usepackage{subfigure}
\usepackage{amsthm, amsmath, mathtools}
\usepackage{upgreek}
\usepackage{amsfonts}
\usepackage{enumitem}
\usepackage[normalem]{ulem}
\usepackage[linesnumbered,vlined,ruled,commentsnumbered]{algorithm2e}
\usepackage{diagbox}
\usepackage[para,online,flushleft]{threeparttable}
\usepackage{multirow}
\usepackage{multicol}
\usepackage{hyperref}
\usepackage{url}
\usepackage{bm}
\usepackage{colortbl}
\usepackage{flushend}
\usepackage{tabularx}
\usepackage[english]{babel}
\usepackage{graphicx}
\graphicspath{{./figs/}}
\usepackage{thmtools}
\usepackage{listings}
\renewcommand{\paragraph}[1]{\medskip \noindent {\bf #1}}
\newcommand{\OUT}{\mathrm{OUT}}
\newcommand{\Q}{\mathcal{Q}}
\newcommand{\V}{\mathcal{V}}
\newcommand{\E}{\mathcal{E}}
\newcommand{\T}{\mathcal{T}}

\newcommand{\dom}{\mathrm{dom}}

\newcommand{\eat}[1]{}

\newcommand{\DCQ}{\textsf{DCQ}}
\newcommand{\SCQ}{\textsf{SCQ}}
\newcommand{\CQ}{\textsf{CQ}}
\newcommand{\NCQ}{\textsf{NCQ}}
\newcommand{\xiao}[1]{{\color{black} #1}}

\newcommand{\x}{\bm{\mathsf{x}}}
\newcommand{\y}{\bm{\mathsf{y}}}

\AtBeginDocument{%
  \providecommand\BibTeX{{%
    \normalfont B\kern-0.5em{\scshape i\kern-0.25em b}\kern-0.8em\TeX}}}

\setcopyright{acmlicensed}
\copyrightyear{2023}
\acmYear{2023}
\acmDOI{10.1145/3589298}
\acmJournal{PACMMOD}
\acmVolume{1}
\acmNumber{2}
\acmArticle{153}
\acmMonth{6}
\acmPrice{15.00}

\received{October 2022}
\received[revised]{January 2023}
\received[accepted]{February 2023}

\lstdefinestyle{mystyle}{
    commentstyle=\color{green},
    keywordstyle=\color{purple},
    stringstyle=\color{purple},
    basicstyle=\small\ttfamily,
    breaklines=true,
    columns=fullflexible,
    frame=single,
}
\newcommand{\revm}[1]{{\color{black} #1}}

\ccsdesc[500]{Information systems~Query optimization}

\begin{document}
\title{Computing the Difference of Conjunctive Queries Efficiently}
\input{abstract}
\author{Xiao Hu}
\email{xiaohu@uwaterloo.ca}
\affiliation{%
  \institution{University of Waterloo}
  \streetaddress{200 University  Avenue West}
  \city{Waterloo}
  \state{Ontario}
  \country{Canada}
  \postcode{N2L 3G1}
}
\orcid{0000-0002-7890-665X}
\department{School of Computer Science}

\author{Qichen Wang}
\email{qcwang@hkbu.edu.hk}
\affiliation{%
  \institution{Hong Kong Baptist University}
  \streetaddress{224 Waterloo Road, Kowloon Tong}
  \city{Hong Kong}
  \country{Hong Kong}
}
\department{Department of Computer Science}
\orcid{0000-0002-0959-5536}
\renewcommand{\shortauthors}{Xiao Hu and Qichen Wang}

\keywords{conjunctive query, query optimization, difference operator}
\maketitle

\input{intro}

\input{preliminary}
\input{easy}
\input{hard}
\input{extension}
\input{experiment}

\input{related}
\newpage
\bibliographystyle{ACM-Reference-Format}
\bibliography{paper}
\input{appendix}
\end{document}

%% file: abstract.tex
\begin{abstract}
We investigate how to efficiently compute the difference result of two (or multiple) conjunctive queries, which is the last operator in relational algebra to be unraveled.  The standard approach in practical database systems is to materialize the results for every input query as a separate set, and then compute the difference of two (or multiple) sets.  This approach is bottlenecked by the complexity of evaluating every input query individually, which could be very expensive, particularly when there are only a few results in the difference.  In this paper, we introduce a new approach by exploiting the structural property of input queries and rewriting the original query by pushing the difference operator down as much as possible.  We show that for a large class of difference queries, this approach can lead to a linear-time algorithm, in terms of the input size and (final) output size, i.e., the number of query results that survive from the difference operator.  We complete this result by showing the hardness of computing the remaining difference queries in linear time.  Although a linear-time algorithm is hard to achieve in general, we also provide some heuristics that can provably improve the standard approach.  At last, we compare our approach with standard SQL engines over graph and benchmark datasets.  The experiment results demonstrate order-of-magnitude speedups achieved by our approach over the vanilla SQL.  
\end{abstract}

%% file: intro.tex
\section{Introduction}
\label{sec:intro}

Conjunctive queries with aggregation, union, and difference (also known as negation) operators \revm{form the full relational algebra} \cite{abiteboul1995foundations}.  While conjunctive queries \cite{ngo2018worst, bagan2007acyclic, yannakakis1981algorithms,amossen2009faster, deep2020fast, 10.14778/3547305.3547326}, with aggregation~\cite{joglekar16:_ajar} and unions \cite{carmeli2019enumeration, christoph2018answering}, have been extensively studied in the literature, the difference operator received much less attention.  In modern database systems, there are several different equivalent expressions for computing the difference between two queries, such as NOT IN, NOT EXIST, EXCEPT, MINUS, DIFFERENCE, and LEFT-OUTER JOIN followed by a non-NULL filter.  In contrast to its powerful expressibility, the execution plan of difference operator in existing database systems or data analytic engines  (e.g., MySQL~\cite{mysql}, Oracle~\cite{oracle}, Postgre SQL~\cite{postgre},  Spark SQL~\cite{spark}) is quite brute-force. Given two (or multiple) conjunctive queries, their difference is simply done by materializing the answers for each participated conjunctive query separately, and then computing the difference of two (or multiple) sets.  Hashing or other indexes may be built on top of the query answers to speed up the computation of the set difference at last.  However, this approach is severely bottlenecked by evaluating every input query individually and materializing a large number of intermediate query results that do not contribute to the final results due to the difference operator. 

Let's consider an example of friend recommendation in social networks (such as Twitter, Facebook, Sina Weibo).  
A friend recommendation is represented as a triple $(a, b, c)$ extracted from the network semantics, such that user $c$ is recommended to user $a$ since user $b$ is a friend of user $a$ and user $c$ is a friend of user $b$, together with other customized constraints.  We also avoid the recommendation when $a$ and $c$ are already friends. The task of finding all valid recommendations can be captured by a SQL query in Example~\ref{exa:intro}, as the difference of two sub-queries. %where the first one ($\Q_1$) is a single relation storing all candidate recommendations and the second one ($\Q_2$) is a triangle join 

\begin{figure*}
    \centering
    \begin{minipage}{0.54\linewidth}
    \caption*{\revm{(a) Execution Plan for Original Query} \qquad \qquad \qquad \qquad \qquad} \vspace{1em}\includegraphics[width=0.9\linewidth]{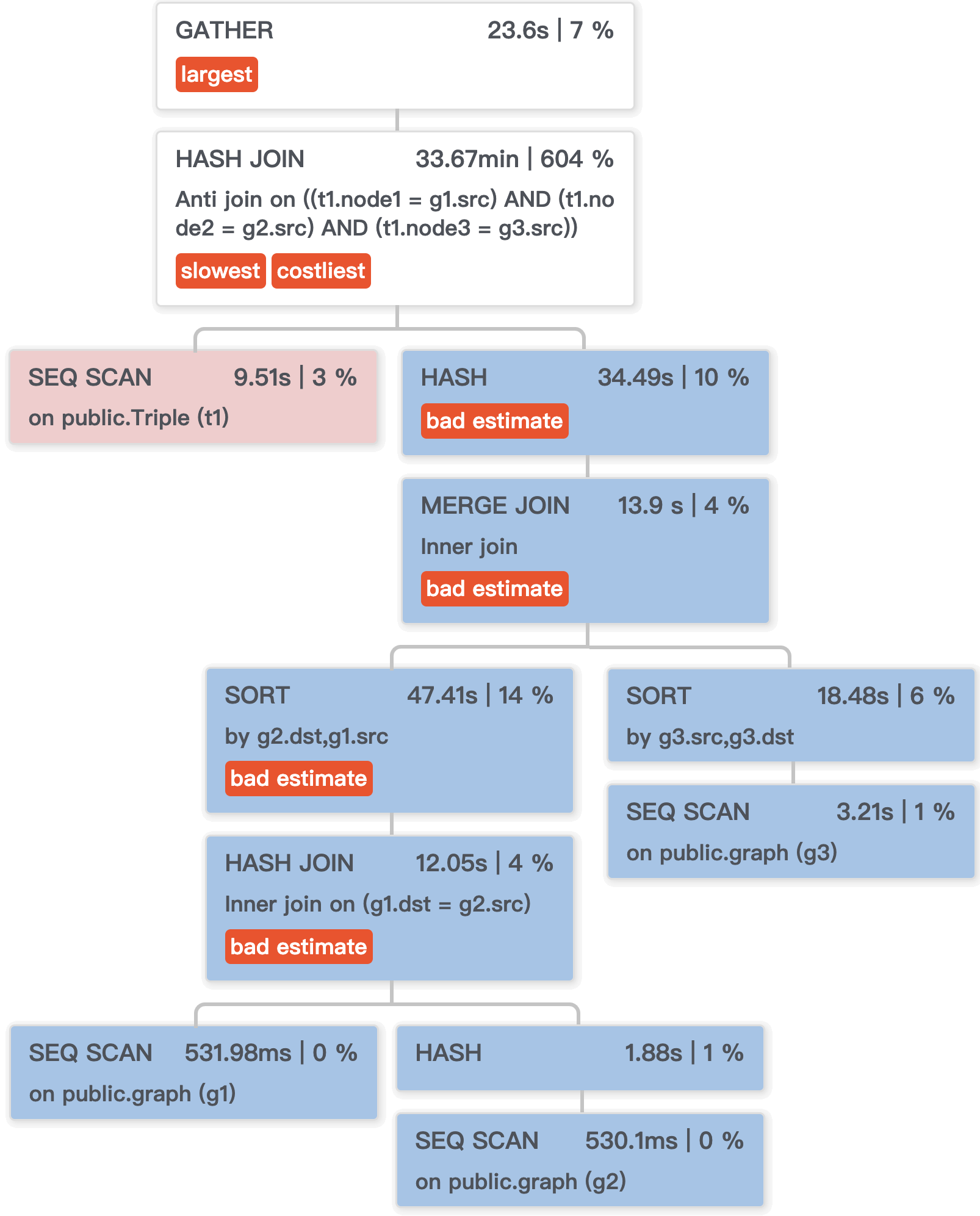}
    \label{fig:plan_vis-1}
    %\vspace{0.4em}
    \caption{Execution plans for SQL queries in Example~\ref{exa:intro} generated by PEV(\url{https://tatiyants.com/pev/}). \revm{Blocks in red indicate $\Q_1$ in $\Q$ and $\Q'$. Blocks in blue of (a) indicate $\Q_2$ in $\Q$, and that of (b) indicate $\Q_3$ in $\Q'$. Blocks in white indicate the difference operator for both $\Q$ and $\Q'$.}}
    %\vspace{1.1em}
    \end{minipage} \ 
    \begin{minipage}{0.42\linewidth}
    \caption*{\revm{(b) Execution Plan for Rewritten Query}}
    \vspace{1em}
    \hspace{-6em}
 \includegraphics[width=1.44\linewidth]{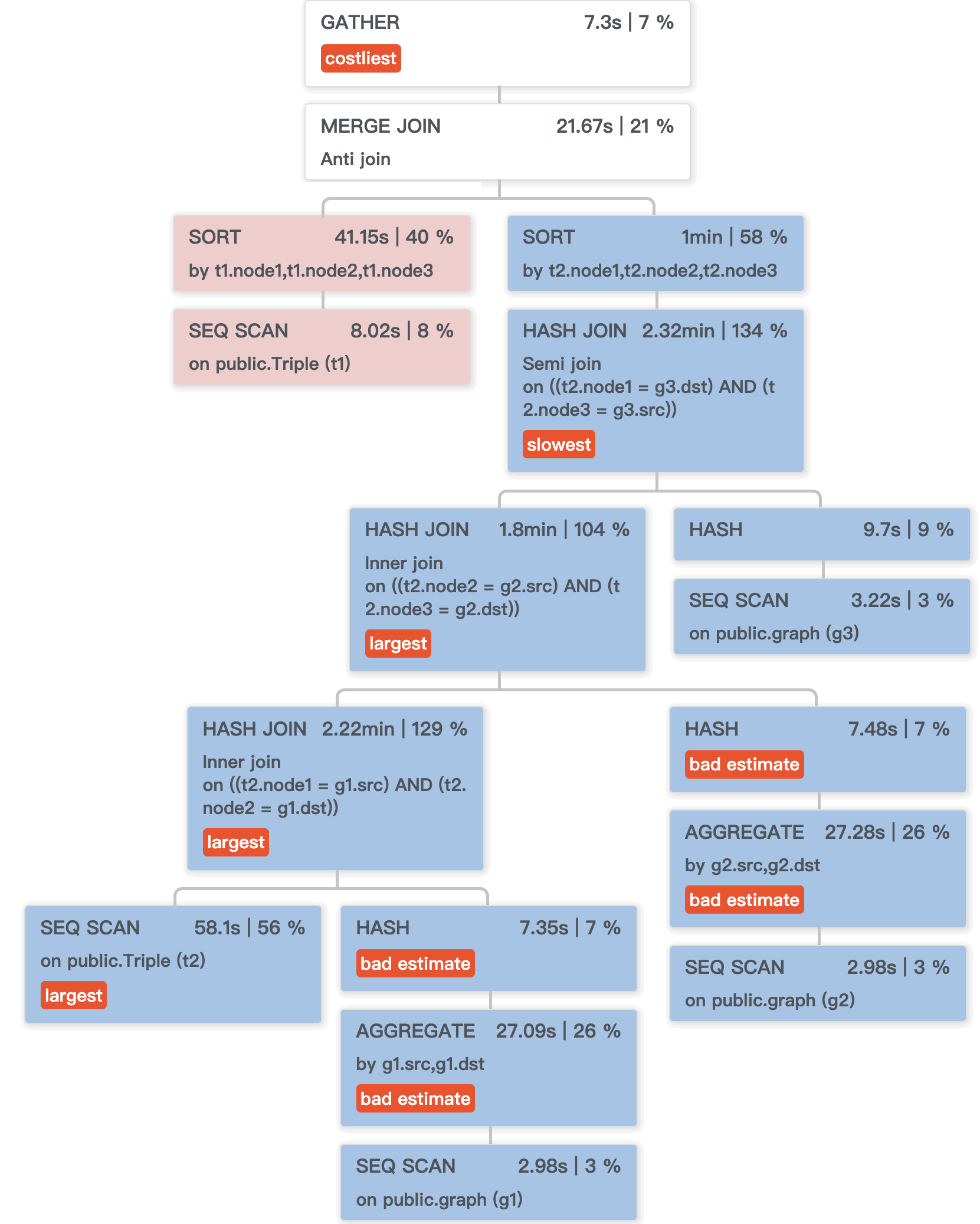}
    \label{fig:plan_vis-2}
    \end{minipage}
    \vspace{1em}
\end{figure*}
\begin{example}
\label{exa:intro}
\emph{
Let $\textsf{Graph}(\textsf{src}, \textsf{dst})$ be a table storing all edges in the social network, and $\textsf{Triple}(\textsf{node1},\\ \textsf{node2}, \textsf{node3})$ be a table storing all candidate recommendations. The following SQL query $(\Q)$ finds all triples from $\textsf{Triple}$ that do not form a triangle in the graph:}
\begin{lstlisting}[language=SQL,
	deletekeywords={IDENTITY},
	deletekeywords={[2]INT},
	morekeywords={clustered},
	mathescape=true,
	xleftmargin=-1pt,
	framexleftmargin=-1pt,
	frame=tb,
	framerule=-1pt, 
        framesep=-1pt]
$\revm{\Q:}$  SELECT node1, node2, node3 FROM Triple t1
    WHERE NOT EXISTS 
    (SELECT * FROM Graph g1, Graph g2, Graph g3 
     WHERE g1.dst = g2.src and g2.dst = g3.src and g3.dst = g1.src and g1.src = t1.node1 and g2.src = t1.node2 and g3.src = t1.node3);
\end{lstlisting}
\revm{\emph{such that $\Q$ is the difference of two sub-queries $\Q_1$ and $\Q_2$, where $\Q_1$ returns all candidate recommendations from \textsf{Triple} and $\Q_2$ returns all triangle friendship in the social network.}}
\begin{lstlisting}[ language=SQL,
	deletekeywords={IDENTITY},
	deletekeywords={[2]INT},
	morekeywords={clustered},
	mathescape=true,
	xleftmargin=-1pt,
	framexleftmargin=-1pt,
	frame=tb,
	framerule=0pt ]
$\revm{\Q_1:}$   SELECT node1, node2, node3 FROM Triple t1
$\revm{\Q_2:}$  SELECT * FROM Graph g1, Graph g2, Graph g3 
     WHERE g1.dst = g2.src and g2.dst = g3.src and g3.dst = g1.src and g1.src = t1.node1 and g2.src = t1.node2 and g3.src = t1.node3;
\end{lstlisting}

\emph{We note that $\Q$ can be rewritten as the following SQL query $\Q'$:}
\begin{lstlisting}[ language=SQL,
	deletekeywords={IDENTITY},
	deletekeywords={[2]INT},
	morekeywords={clustered},
	mathescape=true,
	xleftmargin=-1pt,
	framexleftmargin=-1pt,
	frame=tb,
	framerule=0pt ]
$\revm{\Q':}$     SELECT node1, node2, node3 FROM Triple t1
     WHERE NOT EXISTS 
    (SELECT * FROM Triple t2 
     WHERE EXISTS  (SELECT * FROM graph g1 WHERE t2.node1 = g1.src and t2.node2 = g1.dst) 
     AND EXISTS (SELECT * FROM graph g2 WHERE t2.node2 = g2.src and t2.node3 = g2.dst)
     AND EXISTS (SELECT * FROM graph g3 WHERE t2.node3 = g3.src and t2.node1 = g3.dst)
     AND t2.node1 = t1.node1 and t2.node2 = t1.node2 and t2.node3 = t1.node3)
\end{lstlisting}
% \emph{We note that $\Q$ can be rewritten as the following SQL query $\Q'$: }
% \begin{lstlisting}[ language=SQL,
% 	deletekeywords={IDENTITY},
% 	deletekeywords={[2]INT},
% 	morekeywords={clustered},
% 	mathescape=true,
% 	xleftmargin=-1pt,
% 	framexleftmargin=-1pt,
% 	frame=tb,
% 	framerule=0pt ]
% $\revm{\Q':}$     SELECT node1, node2, node3 FROM Triple 
%      WHERE (node1, node2, node3) NOT IN
%      (SELECT * FROM Triple n2 
%       WHERE EXISTS (SELECT * FROM graph g1 WHERE    n2.node1 = g1.src and n2.node2 = g1.dst) 
%       and EXISTS (SELECT * FROM graph g2 WHERE n2.node2  = g2.src and n2.node3 = g2.dst) 
%       and EXISTS (SELECT * FROM graph g3 WHERE n2.node3 = g3.src and n2.node1 = g3.dst))
% \end{lstlisting}
\revm{\emph{such that $\Q'$ is the difference of $\Q_1$ and another sub-query $\Q_3$, where $\Q_3$ finds all candidate recommendations in \textsf{Triple} that also form a triangle in the social network:}}
\begin{lstlisting}[ language=SQL,
	deletekeywords={IDENTITY},
	deletekeywords={[2]INT},
	morekeywords={clustered},
	mathescape=true,
	xleftmargin=-1pt,
	framexleftmargin=-1pt,
	frame=tb,
	framerule=0pt ]
$\revm{\Q_3:}$       SELECT * FROM Triple t2
     WHERE EXISTS (SELECT * FROM graph g1 WHERE t2.node1 = g1.src and t2.node2 = g1.dst) 
     And EXISTS (SELECT * FROM graph g2 WHERE t2.node2  = g2.src and t2.node3 = g2.dst) 
     And EXISTS (SELECT * FROM graph g3 WHERE t2.node3 = g3.src and t2.node1 = g3.dst)
     AND t2.node1 = t1.node1 and t2.node2 = t1.node2 and t2.node3 = t1.node3
\end{lstlisting}
\end{example}
%\begin{figure*}[t]
%\subfigure[Original query plan]{
%\resizebox{0.37\linewidth}{!}{
%%\centering
%\includegraphics[width=\linewidth]{figure/plan_orig_u.png}
%}
%}~%
%\subfigure[New query plan after rewriting.]{
%\resizebox{0.6\linewidth}{!}{
%\centering
%\includegraphics[width=1.2\linewidth]{figure/plan_new_u.png}
%}
%}
%\caption{Query plan visualization for Example~\ref{exa:intro} generated by PEV(\url{https://tatiyants.com/pev/}). \qichen{Blocks in red represent the subquery for calculating $\Q_1$ in both (a) and (b). Blocks in blue represent the subqueries for calculating $\Q_2$ in (a)  and $\Q_3$ in (b).} \revm{It is predicated to take 33.67 and 11.07 minutes for PostgreSQL to execute (a) and (b) separately. Hence, the optimizer will select query plan (b) in practice.}
%} 
%\label{fig:plan_vis}
%\end{figure*}
Figure 1(a) illustrates the execution plan for $\Q$ generated by PostgreSQL optimizer. It first materializes all triangles in the graph as $\Q_2$, and then computes the difference of \revm{$\Q_1$ and $\Q_2$} by anti-join. Moreover, hashing index is built on top of all triangles of $\Q_2$ so that the anti-join can be executed by checking whether every candidate recommendation in $\Q_1$ appears as a triangle in \revm{$\Q_2$}.  At last, all ``survived'' recommendations are outputted as final answers. \revm{In plan (a), computing the set difference at last is the most time-consuming step, which is predicted to take 33.67 minutes by PostgreSQL optimizer. Although computing the subquery $\Q_2$ is not that expensive, which only takes about 132 seconds, the number of intermediate results materialized for $\Q_2$ is quite large as expected, which finally leads to the inefficiency of the subsequent computation on $\Q_1 - \Q_2$. In Section~\ref{sec:experiment}, plan (a) actually runs in 308.175 seconds in practice.} %We can see that computing the set difference at the last step is predicted to take 33.67 minutes, which is the most time-consuming step, and computing $\Q_2$ is predicted to take 132s.
%Here, the performance of the standard approach is bottlenecked by the complexity of evaluating $\Q_2$, and computing the difference of two answer sets. 
%\begin{figure}
%    \centering
%   \includegraphics[width=\linewidth]{figure/plan_union.png}
%    \caption{Query plan using "or" clause for the example.}
%\end{figure}

To tackle the challenges brought by the difference operator, we take two input sub-queries as a whole into account for algorithm design. %, instead of tackling each individual input query separately. 
We are interested in efficient algorithms with running times linear in the final result size.  This requirement rules out the standard approach of materializing the results for each input sub-query separately and then computing their set difference.  Indeed, the final output size can be many magnitudes smaller than the number of intermediate results that materialized.  To overcome the curse of large intermediate results, we introduce a rewriting-based approach by {\em exploiting the joint structural properties of two input sub-queries} and {\em pushing the difference operator down as far as possible}.  

\revm{In Example~\ref{exa:intro}, we can rewrite the original SQL query $\Q$ into a new one $\Q'$. Instead of computing $\Q_2$, it finds all candidate recommendations that also form a triangle in the social network as $\Q_3$, which is exactly the intersection of $\Q_1$ and $\Q_2$, and then computes the difference of $\Q_1$ and $\Q_3$. Figure 1(b) illustrates the execution plan of this new query. We observe that computing the set difference at last is predicated to only take 21.67 seconds, which is much faster than (a). This is as expected, since the number of intermediate results generated by $\Q_3$ is much smaller than $\Q_2$, after taking $\Q_1$ into consideration, which is the key to the overall improvement. As the price, computing $\Q_3$ is predicated to take a few more minutes than $\Q_2$, but this is totally tolerable. In Section~\ref{sec:experiment}, plan (b) actually runs in 78.918 seconds, which already achieves 4x speedup over (a).} This significant improvement from (a) to (b)  motivates us to further investigate this interesting problem for general queries. %\footnote{Figure~\ref{fig:plan_vis} only includes the predication time made by Postgre SQL. In our experiment, the plan in Figure~\ref{fig:plan_vis}(a) actually runs in 858.452s and plan in Figure~\ref{fig:plan_vis}(b) runs in 25s. }

%Consider the ffar from, or even ba have adopted very simple strategies for evaluating the minus query. We would like to summarize two basic strategies for $\Q_1 - \Q_2$ here. 
%\begin{itemize}[leftmargin=*]
%    \item {\bf (Baseline-1)} $\Q_1, \Q_2$ are evaluated separately with their results materialized as $S_1, S_2$ respectively, and the $\Q_1 - \Q_2$ can be handled by applying the set difference operator to $S_1 - S_2$. Without investigating the interaction between $\Q_1, \Q_2$ in advance, this method suffers from a large number of intermediate results in $S_1, S_2$, while the true results in the difference query are very few. 
%    \item {\bf (Baseline-2)} Query rewriting may also help, based on the fact that $\Q_1 - \Q_2 = \Q_1 - \Q_1 \cap \Q_2$. Then, $\Q_1$ and $\Q_1\cap \Q_2$ will be evaluated separately as {\bf (Baseline-1)}. The benefit comes if it is easier to evaluate $\Q_1 \cap \Q_2$ than $\Q_2$ itself, as some interaction between $\Q_1, \Q_2$ has been exploited under the intersection operator. Moreover, if it is possible to check efficiently for any given tuple $t$ whether $t$ is a query result for $\Q_2$ or not, when indexes are pre-built, the intersection $\Q_1 \cap \Q_2$ can be simply done by checking whether every tuple returned from $\Q_1$ is a valid query result for $\Q_2$. However, this method still suffers from a large number of intermediate results (at least) in $S_1$, while the true results in the difference query are very few. 
%\end{itemize}
\textbf{Our contributions.~} In this paper, we formulate the {\em difference of conjunctive queries} (\DCQ) problem and study the data complexity of this problem.  Our contributions can be summarized as:
\begin{itemize}[leftmargin=*]
	\item \textbf{Complexity Dichotomy:} We give a dichotomy for computing \DCQ s in linear time in terms of input and output size.
	We characterize a class of ``easy'' \DCQ s exploiting the joint properties of two input CQs, and present a linear-time algorithm.  On the other hand, we prove the hardness of obtaining a linear-time algorithm for the remaining ``hard'' \DCQ s via several well-known conjectures. \textbf{ (Section~\ref{sec:easy-algorithm} and \ref{sec:hardness-proof})}
	\item \textbf{Efficient Heuristic:} We propose an efficient heuristic for computing ``hard'' \DCQ s, which does not lead to a linear-time algorithm but still improves the baseline approach greatly. The heuristic investigates the {\em intersection} of two input CQs and incorporates the state-of-the-art algorithms for CQ evaluation. 	\textbf{(Section~\ref{sec:heuristic})}
	\item \textbf{Extension: }
	We explore several interesting extensions.  First, we design a recursive algorithm for computing the difference of {\em multiple} conjunctive queries.  We also extend our algorithm to support other relational operators, such as selection, projection, join, and aggregation.  At last, we investigate the \DCQ\ problem under the bag semantics. 
	\textbf{(Section~\ref{sec:extension})}
    \item \textbf{Experimental Evaluation:} We provide an experimental evaluation of our approach and standard approach on real-world datasets in both centralized and parallel database systems.  The experimental results \revm{show that our approach out-performs the baseline} %demonstrate the out-performance of our approach 
    on different classes of queries and datasets. \textbf{(Section~\ref{sec:experiment})}
\end{itemize}

 {\bf Roadmap.} In Section~\ref{sec:preliminary}, we formally define the \DCQ\ problem and review the literature on evaluating a single CQ.  In Section~\ref{sec:easy-algorithm},  we provide a linear-time algorithm for ``easy'' \DCQ s.  In Section~\ref{sec:hardness}, we prove the hardness for the remaining \DCQ s and show efficient heuristics.  In Section~\ref{sec:extension}, we study several extensions of \DCQ s with other relational operators and bag semantics.  In Section~\ref{sec:experiment}, we present the experimental evaluation.  At last, we review related work in Section~\ref{sec:SCQ} and Section~\ref{sec:related}.

%% file: preliminary.tex
\section{Preliminaries}
\label{sec:preliminary}

\subsection{Problem Definition}
\noindent {\bf Conjunctive Query (CQ).}
We consider the standard setting of multi-relational databases.  Let $\mathbb{R}$ be a database schema that contains $n$ relations $R_1, R_2, \cdots, R_n$.
Let $\V$ be the set of attributes in the database $\mathbb{R}$.  Each relation $R_i$ is defined on a subset of attributes $e_i \subseteq \V$. Let $\E= \{e_1, e_2, \cdots, e_n\}$ be the set of the attributes for all relations. %We also use $R_e$ to denote the relation defined on $e \in \E$. 
Let $\dom(x)$ be the domain of attribute $x \in \V$, and let $\dom(U) = \prod_{x \in U} \dom(x)$ be the domain of attributes $U \subseteq \V$.

\par
Given the database schema $\mathbb{R}$, let an input instance be $D$, and the corresponding instances of $R_1, \cdots, R_n$ be $R_1^{D}, \cdots$, $R_n^{D}$. Where $D$ is clear from the context, we will drop the superscript and use $R_1, \cdots, R_n$ for both the schema and instances.  Any tuple $t \in R_i$ is defined on $e_i$. For any attribute $x \in e_i$, $ \pi_{x} t \in \dom(x)$ denotes the value of attribute $x$ in tuple $t$.	
Similarly, for a set of attributes $U \subseteq e_i$, $\pi_U t$ denotes the values of attributes in $U$ for $t$ with an implicit ordering on the attributes.

We consider the class of \emph{conjunctive queries without self-joins} formally defined as
\begin{equation}
  \label{q1}
  \Q := \pi_{\y} \left( \sigma_{\phi_1} R_1(e_1) \Join \cdots \sigma_{\phi_2} R_2(e_2) \Join \cdots \Join \sigma_{\phi_n} R_n(e_n) \right), 
\end{equation}
where $\sigma$ is the selection operator, $\phi_i$ is the predicate defined over relation $R_i$, $\sigma_{\phi_i} R_i(e_i)$ selects out tuples from $R_i$ passing the predicate $\phi_i$ and $\y \subseteq \V$ denotes the {\em output attributes}.
If $\y = \V$, such a CQ query is known as {\em full join}, which represents the natural join of the underlying relations. We usually use a triple $(\y, \V, \E)$ to represent a CQ $\Q$, and simply use a pair $(\V, \E)$ to represent a full join. Each relation $R_i$ in $\Q$ is distinct, i.e., the CQ does not have a self-join.  As a simplification, we ignore all the selection operators since it just takes $O(1)$ time to decide if a tuple passes the predicate $\phi_i$. \revm{Moreover, we assume every $R_i$ is defined on different subset of attributes; otherwise, we can simply keep the intersection of all relations defining on the same subset of attributes. Hence, we also use $R_e$ to denote the relation defined on $e \in \E$. }

The result of $\Q$ over instance $D$ noted as $\Q(D)$, is defined as:
\[\Q(D) = \left\{t \in \dom(\y): \exists t' \in \dom(\V), s.t. \pi_{\y} t' = t, \pi_{e_i} t \in R_i, \forall i \in [n]\right\},\]
i.e., the projection of all combinations of tuples from every relation onto $\y$, such that tuples in each combination have the same value(s) on the common attribute(s). 
Let $N = |D|$ be the input size, i.e., the total number of tuples in the input instance.
Let $\OUT = |\Q(D)|$ be the output size, i.e., the number of query results of $\Q$ over $D$. %We note that $\OUT$ is the size of underlying join results if $\Q$ is full, and $\OUT$ is the size of the projection of underlying join results onto the output attributes otherwise.

\smallskip \noindent {\bf Difference of Conjunctive Queries (DCQ).} %However, the difference operator makes the problem more complicated, and turns two monotonic queries into non-monotonic one.\footnote{\revm{$\Q$ is monotonic if for any pair of instances $D$,$D'$, if $D \subseteq D'$, $\Q(D) \subseteq \Q(D')$ holds.}}  In this work, we take the first step and study the problem of how to compute the difference of two conjunctive queries efficiently. 
A \DCQ\ $\Q_1 - \Q_2$ consists of two CQs without self-joins $\Q_1, \Q_2$ with the same output attributes. We also assume that the \DCQ\ does not have a self-join, i.e., there exists no pair of relations $R_i$ from $\Q_1$ and $R_j$ from $\Q_2$ such that $R_i, R_j$ are the same. Note that our algorithms presented in this work also applied to the case when self-join exists in \DCQ, but our lower bound assumes that no self-join exists. The input to a DCQ $\Q_1 - \Q_2$ is a pair of database instances $D_1, D_2$ defined for $\Q_1, \Q_2$ respectively\footnote{\revm{We distinguish the input  instances $D_1, D_2$ of $\Q_1, \Q_2$ for simplifying algorithmic description later, which is different from conventional definition. }}. The result of $\Q_1 - \Q_2$ over $D_1,D_2$ is $\Q_1(D_1) - \Q_2(D_2)$, i.e., tuples that appear in the result of $\Q_1$ over instance $D_1$, but not in the result of $\Q_2$ over instance  $D_2$. Let $N = |D_1| + |D_2|$ be the input size, i.e., the total number of tuples in both input instances. Let $\OUT = |\Q_1(D_1) - \Q_2(D_2)|$ be the output size.

In this paper, we adopt standard data complexity \cite{vardi1982complexity}; that is, we measure the complexity of algorithms with input size $N$ and output size $\OUT$, and assume the query size as a constant.

\subsection{Literature Review of CQ Evaluation}
\label{sec:literature}
\begin{figure}[t]
     \centering
    \includegraphics[scale=0.92]{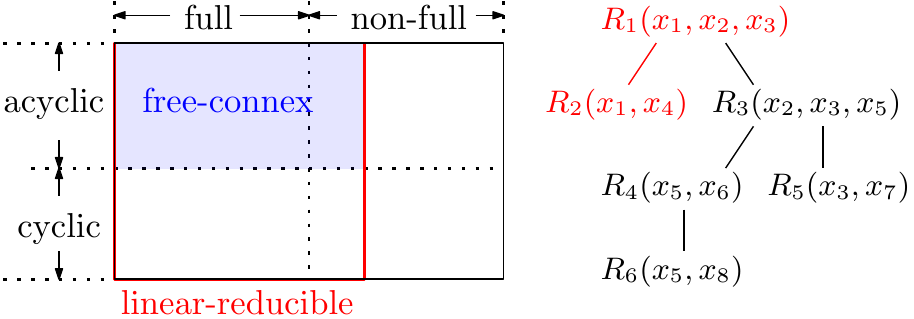}
    \caption{\revm{Left is the classification of CQs via $\alpha$-cyclic/cyclic and full/non-full metrics. Dashed lines indicate the boundaries of different classes.
    Right is a join tree of an $\alpha$-acyclic full CQ $\Q = (\V, \E)$ with $\V = \{x_1,x_2,\cdots, x_8\}$ and $\E = \{e_1 = \{x_1,x_2,x_3\}, e_2 =\{x_1, x_4\}, e_3 = \{x_2, x_3,x_5\}, e_4 = \{x_5, x_6\}, e_5 = \{x_3,$ $x_7\}$, $e_6 = \{x_5, x_8\}\}$. It is also a free-connex join tree for non-full CQ $\Q' = (\y,\V,\E)$ with $\y = \{x_1,  x_2, x_3,$ $x_4\}$, but is not a free-connex join tree with $\y = \{x_1, x_2, x_5\}$, since $\textsf{top}(x_3) = R_1$ is an ancestor of $\textsf{top}(x_5) = R_3$. The subtree in red is the reduced query of $\Q'$ after running Algorithm~\ref{alg:reduce}.
    }}
    \label{fig:classification}
 \end{figure}

Before diving into the massive literature,  we mention two  classes of CQs that play an important role in query evaluation. 
\begin{itemize}[leftmargin=*]
    \item {\bf ($\alpha$-acyclic).} A CQ $\Q = (\y, \V,\E)$ is $\alpha$-acyclic~\cite{beeri1983desirability, fagin1983degrees} if there exists a tree $\T$ (called a {\em join tree}) such that each node in $\T$ corresponds to a relation in $\E$, and for each attribute $x \in \V$, the set of nodes containing $x$ form a connected subtree of $\T$. Moreover, we define $\textsf{top}(x)$ as the highest node of $\T$ that attribute $x$ appears.
    \item {\bf (free-connex).} A CQ $\Q= (\y, \V,\E)$ is {\em free-connex}~\cite{bagan2007acyclic} if there exists a tree $\T$ (called a {\em free-connex join tree}) such that $\T$ is a join tree for $\Q$, and for any pair of attributes $x_1 \in \y, x_2 \in \V - \y$, $\textsf{top}(x_2)$ is not an ancestor of $\textsf{top}(x_1)$. It has been proved equivalently that a CQ $\Q= (\y, \V,\E)$ is {\em free-connex} if $\Q$ is $\alpha$-acyclic and $(\y, \V,\E \cup \{\y\})$ is also $\alpha$-acyclic. 
\end{itemize} 
 Their relationships are illustrated in Figure~\ref{fig:classification}. A free-connex CQ must be $\alpha$-acyclic. An $\alpha$-acyclic full join must be free-connex. Below, when the context is clear, we always refer ``acyclic'' to ``$\alpha$-acyclic''.

 There has been a long line of research on CQ evaluation~\cite{beeri1983desirability, kolaitis1998conjunctive, papadimitriou1997complexity, vardi1982complexity, chekuri2000conjunctive}.  Yannakakis's seminal algorithm~\cite{beeri1983desirability} was proposed for acyclic CQs, whose running time differs over different sub-classes of acyclic CQs. A free-connex CQ can be evaluated in $O(N+\OUT)$ time, which is already optimal since any algorithm needs to read input data and output all query results. On the other hand, an acyclic but non-free-connex CQ can be evaluated in $O(N \cdot \OUT)$ time. Subsequent works have progressively defined different notions of ``width''~\cite{gottlob2002hypertree, gottlob2009generalized}, measuring how far a query is from being acyclic and tackle cyclic queries with decomposition. This line of algorithms run in $O(N^w + \OUT)$ time, where $w$ can be the fractional hypertree width~\cite{gottlob2002hypertree, ngo2018worst}, submodular width~\cite{abo2016faq}, or FAQ-width~\cite{abo2016faq}. In addition, some specific classes of CQs can be speedup by fast matrix multiplication techniques~\cite{amossen2009faster, bjorklund2014listing, deep2020fast}, but we won't go into that direction further. CQ evaluation is still an actively researched problem; any improvement here will also improve \DCQ\ evaluation when plugged into the baseline as well as our approach.
 
 In the remaining, we often use $\texttt{cost}(\Q)$ to denote the time complexity of evaluating a CQ $\Q$.

\smallskip \noindent {\bf Implications to the Baseline Approach of DCQ Evaluation.}  Given a DCQ $\Q_1 - \Q_2$, the baseline approach of computing $\Q_1, \Q_2$ separately and then set difference incurs the following cost:
\begin{corollary}
\label{cor:baseline}
 Given two CQs $\Q_1$ and $\Q_2$, the \DCQ\ $\Q_1 - \Q_2$ can be computed in $O(\texttt{cost}(\Q_1) + \texttt{cost}(\Q_2))$ time.
\end{corollary}
\noindent 
For example, when both $\Q_1$ and $\Q_2$ are free-connex, the baseline approach runs in $O(N+ \OUT_1 + \OUT_2)$ time, where $\OUT_1, \OUT_2$ are the output sizes of $\Q_1, \Q_2$ respectively. %In this case, either $\OUT_1$ or $\OUT_2$ could be much larger than $\OUT$, hence the baseline would be too costly to find a few query results in their difference.  %Later, we will see how to overcome this limitation by exploring the joint structural properties of $\Q_1, \Q_2$.

\subsection{New Results of \DCQ\ Evaluation}
\label{sec:new-result}
Our new complexity results for \DCQ\ evaluation are summarized in Table~\ref{tab:summary}. To help understand these results, we first introduce the class of {\em linear-reducible} CQs, and the {\em reduce} procedure.
\begin{definition}[Linear-reducible]
\label{def:linear-reducible}
A CQ $\Q =(\y, \V, \E)$ is linear-reducible if $(\y, \V, \E \cup \{\y\})$ is free-connex.
\end{definition}
\noindent The relationship between linear-reducible CQs and existing classifications of CQs is illustrated in  Figure~\ref{fig:classification}. Any full or free-connex CQ must be linear-reducible.  In addition, some cyclic but non-full CQs are also linear-reducible, for example, $\Q= \pi_{x_1, x_2, x_3} (R_1(x_1,x_2) \Join R_2(x_2,x_3) \Join R_3(x_1, x_3) \Join R_4(x_3, x_4))$, since adding $R_5(x_1, x_2, x_3)$ will result in a free-connex CQ (see Figure~\ref{fig:classification}). It is also noted that any non-free-connex but acyclic CQ is non-linear-reducible.
 \begin{table}[t]
    \centering
    {\color{black}
    \begin{tabular}{c|c|c}
    \toprule
        $\Q_1 - \Q_2$ &  Baseline & Our Approach \\
        \hline
        \multirow{2}{*}{$\Q_1 - \Q_2$ is difference-linear} & \multirow{4}{*}{$\texttt{cost}(\Q_1)$} & \multirow{2}{*}{$N + \OUT$ \ [Theorem~\ref{the:easy}]} \\
        & \multirow{4}{*}{$+$} & \\
        \cline{1-1} \cline{3-3}
        \multirow{2}{*}{$\Q_2$ is linear-reducible} & \multirow{4}{*}{$\texttt{cost}(\Q_2)$} & \multirow{2}{*}{$\texttt{cost}(\Q_1)$ \ [Corollary~\ref{cor:heuristic1}]} \\
        & & \\
        \cline{1-1} \cline{3-3}
        $\Q_2$ is non-linear-reducible & & 
        \begin{minipage}{21em}
        \begin{displaymath}
        \texttt{cost}(\Q_1) \ + \ \min \left\{ \begin{array}{ll}
        \OUT_1 \cdot \texttt{cost}(\Q^\emptyset_2)  & \textrm{[Theorem~\ref{the:heuristics}]} \\
        \texttt{cost}(\Q^{\oplus}_2) & \textrm{[Theorem~\ref{the:heuristics2}]}
        \end{array} \right. 
        \end{displaymath}
        \end{minipage} \\
        \bottomrule
     \end{tabular}
     }
     \caption{\revm{Summary of complexity results by baseline and our approach. $\Q_1 = (\y,\V_1, \E_1)$ and $\Q_2 = (\y,\V_2, \E_2)$ are two input CQs. $(\y,\E'_1)$ and $(\y,\E'_2)$ are reduced queries of $\Q_1, \Q_2$ respectively. $\Q^\emptyset_2 = (\emptyset, \V - \y, \{e - \y: e \in \E_2\})$ and $\Q^{\oplus}_2 = (\y, \V_2, \{\y\} \cup \E_2)$ are formally defined in Section~\ref{sec:heuristic}. $N$ is the input size. $\OUT_1, \OUT$ are the output sizes of $\Q_1, \Q_1 - \Q_2$ respectively.  $\texttt{cost}(\cdot)$ is the time complexity of evaluating a single CQ.}}
     \label{tab:summary}
\end{table}

Moreover, we introduce a {\em reduce} procedure in Algorithm~\ref{alg:reduce}, that can transform any linear-reducible CQ into a full join query in $O(N)$ time, while preserving the query results. This algorithm is similar to the semi-join phase of Yannakakis algorithm~\cite{yannakakis1981algorithms}. Intuitively, we remove attributes or relations in a bottom-up ordering of nodes in a free-connex join tree. Recall that each node in the join tree corresponds to a relation in $\E$. When a node $e$ is visited, we distinguish two more cases:  (line 4-5) if its output attributes are fully contained in its parent, we remove $e$ and update its parent relation via semi-joins;  (line 6-7) and otherwise, we remove all non-output attributes (if there exists any) in $e$ via projections. The output of Algorithm~\ref{alg:reduce} is a full join query $\Q' = (\y, \E)$ (called the {\em reduced query}) and an instance $D'$ (called the {\em reduced instance}) 
 such that $\Q(D) = \Q'(D')$. An example of reduced query is illustrated in Figure~\ref{fig:classification}. 

We are now ready to present the new results for \DCQ\ evaluation.

\smallskip \noindent {\bf Dichotomy for Linear-time Algorithm.} Our main complexity result is a complete characterization of $\Q_1, \Q_2$ for which a linear algorithm can be achieved for computing $\Q_1 - \Q_2$: 

\begin{definition}[Difference-Linear]
     Given two CQs $\Q_1 = (\y, \V_1, \E_1)$ and $\Q_2 = (\y, \V_2, \E_2)$, the \DCQ\ $\Q_1 - \Q_2$ is difference-linear if $\Q_1$ is free-connex, $\Q_2$ is linear-reducible, and $(\y, \E'_1 \cup \{e\}\})$ is $\alpha$-acyclic for every $e \in \E'_2$, where $(\y,\E'_1)$ and $(\y, \E'_2)$ are the reduced queries of $\Q_1, \Q_2$ respectively.
\end{definition}
  \begin{theorem}[Dichotomy]
    \label{the:dichotomy}
    Given two CQs $\Q_1 = (\y, \V_1, \E_1)$ and $\Q_2 = (\y, \V_2, \E_2)$, the \DCQ\ $\Q_1 - \Q_2$ can be computed in $O(N + \OUT)$ time if and only if it is difference-linear.
\end{theorem}
Our proof of Theorem~\ref{the:dichotomy} consists of two steps.  In Section~\ref{sec:easy-algorithm}, we prove the ``if''-direction by designing a linear algorithm for the class of ``easy'' queries as characterized.  In Section~\ref{sec:hardness-proof}, we prove the ``only-if'' direction by showing the lower bound for the remaining class of ``hard'' queries, based on some well-established conjectures.

\paragraph{Improvement Achieved by Heuristics.} For the class of ``hard'' \DCQ s on which obtaining a linear-time algorithm is hopeless, we further show some efficient heuristics. The complete results are presented in Section~\ref{sec:heuristic} and here we mention an interesting case that our heuristics have strictly improved the baseline: 
 \begin{corollary}
 \label{cor:heuristic1}
    Given two CQs $\Q_1$ and $\Q_2$, if $\Q_2$ is linear-reducible, then \DCQ\ $\Q_1 - \Q_2$ can be computed in $O(\texttt{cost}(\Q_1))$ time.
 \end{corollary}

 %\begin{corollary}
 %\label{cor:heuristic2}
 %   Given two CQs $\Q_1$ and $\Q_2$, the \DCQ\ $\Q_1 - \Q_2$ can be computed in $O(\texttt{cost}(\Q_1) + \min\{\OUT_1 \cdot \texttt{cost}(\Q^\emptyset_2), \texttt{cost}(\Q^\oplus_2)\})$ time.
 %\end{corollary}
 
 We summarize all these results above in Table~\ref{tab:summary}: (1) our approach strictly improves the baseline as long as $\Q_2$ is linear-reducible; (2) furthermore, our approach leads to a linear-time algorithm if $\Q_1$ also satisfies some specific conditions; (3) in the remaining case when $\Q_2$ is non-linear-reducible, the comparison of our approach and baseline depends on specific queries or even input instances.

 \begin{algorithm}[t]
 \caption{\textsc{Reduce}$(\Q=(\y, \V, \E), D)$}
 \label{alg:reduce}
 \SetKwInOut{Input}{Input}
 \SetKwInOut{Output}{Output}
 
 $\T \gets$ the free-connex join tree of $(\y, \V, \E \cup \{\y\})$\;
 \ForEach{$e \in \E$ in a bottom-up way (excluding root) of $\T$}{
    Let $e' \gets $ the (unique) parent node of $e$\;
    \If{$e \cap \y \subseteq e' \cap \y$}{
            $R_{e'} \gets R_{e'} \ltimes R_e$ \textbf{ and } $\E \gets \E - \{e\}$;}
    \Else{\lIf{$e \not \subset \y$}{
        $R_e \gets \pi_{e \cap \y} R_e$ \textbf{ and } $e \gets e \cap \y$}
    }      
 }
 \Return $((\y, \E), D)$\;
 \end{algorithm}

%% file: easy.tex
\section{Easy DCQs}
\label{sec:easy-algorithm}
    In this section, we show a linear-time algorithm for computing the class of ``easy'' \DCQ\ characterized in Theorem~\ref{the:dichotomy}. The main technique we used is simply query rewriting, but by exploiting the structures of two input queries in a non-trivial way.
    \begin{theorem}
    \label{the:easy}
        Given two CQs $\Q_1=(\y, \V_1,\E_1)$ and $\Q_2 = (\y, \V_2, \E_2)$, if $\Q_1 - \Q_2$ is difference-linear, then \DCQ\ $\Q_1 - \Q_2$ can be computed in $O(N + \OUT)$ time.
    \end{theorem}

    We start with a special class of \DCQ s that two input CQs share the same schema. %which is simple enough to illustrate the high-level idea of our approach. %that can be extended to handle the class of all ``easy'' queries. %As mentioned in Section~\ref{sec:intro}, this class of \DCQ s has wide applications in data analytic, such as graph pattern mining, network analysis, etc.
    In Section~\ref{sec:homogeneous}, we introduce an algorithm based on query rewriting, which always pushes the difference operator down to the input relations and avoids materializing a large number of intermediate results that do not participate in the final query result. 
    In Section~\ref{sec:heterogeneous}, we move to general case that $\Q_1$ and $\Q_2$ can have different schemas.

  \subsection{$\Q_1$ and $\Q_2$ share the same \revm{schema}}
  \label{sec:homogeneous}
  
  \revm{We first note that if $\Q_1, \Q_2$ share the same schema, i.e., there is a one-to-one correspondence between the relations/attributes in $\Q_1$ and $\Q_2$,  Theorem~\ref{the:easy} degenerates to the following lemma:}
  \begin{lemma}
  \label{lem:homogeneous}
    Given two CQs $\Q_1= \Q_2 = (\y, \V,\E)$, if $(\y, \V, \E)$ is free-connex, then the \DCQ\ $\Q_1 - \Q_2$ can be computed in $O(N + \OUT)$ time.
  \end{lemma}
  %\noindent %For any free-connex CQ, adding a copy of an existing relation still leads to a free-connex CQ. Hence, the condition in Theorem~\ref{the:easy} is automatically satisfied. % that $(\y, \{e \cap \y: e \in \E_1\} \cup \{e'\cap \y\})$ is $\alpha$-acyclic for every $e \in \E_2$ is automatically satisfied
  %\xiao{$(\y, \E'_1 \cup \{e\}\})$ is $\alpha$-acyclic for every $e \in \E'_2$}, due to the fact that $\E'_1= \E'_2$. 
  
  %Now, given two instance $D, D'$ for $\Q_1, \Q_2$ separately, the target is to output the difference of $\Q_1(D) - \Q_2(D')$. For such a special \DCQ, our algorithm is based on query rewriting to transform the \DCQ\ into the union of multiple join queries while pushing the difference operator down to the base relations. 
  
  Let's start with an example falling into this special case. 
  
  %it suffices to show if $(\y, \V,\E)$ is free-connex, then sFor \DCQ\ $\Q_1- \Q_2$, we first assume that $\Q_1, \Q_2$ are full and share the same schema. For simplicity, denote $\Q_1 = \Q_2 = (\V,\E)$. Then, give two instances $D, D'$, our target is to output the difference of $\Q_1(D) - \Q_2(D')$. For such a special \DCQ, our algorithm is based on query rewriting to transform the difference query into the union of join queries involving the difference of base relations. 

  %\begin{definition}[Homogeneous DCQ]
  %  A DCQ $\Q_1 - \Q_2$ with $\Q_1 = (\V, \E_1)$ and $\Q_2 = (\V, \E_2)$ is homogeneous if $\E_1 = \E_2$. 
  %\end{definition}

 \begin{example}
 \emph{
    Consider a \DCQ\ $\Q_1 - \Q_2$ with $\Q_1 = R_1(x_1,x_2) \Join R_2(x_2,x_3)$ and $\Q_2 = R'_1(x_1,x_2) \Join R'_2(x_2,x_3)$. We can rewrite it as the union of two join queries: $\Q_1 - \Q_2 = (R_1 - R'_1) \Join R_2 + R_1 \Join (R_2 - R'_2)$, where the difference operator is only applied for computing $R_1 - R'_1$ and $R_2 - R'_2$. Intuitively, for every join result $(a,b,c) \in \Q_1 - \Q_2$, it must be $(a,b) \notin R'_1$ or $(b,c) \notin R'_2$; otherwise, $(a,b,c)\in \Q_2$, coming to a contradiction. The correctness of this rewriting will be formally presented in the proof of Lemma~\ref{lem:acylcic-join-rewrite}. In addition, the difference operators can be evaluated in $O(N)$ time, and the join operators can be evaluated in $O(N + \OUT)$ time. }
 \end{example}

\noindent {\bf Rewrite Rule.} We now generalize the rewriting rule in Example~\ref{exp:1} to general \DCQ\ $\Q_1 - \Q_2$ for $\Q_1 = \Q_2 = ( \y, \V, \E)$, where $\y = \V$ and $(\V,\E)$ is $\alpha$-acyclic. In other words, both $\Q_1, \Q_2$ correspond to the same acyclic full join query.  For any $e \in \E$, let $R_e, R'_e$ be the corresponding relations in $\Q_1, \Q_2$ respectively. \revm{Our rule is built on the observation that for any query result $t \in \Q_1 -\Q_2$, $\pi_{e} t \in R_e$ must hold for every $e \in \E$, but $\pi_e t \notin R'_e$ happens for some $e \in \E$.} %In this way, First, we must have $\pi_e t \in R_e$ for every $e \in \E$, since $t \in \Q_1$. By contradiction, assume $\pi_e t \notin R_e$ or $\pi_e t \in R_e$ for every $e \in \E$. If there exists some $e \in E$ such that $t \notin R_e$, then $t \notin \Q_1$, coming to a contradiction. Then, $\pi_e t \in R_e$ holds for every $e \in \E$ by assumption. This way, $t \in \Q_2$, comes to a contradiction.  
Applying this observation, we can rewrite such a \DCQ\ as
%\[\bigcup_{S\subsetneq \E_2} \left(\Join_{e \in S} \wedge R_e\right) \Join \left(\Join_{e \in \E_1 -S} \Delta R_e\right)\]
%\[\Q_1 - \Q_2 = \bigcup_{e \in \E} \left((R_e - R'_e) \Join (\Join_{e' \in \E-\{e\}} R_{e'})\right)\]
the (disjoint) union of a constant number of join queries as follows:

\begin{lemma}
\label{lem:acylcic-join-rewrite}
    Given two CQs $\Q_1 = \Q_2 = (\y,\V,\E)$, if $(\V,\E)$ is $\alpha$-acyclic and $\y = \V$, $\Q_1 - \Q_2 = \bigcup_{e \in \E} \left((R_e - R'_e) \Join (\Join_{e' \in \E -\{e\}} R_{e'}) \right)$.
\end{lemma}

\noindent {\bf Algorithm and Complexity.} An algorithm directly follows the rewriting rule above.  It first computes the difference of every pair of input relations, i.e., $R_e - R'_e$ for each $e \in \E$, and then computes a full join query $(R_e - R'_e) \Join \Q_1$ derived for each $e \in \E$. 
Actually, we can handle a slightly larger class of \DCQ.  For $\Q_1 = \Q_2 = (\y,\V,\E)$, if $(\y,\V,\E)$ is free-connex, we simply remove all non-output attributes for $\Q_1, \Q_2$ separately in the preprocessing step, and then tackle two acyclic full joins, that share the same structure.

\revm{As the pre-processing step and difference operators can be evaluated in $O(N)$ time, this algorithm is bottlenecked by evaluating the join query $(\V,\E)$, which takes $O(N + \OUT)$ time. Putting everything together, we come to Lemma~\ref{lem:homogeneous}.}

\subsection{$\Q_1$ and $\Q_2$ have different \revm{schemas}}
\label{sec:heterogeneous}

 \begin{algorithm}[t]
 \SetAlgoVlined
 \caption{{\sc EasyDCQ}$(\Q_1, \Q_2, D_1, D_2)$}
 \label{alg:easy-dcq}

 \SetKwInOut{Input}{Input}
 \SetKwInOut{Output}{Output}
 
 \revm{\lIf{$\y \neq \V_1$}{$((\y, \E_1), D_1) \gets \textsc{Reduce}(\Q_1, D_1)$}
 \lIf{$\y \neq \V_2$}{$((\y,\E_2), D_2) \gets \textsc{Reduce}(\Q_2, D_2)$}}
 $\mathcal{S} \gets \emptyset$\;
 \ForEach{$e \in \E_2$}{
    $S_e \gets \textsc{Yannakakis}((e, \y, \E_1), D_1)$\;
    $\mathcal{S} \gets \mathcal{S} \cup 
    \textsc{Yannakakis}((\y, \E_1 \cup \{e\}), D_1 \cup \{S_e -R'_e\})$\;
 }
 \Return $\mathcal{S}$\;
 \end{algorithm}

We next move to the general case when these two input CQs have different schemas.  %Recall that the class of \DCQ that can be evaluated in linear time must have $\Q_1 = (\y, \V_1, \E_1)$ as free-connex, $\Q_2 = (\y, \V_2, \E_2)$ as linear-reducible, and \xiao{$(\y, \E'_1 \cup \{e\}\})$ is $\alpha$-acyclic for every $e \in \E'_2$, where $(\y,\E'_1)$ and $(\y, \E'_2)$ are the reduced queries of $\Q_1, \Q_2$ respectively. } %$(\y, \V_1, \E_1 \cup \{e\})$ as free-connex for every $e \in \E_2$ if $e \subseteq \y$. 
%To obtain such a linear algorithm, 
We focus on the case when both $\Q_1, \Q_2$ are full and then extend to non-full case. 

\smallskip \noindent {\bf \DCQ\ with full CQs.} Now, we assume that $\y = \V_1 = \V_2 = \V$. Theorem~\ref{the:easy} simply degenerates to the Lemma~\ref{lem:easy-full}.

\begin{lemma}
\label{lem:easy-full}
    Given two full joins $\Q_1 = (\V, \E_1)$ and $\Q_2 = (\V, \E_2)$, if $\Q_1, \Q_2$ are $\alpha$-acyclic, and $(\V, \E_1 \cup \{e\})$ is $\alpha$-acyclic for every $e \in \E_2$, then $\Q_1 - \Q_2$ can be computed in $O(N + \OUT)$ time.
\end{lemma}

\revm{A straightforward solution is to transform both $\Q_1 = (\V,\E_1)$ and $\Q_2 = (\V, \E_2)$ into one auxiliary query $(\V,\E_1 \cup \E_2)$, and then invoke the algorithm in Section~\ref{sec:homogeneous} to handle the degenerated case. However, this solution does not necessarily lead to a linear-time algorithm. Let's gain some intuition from the example below.}

\begin{example}
\label{exp:q2}
 \emph{Consider a \DCQ\ $\Q_1 - \Q_2$ with $\Q_1 = R_1(x_1,x_2) \Join R_2(x_2,x_3,x_4)$ and $\Q_2= R_3(x_1,x_2,x_3) \\ \Join R_4(x_3,x_4)$. For an auxiliary query, we introduce the following intermediate relations $R_5 = R_1 \Join \pi_{x_2,x_3} R_2$, $R_6 = \pi_{x_3,x_4} R_2$, $R_{7} = \pi_{x_1, x_2} R_3$ and $R_8 = \pi_{x_2, x_3} R_3 \Join R_4$. Then, we can rewrite $\Q_1, \Q_2$ as follows:
    \begin{align*}
        \Q_1 =& R_1(x_1,x_2) \Join R_2(x_2,x_3,x_4) \Join R_5(x_1,x_2,x_3) \Join R_6(x_3,x_4) \\
        \Q_2 =& R_7(x_1,x_2) \Join R_8(x_2,x_3,x_4) \Join R_3(x_1,x_2,x_3) \Join R_4(x_3,x_4)
    \end{align*}
    Then, we are left with two queries that share the same \revm{schema}.
    However, this strategy does not necessarily lead to a linear-time algorithm, since materializing the intermediate relation $R_8$ requires super-linear time, which could be much larger than the final output size $\OUT$. % The reason why materializing $R_5$ does not incur any problem will be explained later.
    }
    
    \emph{Careful inspection reveals that a simpler rewriting rule can avoid materializing $R_8$.  More specifically, we keep $\Q_2$ unchanged and rewrite $\Q_1$ as above.  Then, $\Q_1 - \Q_2$ can be rewritten as $(R_5- R_3) \Join R_1 \Join R_2 + R_1 \Join R_2 \Join (R_6 - R_4)$. Intuitively, for every join result $(a,b,c,d) \in \Q_1 - \Q_2$, it must be $(a,b,c) \notin R_3$ or $(c,d) \notin R_4$; otherwise, $(a,b,c,d) \in \Q_2$, coming to a contradiction. The correctness of this rewriting will be formally presented in the proof of Lemma~\ref{lem:heterogenous-rewrite}.  In this case, materializing $R_6$ only takes $O(N)$ time, but materializing $R_5$ might take super-linear time.  Fortunately, we can bound the size of $R_5$ by $O(N + \OUT)$. The rationale is that every tuple in $R_5 - R_3$ will participate in at least on one join result of $(R_5 - R_3) \Join R_1 \Join R_2$, i.e., the final result of the difference query $\Q_1 - \Q_2$, thus $|R_5 - R_3| \le \OUT$. For the difference operator, $R_5 - R_3$ takes $O(N + \OUT)$ time, and $R_6 - R_4$ takes $O(N)$ time. For the join operator, both simple join queries take linear time in terms of their input size and output size.  Overall, this rewriting rule can compute the example query in $O(N + \OUT)$ time.}
\end{example}

%We next investigate the complexity of this solution. First, both $R_{21}, R_{22}$ have input size bounded by $O(N)$, and can be computed in $O(N)$ time. Moreover, we observe that the subquery $R_1 \Join R_{21}$ has its size bounded by $O(N + \OUT)$ and can be computed in $O(N + \OUT)$ time as well. The argument is that each tuple in $R_1 \Join R_{21} - R_3$ can be joined with at least one tuple in $R_1$ and $R_2$, thus it must participate in at least one query result of $\Q_1 - \Q_2$. This way, we have $|R_1 \Join R_{21} - R_3| \le \OUT$, hence $|R_1 \Join R_{21}| \le N + \OUT$.  Now, we are left with two acyclic queries, where the participated relations have their input sizes bounded by $O(N + \OUT)$. Implied by Theorem~\ref{the:homogeneous-join}, the final query result of $\Q_1- \Q_2$ can be computed in $O(N + \OUT)$ time.

\smallskip \noindent {\bf Rewrite Rule.} Generalizing this observation, we develop the following rewriting rule for arbitrary full joins $\Q_1, \Q_2$. The high-level idea is to introduce an intermediate relation $S_e = \pi_e \Q_1$ for every $e \in \E_2$, i.e., the projection of join results of $\Q_1$ onto attributes $e$. Now we can rewrite $\Q_1- \Q_2$ using input relations in $\Q_1$ and intermediate relations corresponding to $\E_2$, as well as input relations in $\Q_2$, which results in the disjoint union of multiple full joins.

\begin{lemma}
\label{lem:heterogenous-rewrite}
    Given two CQs $\Q_1 = (\y, \V_1, \E_1)$ and $\Q_2 = (\y,\V_2, \E_2)$, if $\y = \V_1 = \V_2$, $\Q_1 - \Q_2 = \bigcup_{e \in \E_2} \left((S_e -R'_{e})\Join \Q_1 \right)$, for $S_e = \pi_e \Q_1$.
\end{lemma}

\begin{proof}
    \revm{\underline{\emph{Direction $\subseteq$}}.} Consider an arbitrary result $t \in \Q_1 - \Q_2$. By definition, $\pi_{e_1} t \in R_{e_1}$ for every $e_1 \in \E_1$, and $\pi_{e_2} t \notin R'_{e_2}$ for some $e_2 \in \E_2$. Moreover, $\pi_{e_2} t \in S_{e_2} = \pi_{e_2} \Q_1$. So, $t \in (S_{e_2} - R'_{e_2}) \Join \Q_1$. \revm{\underline{\emph{Direction $\supseteq$}}.} Consider an arbitrary $e_2 \in \E_2$ and a query result $t \in (S_{e_2} - R'_{e_2}) \Join \Q_1$. By definition, $t \in \Q_1$ and $t \notin R'_{e_2}$, which further implies $t \notin \Q_2$. This way, $t \in \Q_1 - \Q_2$.
\end{proof}

  \begin{algorithm}[t]
 \caption{\textsc{Yannakakis}$(\Q, D)$~\cite{yannakakis1981algorithms}}
  \label{alg:yannakakis}
 \SetKwInOut{Input}{Input}
 \SetKwInOut{Output}{Output}
  $(\Q,D) \gets \textsc{Reduce}(\Q, D)$\;
  $\T \gets$ the free-connex join tree of $\Q$ rooted at $r$\;
   \ForEach{$v \in \T$ in a bottom-up way (excluding root)}{
    %Let $u \gets $ the (unique) parent node of $v$\; 
    $R_u \gets R_u \ltimes R_v$ for the parent node $p(u)$ of $u$\;
   }
   \ForEach{$u \in \T$ in a top-down way (excluding leaves)}{
    \lForEach{$v$ is a child of $u$}{
       $R_u \gets R_u \ltimes R_v$}
 }
 
 \ForEach{$v \in \T$ in a bottom-up way (excluding root)}{
    %Let $u \gets $ the (unique) parent node of $v$\; 
    $R_u \gets R_u \Join R_v$ for the parent node $p(u)$ of $u$\;
}
 \Return $R_r$\;
 \end{algorithm}

\noindent {\bf Algorithm and Complexity.} An algorithm for computing the difference of two full join queries follows the rewriting rule above. %As a preprocessing step, we first remove dangling tuples in $Q_1$ by invoking a standard reduce procedure (as described in the Algorithm). 
For each $e \in \E_2$, it first materializes the query results of $\pi_e \Q_1$, then computes the difference operator $\pi_e \Q_1 - R'_e$, and finally the full join $(\pi_e \Q_1 - R'_e) \Join \Q_1$ by invoking the classical Yannakakis algorithm. We next analyze the complexity of the algorithm above. To establish the complexity, we first show an upper bound on the size of any intermediate relation constructed:

\begin{lemma}
\label{lem:new-relation}
    $|S_e| = O(N+\OUT)$ for any $e \in \E_2$, where $S_e = \pi_e \Q_1$.
\end{lemma}
    
\begin{proof}
    Consider an arbitrary tuple $t \in S_e - R'_e$. First, $t$ participates in at least one query result of $\Q_1$.  As $t \in S_e$, $t \in \pi_{e} \Q_1$ by definition. There must exist some tuple $t' \in \Q_1$ such that $\pi_{e} t' =t$. Thus, $t$ participates in some query results of $\Q_1$.  Meanwhile, $t$ does not participate in any query result of $\Q_2$, since $t \notin R'_e$. In this way, $t$ participates in at least one result in $\Q_1 - \Q_2$, thus $|S_e - R'_e| \le \OUT$. Moreover, $|R'_e| \le N$. Together, we obtain $|S_e| = O(N+\OUT)$.
\end{proof} 
 
 Let $\V = \V_1 = \V_2$. If $(\V,\E_1)$ is $\alpha$-acyclic, and $(\V,\E_1 \cup \{e\})$ is also $\alpha$-acyclic for every $e \in \E_2$, then the constructed CQ $\pi_e \Q$ is free-connex. Implied by the existing result on CQ evaluation, $S_e$ can be computed in $O(N+|S_e|) = O(N + \OUT)$ time by the classic Yannakakis algorithm, where $\OUT$ is the output size of the difference query!  The invocation of Yannakakis algorithm here is crucial for achieving linear complexity.  For example, if $S_e$ is computed by first materializing the query results of $\Q_1$ and then computing their projection onto $e$, the time complexity would be as large as $O(N + \OUT_1)$, where $\OUT_1$ is the output size of $\Q_1$.  Now, each full join $(S_e -R'_e) \Join \Q_1$ is $\alpha$-acyclic with input size $O(N+\OUT)$ and output size $\OUT$, thus can be computed in $O(N + \OUT)$ time. Therefore, the total time complexity is bounded by $O(N+\OUT)$, since there are $O(1)$ sub-queries in $\Q_1 - \Q_2$.  Putting everything together, we come to Lemma~\ref{lem:easy-full}.

 \smallskip \noindent {\bf \DCQ\ with general CQs.} Now, we are ready to present an linear-time algorithm for computing $\Q_1 -\Q_2$, such that $\Q_1$ is free-connex, $\Q_2$ is linear-reducible, and \xiao{$(\y, \E'_1 \cup \{e\}\})$ is $\alpha$-acyclic for every $e \in \E'_2$, where $(\y,\E'_1)$ and $(\y, \E'_2)$ are the reduced queries of $\Q_1, \Q_2$ respectively.} %$(\y, \{e \cap \y: e \in \E_1\} \cup \{e'\cap \y\})$ is $\alpha$-acyclic for every $e' \in \E_2$. 
 As described in Algorithm~\ref{alg:easy-dcq}, we first apply a preprocessing step to $\Q_1$ and $\Q_2$ (line 1-4), which removes non-output attributes in $\Q_1$ and $\Q_2$ if they are non-full. % as well as dangling tuples that do not participate in the final query results of $\Q_1$. 
 
 As shown in Algorithm~\ref{alg:reduce}, this reduce step is quite standard by first building a free-connex join tree for the derived query $(\y, \V, \E \cup \{\y\})$, and then traversing the tree in a bottom-up way.  In the traversal, when a relation is visited and contains some non-output attributes, we just update its parent relation by applying a semi-join and removing it.  Note that if a relation does not contain any non-output attribute, then its ancestor also does not contain any, implied by the property of the free-connex join tree.  Thus, the residual tree is a connected subtree that contains the root.  Note that no physical relation is defined to $\y$, but this is not an issue since when such a relation is visited, Algorithm~\ref{alg:reduce} simply skips it (line 4) as well as its ancestors.  This algorithm only takes $O(N)$ time.
 
 Then, we are left with two full joins, and invoke our rewriting rule proposed in Section~\ref{sec:heterogeneous} (line 6-8).  As the reduce procedure takes $O(N)$ time, and the join phase takes $O(N+ \OUT)$ time implied by Lemma~\ref{lem:easy-full}, we can obtain the complexity result in Theorem~\ref{the:easy}. 
 
 \revm{\paragraph{Improvement over Baseline.} When $\Q_1,\Q_2$ fall into the class of ``easy'' \DCQ s as characterized by Theorem~\ref{the:easy}, our algorithm only takes $O(N+ \OUT)$ time for computing $\Q_1 - \Q_2$, while the baseline takes $O\left(N+ \OUT_1 + \texttt{cost}(\OUT_2)\right)$ time, since $\texttt{cost}(\Q_1) = O(N+ \OUT)$ for free-connex $\Q_1$.} We next use a few examples of ``easy'' \DCQ s to illustrate the improvement achieved by our approach.

\begin{example}
\label{exp:1}
\emph{Consider a \DCQ\ with $\Q_1 = R_1(x_1,x_2,x_3)$ and $\Q_2 = R_2(x_1,x_2) \Join R_3(x_2,x_3) \Join R_4(x_1,x_3)$. The baseline takes $O\left(N^{\frac{2 \cdot \omega}{\omega+1}} + N^{\frac{3(\omega-1)}{\omega+1}} \cdot \OUT_2^{\frac{3-\omega}{\omega+1}}\right)$ time to compute the triangle join $R_2 \Join R_3 \Join R_4$ in $\Q_2$, where $\omega$ is the exponent of fast matrix multiplication. In contrast, our approach only takes $O(N)$ time since $\OUT \le N$, improving the baseline by a factor of $O\left(N^{\frac{\omega-1}{\omega+1}} + N^{\frac{2\omega-4}{\omega+1}} \cdot \OUT_2^{\frac{3 - \omega}{\omega+1}}\right)$.}
\end{example}

\begin{example}
\label{exp:3}
    \emph{Consider a \DCQ\ with $\Q_1 = R_1(x_1,x_2) \Join  R_2(x_3,x_4)$ and $\Q_2 = R_3(x_1,x_2) \Join R_4(x_2,x_3) \Join R_5(x_1,x_3)$. The baseline takes $O(N^2)$ time to materialize $\Q_1$, which degenerates to the Cartesian product of $R_1$ and $R_2$.  In contrast, our approach only requires $O(N + \OUT)$ time, improving the baseline by a factor of $O\left(\frac{N^2}{\OUT}\right)$, since $\OUT$ can be much smaller than $N^2$. }
\end{example}

\begin{example}
\label{exp:2}
    \emph{Consider a \DCQ\ with $\Q_1 = \Join_{e \subseteq U: |e|=1} R_e(\{x_1\} \cup e)$ and $\Q_2 = \Join_{e' \subseteq U: |e'|=2} R_{e'}(\{x_1\} \cup e')$ for $U = \{x_2,\cdots, x_{k+1}\}$. The baseline takes $O(N)$ time to materialize $\Q_1$, and $O(N^{\frac{k}{2}})$ time to materialize $\Q_2$. In contrast, our approach can compute it in $O(N + \OUT)$ time, improving the baseline by a factor of $O\left(\frac{N^{k/2}}{\OUT}\right)$, since $\OUT$ can be much smaller than $N^\frac{k}{2}$. }
\end{example}
  

%% file: hard.tex
\section{Hard \DCQ s}
\label{sec:hardness}

In this section, we turn to the class of ``hard'' \DCQ s characterized by Theorem~\ref{the:dichotomy}. We first prove the hardness of computing \DCQ s in linear time via some well-known conjectures, and then show an efficient heuristic for hard \DCQ s by further exploiting the query structures. 

\subsection{Hardness}
\label{sec:hardness-proof}

We will prove the hardness of computing a hard \DCQ\ $\Q_1 - \Q_2$, in particular: (1) $\Q_1$ is non-free-connex; or (2) $\Q_1$ is free-connex but $\Q_2$ is non-linear-reducible; or (3) $\Q_1$ is free-connex, $\Q_2$ is linear-reducible, but there exists some $e \in \E'_2$ such that \xiao{$(\y, \E'_1 \cup \{e\}\})$ is cyclic, where $(\y,\E'_1)$ and $(\y, \E'_2)$ are the reduced queries of $\Q_1 = (\y,\V_1, \E_1), \Q_2 = (\y,\V_2, \E_2)$ respectively. }
We will prove the hardness for each class of hard \DCQ s separately.

\smallskip \noindent {\bf Hardness-(1).} The hardness of computing \DCQ s in case (1) comes from computing a non-free-connex \CQ~\cite{bagan2007acyclic}. %which is built on some well-known conjectures such as {\em matrix multiplication conjecture}\footnote{Matrix Multiplication Conjecture: Given two Boolean matrix $M_1, M_2$ of size $n \times n$, any algorithm computes $M_1 \times M_2$ requires $\Omega(n^2)$ time.} and {\em listing triangle conjecture}\footnote{Listing Triangle Conjecture: Listing $m$ triangles in an $m$-edge graph requires $\Omega(m^{4/3})$ time, assuming the 3SUM conjecture.}~\cite{patrascu2010towards}.
By setting the result of $\Q_2$ as $\emptyset$, $\Q_1 - \Q_2$ simply degenerates to $\Q_1$, hence we obtain: 

\begin{lemma}
\label{lem:hardness-1}
    For any \DCQ\ $\Q_1 -\Q_2$, if $\Q_1$ is non-free-connex, any algorithm computing $\Q_1 - \Q_2$ requires at least $\Omega(N + \OUT)$ time. 
\end{lemma}

The hardness of case (2) and (3) is built on the {\em strong triangle conjecture} in the literature:

\begin{conjecture}[Strong Triangle conjecture~\cite{abboud2014popular}]
Detecting whether an $n$-node $m$-edge graph contains a triangle requires $\Omega\left(\min\left\{n^{\omega - o(1)}, m^{2\omega/(\omega+1)-o(1)}\right\}\right)$ time in expectation, where $\omega = 2+ o(1)$ is assumed as the exponent of fast matrix multiplication.
\end{conjecture}

\smallskip
\noindent {\bf Hardness-(2).} We start with  two hardcore \DCQ s in Lemma~\ref{lem:hard-2} and Lemma~\ref{lem:hard-3}. The proof of Lemma~\ref{lem:hardness-2} for general \DCQ s in case (2) is given in Appendix~\ref{appendxi:hardness-proof}.

\begin{lemma}
\label{lem:hard-2}
    Any algorithm for computing the following \DCQ: \[\Q_1 - \Q_2  = R_1(x_1,x_3)- \pi_{x_1,x_3} \left(R_2(x_1,x_2) \Join R_3(x_2,x_3)\right)\] 
    requires $\Omega(N)$ time, assuming the strong triangle conjecture.
\end{lemma} 

\begin{proof}
    For a graph $G=(V,E)$, we denote $m = |E|$ and $n = |V|$. Note that $n<m<n^2$; otherwise, we simply remove vertices that do not incident to any edges in $G$. We then construct an instance $D_1, D_2$ for $\Q_1, \Q_2$ by setting $R_1 = R_2 = R_3 = E$. Hence, $N = m$. Note that there exists some triangle in $G$ if and only if $\Q_1 \cap \Q_2$ is non-empty. Together with $\Q_1 \cap \Q_2 = \Q_1 - (\Q_1 - \Q_2)$, we output ``a triangle is detected in $G$'' if and only if $|\Q_1 - \Q_2| < N$. 
    If $\Q_1 - \Q_2$ can be computed in $O(N)$ time, whether there exists a triangle in $G$ can be detected in $O(\min\{n^2, m^{4/3}\})$ time, coming to a contradiction of strong triangle conjecture.
\end{proof}

\begin{lemma}
\label{lem:hard-3}
    Any algorithm for computing the following \DCQ:  \[\Q_1 - \Q_2  = R_1(x_1)- \pi_{x_1} \left(R_2(x_1,x_3) \Join R_3(x_2,x_3) \Join R_4(x_1, x_3)\right),\] requires $\Omega(N)$ time, assuming the strong triangle conjecture.
\end{lemma}

\begin{proof}
    This is similar to the proof of Lemma~\ref{lem:hard-2}. For a graph $G = (V, E)$, we construct $R_2 = R_3 = R_4 = E$ and $R_1 = V$, with $m = |E| = N$ and $n = |V|$. Note that there exists some triangle in $G$ if and only if $\Q_1 \cap \Q_2$ is non-empty. Together with $\Q_1 \cap \Q_2 = \Q_1 - (\Q_1 - \Q_2)$, we output ``a triangle is detected in $G$'' if and only if $|\Q_1 - \Q_2| < |R_1|$. If $\Q_1 - \Q_2$ can be computed in $O(N)$, whether there exists a triangle in $G$ can be detected in $O(\min\{n^2, m^{4/3}\})$ time, coming to a contradiction of strong triangle conjecture. 
\end{proof}

\begin{lemma}
\label{lem:hardness-2}
    Given two CQs $\Q_1, \Q_2$, if $\Q_1$ is free-connex and $\Q_2$ is non-linear-reducible, any algorithm computing $\Q_1 - \Q_2$ requires $\Omega(N + \OUT)$ time, assuming the strong triangle conjecture.
\end{lemma} 

\noindent {\bf Hardness-(3).} The hardness of evaluating a \DCQ\ in case (3) inherits the hardness of {\em deciding} a \DCQ: given a \DCQ\ $\Q_1-\Q_1$ and input databases $D_1,D_2$, the {\em decidability} problem asks to decide whether there exists a query result in $\Q_1- \Q_2$. We identify a few hardcore \DCQ s in Lemma~\ref{lem:hard-4}. The proof of Lemma~\ref{lem:hardness-3} for general \DCQ s in case (3) is given in Appendix~\ref{appendxi:hardness-proof}.

\begin{lemma}
\label{lem:hard-4}
	Any algorithm for deciding the following \DCQ        \begin{align*}
	\Q_1 -\Q_2 &= R_1(x_1, x_2) \Join R_2(x_2, x_3)
        - R_3(x_1, x_3) \Join R_4(x_2) \\
        \Q_1 -\Q_2 &= R_1(x_1, x_2) \Join R_2(x_2, x_3) - R_3(x_1, x_3) \Join R_4(x_2,x_3)\\
        \Q_1 -\Q_2 &= R_1(x_1, x_2) \Join R_2(x_2, x_3) - R_3(x_1, x_3) \Join R_5(x_1, x_2)\\
        \Q_1 -\Q_2 &= R_1(x_1, x_2) \Join R_2(x_2, x_3) -R_3(x_1, x_3) \Join R_4(x_2,x_3) \Join R_5(x_1, x_2)
	\end{align*}
	requires $\Omega(N)$ time, assuming the strong triangle conjecture.
\end{lemma}

\begin{proof}
    We first focus on the first \DCQ\, and the remaining ones can be proved similarly.
    Given an arbitrary graph $G = (V,E)$ with $m = |E|$ and $n = |V|$, we develop an algorithm to detect whether there exists a triangle in $G$. Note that $n<m<n^2$; otherwise, we simply remove vertices that do not incident to any edges in $G$. The degree $\textrm{deg}(u)$ of a vertex $u \in V$ is defined as the size of neighbors of $u$, i.e., those incident to $u$ with an edge in $E$. We partition vertices in $V$ into two subsets: $V^H = \{v \in V: \textrm{deg}(v) > m^{1/3}\}$ and $V^L=V-V^H$. From $G$, we construct following relations: $R=E$, $R_0 =\{(u,v) \in E: u \in V^L \textrm{ or } v \in V^L\}$, $R_1 = \{(u,v) \in E: u \in V^H\}$, $R_2 = \{(u,v) \in E: v \in V^H\}$ and $R_3 = V^H \times V^H - E$. Set $N = m^{4/3}$. It can be easily checked that each relation contains at most $m^{4/3}$ tuples, hence $N = m^{4/3}$. We further define a \CQ\ $\Q$ as follows: 
    \begin{align*}
	& \Q = R(x_1, x_2) \Join R(x_2, x_3) \Join R_0(x_1, x_3) 
    \end{align*}
    
    For $\Q_1 -\Q_2 = R_1(x_1, x_2) \Join R_2(x_2, x_3) - R_3(x_1, x_3) \Join R_4(x_2)$, we set $R_4 = V$ and output ``a triangle is detected'' if and only if $\Q$ or $\Q_1 - \Q_2$ is not empty. We first prove the correctness of this algorithm, i.e., a triangle exists in $G$ if and only if $\Q$ or $\Q_1-\Q_2$ is not empty. \revm{\underline{\emph{Direction Only-If}}.} Consider an arbitrary triangle $(u,v,w)$ in $G$. We distinguish two cases: (\romannumeral 1) at least one of $u,w$ is light; (\romannumeral 2) both $u$ and $w$ are heavy. In (\romannumeral 1), assume $u$ is light. Then, $(u,v), (u,w) \in R_0$. We come to $(u,v,w) \in Q$. In (\romannumeral 2), $(u,v) \in R_1$,$(v,w) \in R_2$, $(u,w) \notin R_3$, so we come to $(u,v,w) \in \Q_1-\Q_2$. \revm{\underline{\emph{Direction If}}.} If $\Q \neq \emptyset$, say $(u,v,w) \in \Q$, then $(u,v), (v,w), (u,w) \in E$ and therefore $(u,v,w)$ is a triangle in $G$. If $\Q_1-\Q_2 \neq \emptyset$, say $(u,v,w) \in \Q_1-\Q_2$, then $(u,v) \in R_1, (v,w) \in R_2, (u,w) \notin R_3$, i.e., $(u,v), (v,w), (u,w) \in E$, and therefore $(u,v,w)$ is a triangle in $G$. 
    
    We next turn to the time complexity. All statistics and relations $R_1, R_2, R_4$ can be computed in $O(m)$ time. Moreover, relation $R_3$ can be constructed in $O(m^{4/3})$ time since $|V^H| = O(m^{2/3})$. $\Q$ can be evaluated in $O(m^{4/3})$ time, since each of $R_0(x_1, x_2) \Join R(x_2, x_3)$ generates at most $O(m^{4/3})$ intermediate join results if $x_2 \in V^L$, and each of $R_0(x_1, x_2) \Join R(x_1, x_3)$ generates at most $O(m^{4/3})$ intermediate join results if $x_1 \in V^L$. If $\Q_1-\Q_2$ can be decided in $O(N)$ time, whether there exists a triangle in $G$ can be decided in $O(m^{4/3})$ time, coming to a contradiction to strong triangle conjecture.

    For $\Q_1 - \Q_2 = R_1(x_1, x_2) \Join R_2(x_2, x_3) - R_3(x_1, x_3) \Join R_4(x_2, x_3)$, we set $R_4 = E$ and output ``a triangle is detected'' if and only if $\Q$ or $\Q_1 - \Q_2$ is not empty. For $\Q_1 - \Q_2 = R_1(x_1, x_2) \Join R_2(x_2, x_3) - R_3(x_1, x_3) \Join R_5(x_1, x_2)$, we set $R_5 =R$ and output ``a triangle is detected'' if and only if $\Q$ or $\Q_1 - \Q_2$ is not empty. For $\Q_1 - \Q_2 = R_1(x_1, x_2) \Join R_2(x_2, x_3) - R_3(x_1, x_3) \Join R_4(x_2, x_3) \Join R_5(x_1, x_2)$, we set $R_4 = R_5 = E$ and output ``a triangle is detected'' if and only if $\Q$ or $\Q_1 - \Q_2$ is not empty. Similarly, $\Q$ can be computed in $O(m^{4/3})$ time. This way, if $\Q_1-\Q_2$ can be decided in $O(N)$ time, whether there exists a triangle in $G$ can be decided in $O(m^{4/3})$ time, coming to a contradiction to strong triangle conjecture. Together, we have completed the proof. 
\end{proof}

\begin{lemma}
\label{lem:hardness-3}
   Given two CQs $\Q_1 = (\y, \V_1, \E_1), \Q_2 = (\y, \V_2, \E_2)$, if $\Q_1$ is free-connex, $\Q_2$ is linear-reducible, and there exists some $e \in \E'_2$ such that \xiao{$(\y, \E'_1 \cup \{e\}\})$ is cyclic where $(\y,\E'_1)$ and $(\y, \E'_2)$ are the reduced queries of $\Q_1, \Q_2$ respectively, }
   any algorithm computing $\Q_1 - \Q_2$ requires $\Omega(N + \OUT)$ time, assuming the strong triangle conjecture. 
\end{lemma}

\subsection{Efficient Heuristics}
\label{sec:heuristic}

Although the hardness results in Section~\ref{sec:hardness} have ruled out a linear-time algorithm for the ``hard'' \DCQ s, we find that it is still possible to explore efficient heuristics that can outperform the baseline approach. Our heuristic is based on a simple fact that $\Q_1 - \Q_2 = \Q_1 - \Q_1 \cap \Q_2$. After computing the query results for $\Q_1$, a straightforward way of deciding $\Q_1 \cap \Q_2$ is to decide for each result $t \in \Q_1$, whether $t\in \Q_2$ nor not. This decidability query can be viewed as a special Boolean query by replacing every output attribute $x \in \y_2$ with a constant $\pi_{x} t$. More specifically, for $\Q_2 = (\y, \V_2, \E_2)$, the derived a Boolean query can be represented as $(\emptyset, \V_2 - \y, \{e - \y: e\in \E_2\})$. Putting everything together, we come to Theorem~\ref{the:heuristics}. 
 \begin{theorem}
 \label{the:heuristics}
    Given two CQs $\Q_1 = (\y, \V_1, \E_1)$ and $\Q_2 = (\y, \V_2, \E_2)$, $\Q_1 - \Q_2$ can be computed in $O(\texttt{cost}(\Q_1) + \OUT_1 \cdot \texttt{cost}(\revm{\Q^\emptyset_2}))$ time, where $\revm{\Q^\emptyset_2} = (\emptyset, \V_2 - \y, \{e - \y: e \in \E_2\})$.
 \end{theorem}
 
 \noindent {\bf Remark.} If $\Q_2$ is linear-reducible, $\Q_2$ can be reduced to a full join in $O(N)$ time by Algorithm~\ref{alg:reduce}. Then, $\Q^\emptyset_2$ becomes empty.  A faster solution is to build hashing indexes on every relation in the reduced $\Q_2$.  For each tuple $t \in \Q_1$, it suffices to check for every $e \in \E_2$ whether $\pi_{e} t \in R'_e$, which only takes $O(1)$ time. We note that the rewriting rule in Lemma~\ref{lem:heterogenous-rewrite} can also apply to this case and lead to the same complexity. Suppose $\Q_2$ is reduced. Each $e \in \E_2$ induces a CQ $(\pi_e \Q_1 - R'_e) \Join \Q_1$. After materializing the results of $\Q_1$, it suffices to check for each tuple $t \in \Q_1$, whether $\pi_e t \in R'_e$ or not.  This is exactly how our heuristic proceeds. Hence, Corollary~\ref{cor:heuristic1} follows.
 \begin{example}
 \label{exp:h1}
 \emph{Consider a \DCQ\ with $\Q_1 = \pi_{x_1, x_2, x_3} R_1(x_1,x_4)\Join R_2(x_4, x_2,x_3)$ and $\Q_2 = \pi_{x_1, x_2, x_3} R_3(x_1,\\x_2) \Join R_4(x_2,x_3) \Join R_5(x_1,x_3) \Join R_6(x_3,x_4)$. The baseline spends $O(N^{\frac{2 \cdot \omega}{\omega+1}} + N^{\frac{\omega-1}{\omega+1}}\cdot \OUT_1)$ time computing
 $Q_1$ and $O(N^{\frac{2 \cdot \omega}{\omega+1}} + N^{\frac{3(\omega-1)}{\omega+1}} \cdot \OUT_2^{\frac{3-\omega}{\omega+1}})$ time computing the hidden triangle join $R_3 \Join R_4 \Join R_5$ in $\Q_2$, where $\omega$ is the exponent of fast matrix multiplication. In contrast, our algorithm only spends $O(N^{\frac{2 \cdot \omega}{\omega+1}} + N^{\frac{\omega-1}{\omega+1}} \cdot \OUT_1)$ time for computing $\Q_1$, without computing the expensive $\Q_2$, hence can improve the baseline by a factor of $O\left(N^{\frac{2(\omega-1)}{\omega+1}} \cdot \OUT_2^{\frac{3 - \omega}{\omega+1}} /\OUT_1\right)$ when $N^{\frac{2(\omega-1)}{\omega+1}} \cdot \OUT_2^{\frac{3 - \omega}{\omega+1}} \ge \OUT_1$. }
 \end{example}

 We can show some further improvement when $\Q_2$ is non-linear-reducible. Instead of issuing an individual Boolean query for every query result $t \in \Q_1$, we take all the Boolean queries into account as whole. To do so, we further explore the structural property of the intersection query $\revm{\Q^{\oplus}_2} = (\y, \V_2, \{\y\} \cup \E_2)$, by treating the query results of $\Q_1$ as a single relation over attributes $\y$. It is unclear how $\texttt{cost}(\Q_2)$ compares with $\texttt{cost}(\Q^{\oplus}_2)$, since $\Q^{\oplus}_2$ involves an extra relation (over attributes $\y$) of input size as large as the output size of $\Q_1$.
 
 \paragraph{Remark.}  We note that if $\Q_1$ only produces $O(N)$ query results, then it is always cheaper (or at least not more expensive) to compute $\Q^{\oplus}_2$ than $\Q_2$. The observation is that we can always materialize the query results of $\Q_2$, and then check for every result whether it is in the extra relation of input size $N$, which does not increase the complexity of computing $\Q_2$ asymptotically.

 \begin{theorem}
 \label{the:heuristics2}
    Given two CQs $\Q_1 = (\y, \V_1, \E_1)$ and $\Q_2 = (\y, \V_2, \E_2)$, $\Q_1 - \Q_2$ can be computed in $O(\texttt{cost}(\Q_1) + \texttt{cost}(\revm{\Q^{\oplus}_2}))$ time, where $\revm{\Q^{\oplus}_2} = (\y, \V_2, \{\y\} \cup \E_2)$.
 \end{theorem}

 \begin{example}
  \label{exp:4}
  \emph{Consider a \DCQ\ with $\Q_1 = R_1(x_1,x_3)$ and $\Q_2 = \pi_{x_1,x_3} (R_3(x_1,x_2) \Join R_4(x_2,x_3))$. The baseline takes $O(N + N\cdot \sqrt{\OUT_2})$ time to materialize $\Q_2$. 
  The first heuristics of issuing $\Q^\emptyset_2$ for each tuple $t \in R_1$ takes $O(N^{3/2})$ time. We note that $\Q_{12}=
  \pi_{x_1, x_3} \left(R_1(x_1,x_3) \Join R_2(x_1,\\ x_2) \Join R_3(x_2, x_3) \right)$ lists edges that participate in at least one triangle. The existing best algorithm takes $O(N^{\frac{2\omega}{\omega+1}})$ time to compute $\Q_{12}$, where $\omega$ is the exponent of fast matrix multiplication, dominating the overall complexity. Our approach will improve the baseline if $\OUT_2 > N^{\frac{2(\omega-1)}{\omega+1}}$, and strictly outperforms the naive heuristic.}
\end{example}

\begin{example}
\label{exp:5}
    \emph{Consider a \DCQ\ with $\Q_1 = \pi_{x_1,x_3} R_1(x_1, x_2) \Join R_2(x_2, x_3)$ and $\Q_2 = \pi_{x_1,x_3} R_3(x_1,x_2) \Join R_4(x_2, x_3)$. Let $R_5(x_1, x_3) = \Q_1$. Here, $\Q_{12} = \pi_{x_1,x_3} R_3(x_1, x_2) \Join R_4(x_2, x_3) \Join R_5(x_1, x_3)$ with $|R_5| = \OUT_1$. Similarly, the existing best algorithm takes $O(\OUT^{\frac{\omega}{\omega+1}}_1 \cdot N^{\frac{\omega}{\omega+1}})$ time to compute $\Q_{12}$. It is worth mentioning that $\Q_{12} \neq \pi_{x_1,x_3} R_1(x_1, x_2) \Join R_2(x_2, x_3) \Join R_3(x_1, x_2) \Join R_4(x_2, x_3)$. Suppose $(a,c) \in \Q_{12}$, is witnessed by $(a,b_1, c) \inf R_1\Join R_2$ and $(a,b_2, c) \in R_3 \Join R_4$, but $(a,b_1, c) \notin R_3 \Join R_4$ and $(a,b_2, c) \notin R_1 \Join R_2$. Then, $(a,b_1,c), (a,b_2, c) \notin R_1 \Join R_2 \Join R_3 \Join R_4$, hence the result $(a,c)$ will be missed in this rewriting.}
\end{example}

%% file: extension.tex
\section{Extensions}
\label{sec:extension}

Based on the basic \DCQ\ over two CQs discussed so far, we next consider several interesting extensions of \DCQ\ with rich interaction between difference and other relational algebra operators. 

\subsection{Difference of Multiple CQs} 
\label{sec:multiple-dcq}

The first extension is adapting our result for computing DCQ involving two CQs to multiple CQs, say $\Q = \Q_1 - \Q_2 - \cdots - \Q_k$.  Suppose $\Q_i = (\V,\E_i)$ for $i \in \{1,2,\cdots, k\}$. We next introduce a recursive algorithm for tackling the general case with $k > 2$.

The base case with $k=2$ is tackled by our previous algorithm {\sc EasyDCQ} in Section~\ref{sec:easy-algorithm}. We rewrite a general \DCQ\ with $k$ CQs into a union of multiple \DCQ s, each consisting of $k-1$ CQs.  We start from the first two CQs and apply a similar strategy in Section~\ref{sec:easy-algorithm}.  Suppose $\Q_1$ and $\Q_2$ are full; otherwise, we just invoke the reduce procedure to remove all non-output attributes via semi-joins.  More specifically, we define an auxiliary relation $S_{e} = \pi_{e} \Q_1$ for each $e \in \E_2$, and rewrite the input \DCQ\ as  $\displaystyle{\Q = \bigcup_{e \in \E_2} \left((\pi_e \Q_1 - R'_e) \Join \Q_1 - \Q_3 - \cdots - \Q_k\right)}$.
If unwrapping the recursions, we can give a complete form for $\Q$: 
    \begin{align*}
     \Q = \bigcup_{(e_2,e_3, \cdots, e_k) \in \E_2 \times \E_3 \times \cdots \E_k}  \left((S_{I_k} - R_{e_k})  \Join \cdots \Join (S_{I_3}- R_{e_3}) \Join (S_{I_2} - R_{e_2}) \Join \Q_1 \right)
    \end{align*}
    where $I_j = \{e_2, e_3, \cdots,e_j\}$ for any $j \in \{2,3,\cdots, k\}$, and 
    \begin{equation}
    \label{eq:S-I-j}
     S_{I_j} = \pi_{e_j} \left\{ (S_{I_{j-1}} - R_{e_{j-1}}) \Join \cdots \Join (S_{I_2} - R_{e_2}) \Join \Q_1\right\}
    \end{equation}
    for any $j \ge 2$. An algorithm follows this rewriting directly. 
% \end{lemma}
%
%
%
%Let's start with an simple example of three CQs to gain some intuition. Given three input CQs $\Q_1, \Q_2, \Q_3$, we adapt our algorithm in Section~\ref{sec:easy-algorithm} to their difference query $\Q_1 -\Q_2 -\Q_3$. Similarly, we define an auxiliary relation $S_{e_2}= \pi_{e_2} \Q_1$ for each $e_2 \in \E_2$, and  and $T_{e_2, e_3} = \pi_{e_3} \left((S_{e_2} - R_{e_2}) \Join \Q_1\right)$ for each $e_3 \in \E_3$ subject to $e_2 \in \E_2$. Then, we can rewrite $\Q_1 - \Q_2 - \Q_3$ as: 
%\begin{align*}
%    \Q_1 - \Q_2 - \Q_3 &= \cup_{e_2 \in \E_2} \left( (S_{e_2} - R_{e_2}) \Join \Q_1 \right) - \Q_3 \\
%    &= \cup_{e_2 \in \E_2}  \left((S_{e_2} - R_{e_2}) \Join \Q_1 - \Q_3 \right) \\
%    &= \cup_{e_2 \in \E_2}  \left( \cup_{e_3 \in \E_3} (T_{e_2, e_3}- R_{e_3}) \Join (S_{e_2} - R_{e_2}) \Join \Q_1\right) \\ 
%    &= \cup_{(e_2,e_3) \in \E_2 \times \E_3}  \left((T_{e_2, e_3}- R_{e_3}) \Join (S_{e_2} - R_{e_2}) \Join \Q_1 \right)
%\end{align*}
%We get some immediate result that $\Q_1 - \Q_2 - \Q_3$ can be evaluated in linear time, if $\Q_1$ is free-connex, $\Q_2$ is free-connex or full, $\Q_3$ is free-connex or full, and $\E_1 \cup \{e_2\} \cup \{e_3\}$ is $\alpha$-acyclic for every $e_2 \in \E_2$ and $e_3 \in \E_3$. Generalizing this observation, we obtain the following result:
%
%
Now, we come to the complexity of this algorithm.  We can first bound $S_{I_j} = O(N+k)$ for each $j \in \{2,3,\cdots, k\}$, since every tuple from $S_{I_j} - R_{e_j}$ must participate in at least query result. Moreover, if $\E_1 \cup I_j$ is free-connex, $S_{I_{j}}$ is free-connex from (\ref{eq:S-I-j}). As the subquery corresponding to $S_{I_{j}}$ has input size $O(N+\OUT)$ and output size $O(\OUT)$, it can be evaluated in $O(N + \OUT)$ time. 

\begin{theorem}
\label{the:multiple-dcq}
    Given a \DCQ $\Q = \Q_1-\Q_2-\cdots-\Q_k$, $\Q$ can be evaluated in $O(N + \OUT)$ time if $\Q_1$ is free-connex, $\Q_i$ for $i \ge 2$ is linear-reducible, and for every $j \in \{2,3,\cdots, k\}$, the subquery induced by $\E'_1 \cup \{e_2,e_3,\cdots, e_j\}$ is $\alpha$-acyclic for any $(e_2, e_3, \cdots, e_j) \in \E'_2 \times \E'_3 \times \cdots \times \E'_j$, where $(\y, \E'_i)$ is the reduced query of $\Q_i = (\y,\V_i, \E_i)$.
\end{theorem}

\subsection{Select, Project and Join}
%We next consider how to equip the basic \DCQ\ with other relational algebra operators. 
\begin{itemize}[leftmargin=*]
    \item If there is a selection operator $\sigma_\phi$ over $\Q = \Q_1 - \Q_2$, we can push it down such that $\Q = \sigma_\phi \Q_1 - \sigma_\phi \Q_2$. If $\phi$ is a predicate on a base relation $R_e$ of $\Q_1$ (resp. $\Q_2$), we can we simply check if $\phi(t)$ is true for each tuple $t \in \sigma_\phi R_e$, and discard it if not.  This only takes $O(N)$ time.  It is challenging that $\phi$ is a predicate not on any base relation, even for a single CQ evaluation. 
    
   \item If there is a projection operator $\pi_\theta$ over $\Q = \Q_1 - \Q_2$, we can push it down such that $\Q = \pi_\theta \Q_1 - \pi_\theta \Q_2$, and handle a new \DCQ\ $\Q'_1 - \Q'_2$ with $\Q'_1 = \pi_\theta \Q_1$ and $\Q'_2 = \pi_\theta \Q_2$.
   
   \item If there is a join operator over multiple \DCQ s, we first rewrite the join into a \DCQ over multiple CQs and invoke our previous algorithm in Section~\ref{sec:multiple-dcq}.  More specifically, given $k$ \DCQ s $\Q^1, \Q^2,\cdots, \Q^k$ with $\Q^i = \Q^i_{1} -\Q^{i}_2$ for any $i \in [k]$, we can rewrite $\Q^1 \Join \Q^2 \Join \cdots \Join \Q^k$ as
   \[ \ \ \ \ \ \ (\Join_{i\in[k]}\Q^i_{1}) - \{(\Join_{i \in I} \Q^i_1) \Join (\Join_{j \in J} \Q^j_2): I \subsetneq [k], J = [k]-I\}.\]
   The characterization of input CQs for which a linear algorithm exists follows Theorem~\ref{the:multiple-dcq}.

   %
   %\item If there is a union operator over multiple \DCQ s, 
   %
   %\item If there is a different operator over two \DCQ s, 
   %More specifically, suppose we are given $\Q^1 = \Q^1_1 - \Q^1_2$ and $\Q^2 = \Q^2_1 - \Q^2_2$, 
   %\[\Q^1 \cup \Q^2 = (\Q^1_1 - \Q^1_2) \cup \Q^2_1 -  \Q^2_2 \]
\end{itemize}

 %\begin{theorem}
 %   \label{the:join-dcq}
 %      Given $k$ \DCQ s $\Q^1, \Q^2, \cdots, \Q^k$ with $\Q^i = \Q^i_1 - \Q^i_2$ for any $i \in [k]$, if $\Join_{i \in [k]} \Q^i_1$ is free-connex, $(\Join_{i \in I} \Q^i_1) \Join (\Join_{j \in J} \Q^j_2)$ is linear-reducible for any $I \subsetneq [k]$ and $J = [k]-I$, and $\Join_{i \in [k]} \Q'^i_1 \Join (\pi_{\y} R_e)$ is free-connex for every $e \in \Q^j_2$, then $\Q^1 \Join \cdots \Join \Q^k$ can be computed in $O(N + \OUT)$ time.
 %   \end{theorem}

\subsection{Aggregation}
\label{sec:aggregation}
Our algorithm for \DCQ\ can also be extended to support aggregations over annotated relations~\cite{abo2016faq,joglekar16:_ajar}. Let $(S, \oplus, \otimes)$ be a commutative ring. For a CQ $\Q$ over an annotated instance $D$, every tuple $t \in R_e$ has an annotation $w(t) \in S$.  For a full query $\Q = (\V,\E)$, the annotation for any join result $t \in \Q(D)$ is defined as 
$w(t) := \mathop{\otimes}\limits_{e \in \E} w(\pi_e t)$.
For a non-full query $\Q= (\y, \V,\E)$, the aggregation becomes {\tt GROUP BY} $\y$, and the annotation for each result $t \in \Q$ (i.e., the aggregate of each group) is 
$w(t) := \mathop{\oplus}\limits_{t' \in \Join_{e \in \E} R_e: \pi_{\y} t'= t} w(t')$.  
Below, we introduce two commonly-used formulations. \revm{Given $\Q_1 = (\y, \V_1, \E_1)$, $\Q_2 =(\y, \V_2, \E_2)$ and instances $D_1, D_2$, let $w_1, w_2$ be the annotations of tuples in $\Q_1, \Q_2$ respectively. For completeness, we set $w_1(t) = 0$ if $t \notin \Q_1(D_1)$ and $w_2(t) = 0$ if $t \notin \Q_2(D_2)$.}

\begin{figure}[t]
    \centering
    \includegraphics[scale=0.95]{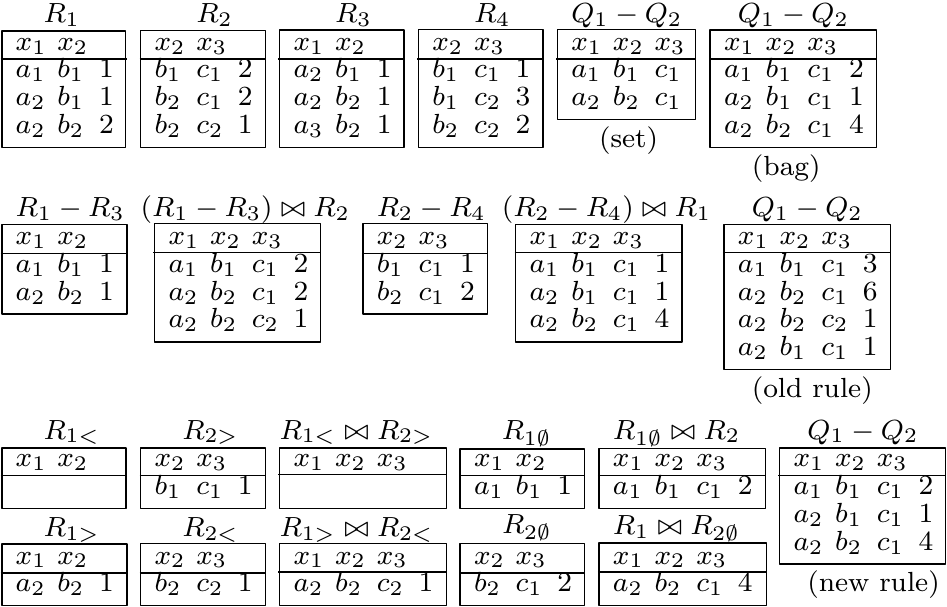}
    \caption{An example of \DCQ\ $\Q_1 - \Q_2$ with $\Q_1 = R_1(x_1, x_2) \Join R_2(x_2, x_3)$ and $\Q_2 = R_3(x_1, x_2) \Join R_4(x_2, x_3)$.
    In the bag semantics, the input size is $18$, and the output size is $7$. $(a_1, b_1, c_1)$ is a result under bag semantics, but not a result under set semantics. The results of $\Q_1- \Q_2$ computed by the old and new rewriting rules are also illustrated separately.}
    \label{fig:bag}
\end{figure}

\revm{
\smallskip \noindent {\bf Relational difference.} For \DCQ s defined on relational difference, a tuple $t$ appears in the query results of $\Q_1 - \Q_2$ if and only if $t \in \Q_1(D_1)$ and $t \notin \Q_2(D_2)$.
For $t \in \pi_{\y'} (\Q_1(D_1) - \Q_2(D_2))$, the annotation of $t$ is defined as $\displaystyle{w(t) = \oplus_{t' \in \Q_1(D_1) - \Q_2(D_2): \pi_{\y'} t'= t} w(t')}$. The input size is defined as $N =|D_1| + |D_2|$, and the output size is $\OUT =|\pi_{\y'} \left(\Q_1(D_1) - \Q_2(D_2)\right)|$. Again, our target is to find a linear-time algorithm in terms of $N$ and $\OUT$.  Our algorithms can be applied directly, followed by aggregation, and its complexity is bottlenecked by the output size of the difference query, i.e., $|\Q_1(D_1) - \Q_2(D_2)|$, which could be much larger than $\OUT$.

\smallskip \noindent {\bf Numerical difference.}
For \DCQ s defined on numerical difference, a tuple $t$ appears in the query results of $\Q_1 - \Q_2$ if and only if $t \in \Q_1(D_1)$ or $t \in \Q_2(D_2)$, with annotation $w(t) = w_1(t) - w_2(t)$. Then, the aggregation operator defined over attributes $\y'$ on top of $\Q_1 - \Q_2$ can be rewritten as the numerical difference of two new annotated queries, i.e.,  $\pi_{\y'} \Q_1 - \pi_{\y'} \Q_2$. 
The input size is defined as $N =|D_1| + |D_2|$, and the output size is $\OUT =|\pi_{\y'} \Q_1(D_1) \cup \pi_{\y'} \Q_2(D_2)|$. 
Again, our target is to find an linear-time algorithm in terms of $N$ and $\OUT$. Here, any algorithm with time complexity $O(N + |\pi_{\y'} \Q_1(D_1)| + |\pi_{\y'} \Q_2(D_2)|)$ is already optimal, since $|\pi_{\y'} \Q_1(D_1) \cup \pi_{\y'} \Q_2(D_2)| \ge \frac{1}{2}\left(|\pi_{\y'}\Q_1(D_1)| + |\pi_{\y'} \Q_2(D_2)|\right)$. Hence, if $\pi_{\y'} \Q_1$ and $\pi_{\y'} \Q_2$ are free-connex, both our algorithm and baseline are optimal.
}
\begin{theorem}
    Given two CQs $\Q_1=(\y,\V_1,\E_1)$ and $\Q_2 = (\y,\V_2,\E_2)$, and a subset of aggregation attributes $\y' \subseteq \y$, if $(\y',\V_1,\E_1)$ and $(\y', \V_2, \E_2)$ are free-connex, $\pi_{\y'}(\Q_1 - \Q_2)$ with numerical difference can be computed in $O(N + \OUT)$ time.
\end{theorem}
\begin{example}
\label{exp:aggregate}
   \emph{
    Consider an example \DCQ\ $\Q = \pi_{x_1} (R_1(x_1, x_2)  \Join R_2(x_2, x_3) -  R_3(x_1, x_2) \Join R_4(x_2, x_3))$ over an instance in Figure~\ref{fig:bag}. This \DCQ\ can capture Q16 in the TPC-H benchmark~\cite{tpch} as a special case. 
    For relational difference, the query result of $\Q$ includes 2 tuples as $\{(a_1,1), (a_2, 1)\}$. For numerical difference, the query result of $\Q$ includes 3 tuples as $\{(a_1,1), (a_2, 2), (a_3, -2)\}$. }
\end{example}

\subsection{Bag Semantics}
\label{sec:bag}

We consider the bag semantics that the set of query result is a multi-set. For simplicity, each distinct tuple $t$ is annotated with a positive integer $w(t)$ to indicate the number of copies. In a full CQ $\Q=(\V,\E)$, the annotation of $t \in \Q$ is defined as $\displaystyle{w(t) = \times_{e \in \E} w(\pi_e t)}$. For a projection of $R_e$ onto attributes $e'$, the annotation of $t \in \pi_{e'} R_e$ is defined as $w(t) = \sum_{t' \in R_e: \pi_{e'} t' = t} w(t')$. 
\revm{Given two CQs $\Q_1 = (\y, \V_1, \E_1), \Q_2 =(\y, \V_2, \E_2)$ and two input instances $D_1, D_2$, let $w_1, w_2$ be the annotations of tuples in $\Q_1, \Q_2$ respectively. For completeness, we set $w_1(t) = 0$ if $t \notin \Q_1(D_1)$ and $w_2(t) = 0$ if $t \notin \Q_2(D_2)$. A tuple $t$ is a query result of $\Q_1 - \Q_2$ if and only if $t \in \Q_1(D_1)$ and $w_1(t) > w_2(t)$. 
An example is given in Figure~\ref{fig:bag}.
%

%Let $D_1, D_2$ be two input instances of $\Q_1, \Q_2$ respectively. 
The input size is %$N = \sum_{e \in \E_1} \sum_{t \in R_e} w_1(t) + \sum_{e' \in \E_2} \sum_{t \in R_{e'}} w_2(t)$, 
$N = |D_1| + |D_2|$,
and the output size is $\OUT = \sum_{t \in \Q_1(D_1)} \max\{0, w_1(t) - w_2(t)\}$.} Again, our target is to find an linear-time algorithm in terms of $N$ and $\OUT$. Unfortunately, our rewriting rule in Section~\ref{sec:easy-algorithm} cannot be adapted here. Figure~\ref{fig:bag} shows several incorrect behaviors: some tuple has a much higher annotation (e.g., $(a_1, b_1, c_1)$); and some tuple should not appear (e.g., $(a_2, b_2, c_2)$), which motivates us to explore new rules here.

\begin{example}
\label{exp:bag}
    \emph{Consider a \DCQ\ with $\Q_1 = R_1(x_1, x_2) \Join R_2(x_2, x_3)$ and $\Q_2 = R_3(x_1, x_2) \Join R_4(x_2, x_3)$ under the bag semantics. Any result $(a,b,c) \in \Q_1- \Q_2$ falls into one of the three cases: (1) $(a,b) \notin R_2$ or $(b,c) \notin R_4$; (2) $w_1(a,b) > w_2(a,b)$ and $w_1(b,c) > w_2(b,c)$; (3) either $w_1(b,c) \le w_2(b,c)$ or $w_1(a,b) \le w_2(a,b)$, but $w_1(a,b) \cdot w_1(b,c) > w_2(a,b) \cdot w_2(b,c)$. We partition $R_{1}$ into three subsets, $R_{1\emptyset} = \{t \in R_1: t \notin R_3\}$, $R_{1<} = \{t \in R_1: t \in R_3, w_1(t) \le w_2(t)\}$ and $R_{1>} = \{t \in R_1: t \in R_3: w_1(t) > w_2(t)\}$. Similarly, we partition $R_{2}$ into  $R_{2\emptyset}$, $R_{2<}$, $R_{2>}$ with respect to $R_4$.  Results falling into (1) can be found by $(R_{1\emptyset} \Join R_2) \cup (R_1 \Join R_{2\emptyset})$. Results falling into (2) can be found by $R_{1>} \Join R_{2>}$. Results falling into (3) can be found by two new $\theta$-joins $\left(R_{1>} \Join_{\theta} R_{2<}\right) \cup \left(R_{1<} \Join_{\theta} R_{2>}\right)$, where a pair of tuples $(t_1, t_2 )$ can be $\theta$-joined if and only if $w_1(t_1) \cdot w_1(t_2) > w_2(t_1) \cdot w_2(t_2)$. }
    
    \emph{All auxiliary relations as well as $(R_{1\emptyset} \Join R_2) \cup (R_1 \Join R_{2\emptyset})$ and $R_{>} \Join R_{2>}$ can be computed efficiently. We consider $R_{1>} \Join_{\theta} R_{2<}$ ($\left(R_{1<} \Join_{\theta} R_{2>}\right)$ is symmetric). The solution of checking $\theta$-condition for all combinations of tuples in $R_{1>}$ and $R_{2<}$ incurs quadratic complexity. A smarter way is to sort $R_{1>}$ and $R_{2<}$ by $B$ first, and then by the ratio of $\frac{w_1(\cdot)}{w_2(\cdot)}$ decreasingly. Then, we start with  $(a,b) \in R_{1>}$ with maximum ratio, and linearly scan tuples in $R_{2<}$ with the join value $b$ until we meet some tuple $(b,c)$ such that $\frac{w_1(b,c)}{w_2(b,c)} \le \frac{w_2(a,b)}{w_1(a,b)}$. We then stop and proceed with the next tuple in $R_{1>}$. If no join result is produced by $(a,b)$, we skip the subsequent tuples with the same join value $b$ and continue. Overall, this algorithm takes $O(N \log N + \OUT)$ time.}
\end{example}
\begin{theorem}
\label{the:bag}
    Given two CQs $\Q_1 = \Q_2 = (\y,\V,\E)$, if $(\y,\V,\E)$ is free-connex, then $\Q_1 - \Q_2$ under the bag semantics can be computed in $O(N\log N+ \OUT)$ time.
\end{theorem}
Our observation above can be extended to the case when both $\Q_1,\Q_2$ correspond to the same free-connex query. The proof of Theorem~\ref{the:bag} is given in Appendix~\ref{appendix:extension}. The case when $\Q_1, \Q_2$ have different schema is left as future work.

%% file: experiment.tex
\section{Experiments}
\label{sec:experiment}

\subsection{Experimental Setup}
\noindent  {\bf Prototype implementation.}  Our newly developed algorithms can be easily integrated into any SQL engine by rewriting the original \revm{SQL query}.  It can be further optimized if we directly integrate the rewrite procedure into the SQL parser and have customized index support.  Our ultimate goal is to implement our algorithms into a system prototype with three components: a SQL parser, a query optimizer, and new indices. % However, it went out of the scope of this paper.  
At the current stage, we choose to manually rewrite all \revm{SQL queries} and demonstrate the power of our optimizations via the comparison with vanilla \revm{ \revm{SQL queries} }.

\smallskip \noindent  {\bf Query processing engines compared.} To compare the performance of all optimized techniques we proposed in the paper, we choose PostgreSQL~\cite{postgre}, \revm{DuckDB\cite{duckdb}, SQLite\cite{sqlite}, MySQL\cite{mysql} running in centralized settings, and Spark SQL~\cite{spark} running in parallel/distributed settings, as the query processing engines.  All of them are widely used in academia and industry.}  \revm{In the experiments, we observed that SQLite and MySQL show very poor performance, with most of the test points being timed out.  Hence, we built full indices on these systems to expedite the execution. Moreover, DuckDB is a columnar-vectorized query execution engine, and indices are built when importing input data.} During the experiments, we test the single-thread performance of our new optimization techniques over PostgreSQL, \revm{DuckDB, SQLite and MySQL},
and parallel performance over Spark SQL. In order to separate the I/O cost from the total execution time, we load all data into the memory in advance by using pg-prewarm in PostgreSQL and cache in Spark SQL. \revm{For DuckDB and SQLite, the data need to be loaded into memory before execution, so we only count the query execution time.}

\smallskip \noindent  {\bf Experimental environment.}  We perform all experiments in two machines.  For experiments conducted on PostgreSQL \revm{and MySQL}, we use a machine equipped with two Intel Xeon 2.1GHz processors, each having 12 cores/24 threads and 416 GB memory.  For all experiments on Spark SQL, \revm{DuckDB and SQLite}, we use a machine equipped with two Xeon 2.0GHz processors, each having 28 cores / 56 threads and 1TB of memory.  All machines run Linux, with Scala 2.13.9 and JVM 1.8.0.  We use Spark 3.3.0 and PostgreSQL 16.0.  We assign 8 cores for Spark and 1 core for the rest platforms during the experiments.  Each query is evaluated 10 times with each engine, and we report the average running time. Each query runs at most 10 hours to obtain meaningful results. 

\subsection{Datasets and Queries}
The experiments consist of graph queries and benchmark queries.  

{\bf Benchmark queries.} For relational queries, we adopt two standard benchmarks (TPC-DS~\cite{tpcds} and TPC-H~\cite{tpch}) in industry and select 3 queries with difference operator (TPC-H Q16, TPC-DS Q35, and TPC-DS Q69).  These three benchmark queries connect DCQ with other relational operators like selection, projection, join, and aggregation.  All benchmark queries can be captured by a common schema $\Q = R_1(x_1, x_2) \Join (\pi_{x_2} R(x_1, x_2) - \pi_{x_2} R_2(x_2, x_3) \Join R_3(x_3, x_4))$ and the joins are all primary-key foreign-key joins. 	

\begin{table*}
\centering
\resizebox{\linewidth}{!}{
%\begin{scriptsize}
\setlength{\tabcolsep}{4pt}
\begin{tabular}{|c|r|r|r|r|r|r|r|r|r|r|}
\hline
\rowcolor[HTML]{C0C0C0} 
Graph & \multicolumn{1}{c|}{\cellcolor[HTML]{C0C0C0}\#edge} & \multicolumn{1}{c|}{\cellcolor[HTML]{C0C0C0}\#vertex} & \multicolumn{1}{c|}{\cellcolor[HTML]{C0C0C0}\#l2 path} & \multicolumn{1}{c|}{\cellcolor[HTML]{C0C0C0}\#triangle} & \multicolumn{1}{c|}{\cellcolor[HTML]{C0C0C0}\#$\Q_{G1}$} & \multicolumn{1}{c|}{\cellcolor[HTML]{C0C0C0}\#$\Q_{G2}$} & \multicolumn{1}{c|}{\cellcolor[HTML]{C0C0C0}\#$\Q_{G3}$}& \multicolumn{1}{c|}{\cellcolor[HTML]{C0C0C0}\#$\Q_{G4}$} & \multicolumn{1}{c|}{\cellcolor[HTML]{C0C0C0}\#$\Q_{G5}$} & \multicolumn{1}{c|}{\cellcolor[HTML]{C0C0C0}\#$\Q_{G6}$} \\ \hline
Bitcoin & 24,186 & 3,783 & 1,256,332 & 88,753 & 820 & $1.0 \times 10^7$ & 585,958 & 331,497 & $3.8 \times 10^7$ & $5.7 \times 10^8$ \\ \hline
Epinions & 508,837 & 75,879 & $3.9\times 10^7$ & 3,586,405 & 25,947 &  $9.3\times 10^8$ & $1.8 \times 10^7$ & $1.0 \times 10^7$ & $3.5 \times 10^9$ & $2.5 \times 10^{11}$ \\  \hline
DBLP & 1,049,866 & 317,080 & 7,064,738 & 2,224,385 & 466,646 & $1.6\times 10^8$ & 3,532,369 & 2,203,597 & $6.7 \times 10^7$ & - \\ \hline
Google & 5,105,039 & 875,713 & $6.0\times 10^7$ & $2.8\times 10^7$ & 372,042 & $2.1\times 10^8$ & $2.4 \times 10^7$ & $1.5 \times 10^7$ &  $7.8 \times 10^8 $ & - \\ \hline
Wiki & $2.8\times 10^7$ & 2,394,385	 & $2.6\times 10^9$ & $8.1 \times 10^7$ & 0 & $1.1 \times 10^{10}$ & $1.3  \times 10^8$ & $6.6 \times 10^7$ & - & - \\ \hline
\end{tabular}
%\end{scriptsize}
}
\caption{\revm{Graph datasets and their statistics. \#edge is the input size of graph datasets. \#$\Q_{Gi}$ is the output size of $\Q_{Gi}$ over the corresponding graph datasets. `-' indicates that the output size is too huge such that all systems cannot report the output size within the time limit.}}
\label{tbl:graph}
\end{table*}

{\bf Graph queries.}  For graph pattern queries, we use real-world graphs (such as BitCoin, DBLP, Eponions, Google, and Wiki) from SNAP (Stanford Network Analysis Project) \cite{SNAP}, summarized in  Table~\ref{tbl:graph}.  We store edge information as a relation $\textsf{Graph}(\textsf{src}, \textsf{dst})$ and manually create a triple relation $\textsf{Triple}(\textsf{node1}, \textsf{node2}, \textsf{node3})$ from the graph. Tuples in $\textsf{Triple}$ are generated by following rules: (rule 1) a random length-2 path in the graph as $(\textsf{node1}, \textsf{node2}, \textsf{node3})$; or (rule 2) a random edge in the graph as $(\textsf{node1}, \textsf{node2})$, together with a random vertex in the graph as $\textsf{node3}$; or (rule 3) a triple $(\textsf{node1}, \textsf{node3}, \textsf{node5})$ from a random length-4 path $(\textsf{node1}, \textsf{node2}, \textsf{node3}, \textsf{node4}, \textsf{node5})$ in the graph. $\textsf{Triple}$ may involve different portions of tuples generated by three rules in different queries.  For a graph with $n$ length-2 paths, we set the size of $\textsf{Triple}$ to be $0.05n$ for Wiki (since it is too large to process as shown in Table~\ref{tbl:graph}), and $0.5n$ for the remaining graphs.  We evaluate 6 graph queries as described in Figure 5, whose original \revm{SQL queries} as well as optimized \revm{SQL queries} after rewriting are given in the full version~\cite{full}. 
\begin{figure*}
\label{fig:query}
\centering
{\small $
\begin{aligned}
    \Q_{G1} &= \textsf{Graph}(\textsf{node}_1, \textsf{node}_2) - \pi_{\textsf{node}_1, \textsf{node}_2} (\textsf{Graph}(\textsf{node}_1, \textsf{node}_2) \Join \textsf{Graph}(\textsf{node}_2, \textsf{node}_3)) \\
    \Q_{G2} &= \textsf{Graph}(\textsf{node}_1, \textsf{node}_2) \Join \textsf{Triple}(\textsf{node}_2, \textsf{node}_3, \textsf{node}_4) \\ 
    &- \textsf{Triple}(\textsf{node}_1, \textsf{node}_2, \textsf{node}_3) \Join \textsf{Graph}(\textsf{node}_3, \textsf{node}_4) \\
    \Q_{G3} &= \textsf{Triple}(\textsf{node}_1, \textsf{node}_2, \textsf{node}_3) - \textsf{Graph}(\textsf{node}_1, \textsf{node}_2) \Join \textsf{Graph}(\textsf{node}_2, \textsf{node}_3) \Join \textsf{Graph}(\textsf{node}_3, \textsf{node}_1) \\
    \Q_{G4} &= \textsf{Triple}(\textsf{node}_1, \textsf{node}_2, \textsf{node}_3)\\ 
    &- \pi_{\textsf{node}_1, \textsf{node}_2, \textsf{node}_3}\textsf{Graph}(\textsf{node}_1, \textsf{node}_2) \Join \textsf{Graph}(\textsf{node}_2, \textsf{node}_3) \Join \textsf{Graph}(\textsf{node}_3, \textsf{node}_4) \\
    \Q_{G5} &= \textsf{Graph}(\textsf{node}_1, \textsf{node}_2) \Join \textsf{Graph}(\textsf{node}_2, \textsf{node}_3) \Join \textsf{Graph}(\textsf{node}_3, \textsf{node}_4) \\
    &- \textsf{Graph}(\textsf{node}_2, \textsf{node}_3) \Join \textsf{Graph}(\textsf{node}_3, \textsf{node}_4) \Join \textsf{Graph}(\textsf{node}_4, \textsf{node}_1) \\
    \Q_{G6} &= \textsf{Graph}(\textsf{node}_1, \textsf{node}_2) \Join \textsf{Graph}(\textsf{node}_3, \textsf{node}_4) \\
    &- \textsf{Graph}(\textsf{node}_1, \textsf{node}_2) \Join \textsf{Graph}(\textsf{node}_2, \textsf{node}_3) \Join \textsf{Graph}(\textsf{node}_3, \textsf{node}_1) \Join \textsf{Graph}(\textsf{node}_3, \textsf{node}_4)
\end{aligned}$}
\vspace{-1em}
\caption{Graph queries.}
\end{figure*}

More specifically, $\Q_{G1}$ finds all edges in the graph that do not participate in any length-2 path. %For $\Q_{G2}$, the two \textsf{Triple} relations only include tuples generated by rule 1 and 2 separately, so
$\Q_{G2}$ finds all length-3 paths that the third node ($\textsf{node}_3$) is not sampled together with the edge $(\textsf{node}_1, \textsf{node}_2)$. %For $\Q_{G3}$ and $\Q_{G4}$, the \textsf{Triple} relation only includes tuples generated by rule 1. 
$\Q_{G3}$ finds length-2 paths that do not form a triangle.  $\Q_{G4}$ finds all generated triples that cannot extend to a length-4 path. $\Q_{G5}$ finds all length-4 paths that do not form a length-4 cycle.  $\Q_{G6}$ finds all pairs of edges in the graph, which do not form a length-4 cycle. 
\subsection{Experiment Results}
\begin{figure*}
\centering
\includegraphics[width=\linewidth]{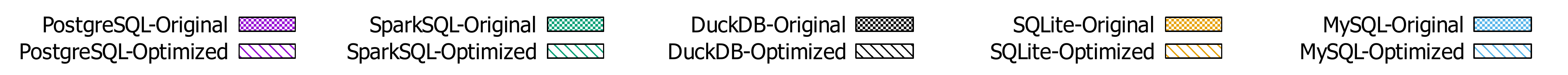}
\includegraphics[width=\linewidth]{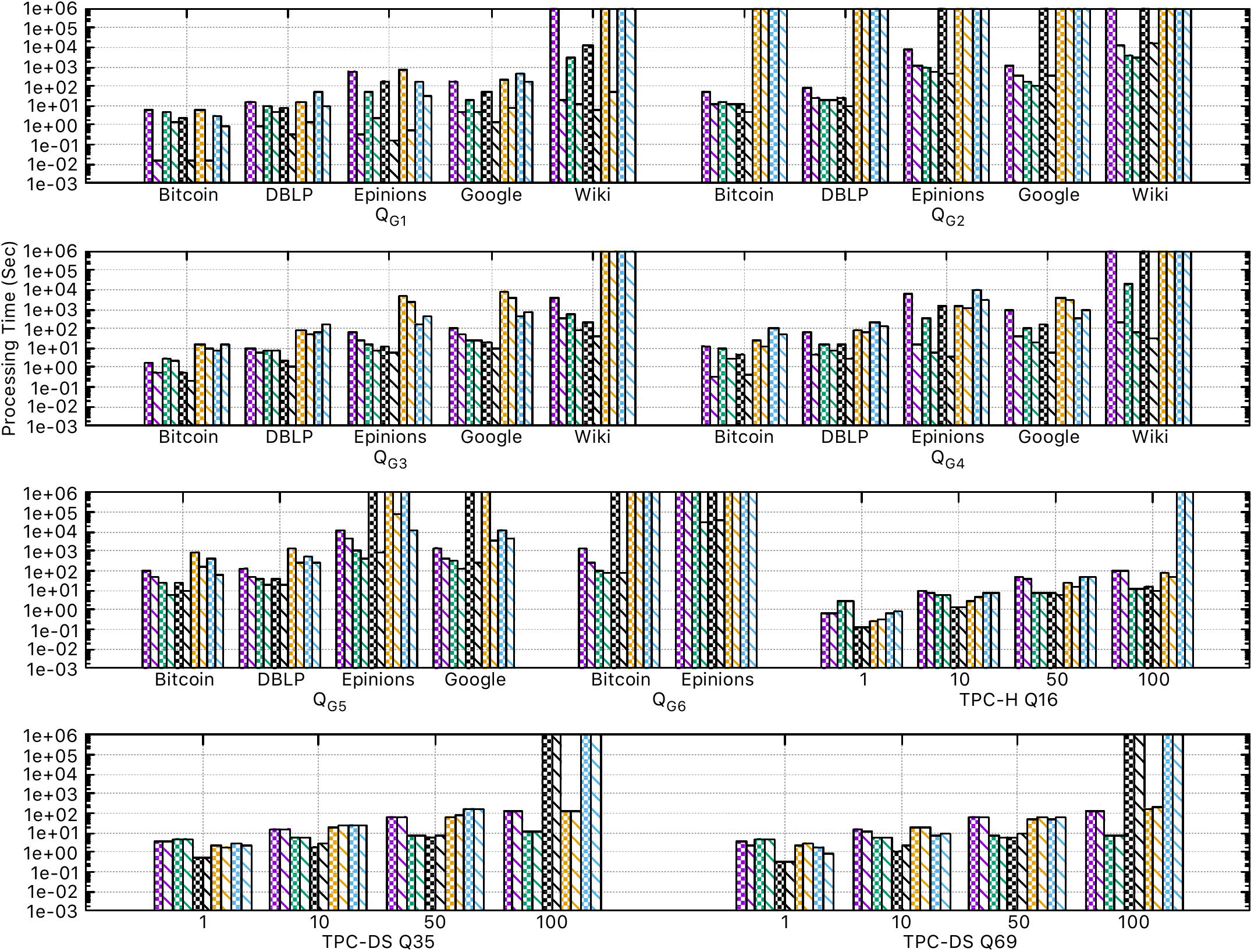}

\caption{Running time of graph and benchmark queries.}
\label{fig:running time}
\end{figure*}

\noindent  {\bf Running time.}  Figure~\ref{fig:running time} shows the running time of different engines on graph queries.  \revm{The input and output size of all graphs queries are given in Table~\ref{tbl:graph}.} All bars reaching the axis boundary indicate that the system did not finish within the 8-hour limit, or ran out of memory. As $\Q_{G6}$ contains an expensive Cartesian product as sub-query,  materializing its query result exceeds the memory capacity of our machines on most datasets.  PostgreSQL can only evaluate the original \revm{SQL query} of $\Q_{G6}$ on Bitcoin dataset.  By adding the parallelism from 8 to 80, our optimized Spark SQL can evaluate $\Q_{G6}$ on Epinions dataset within the time limit, while the vanilla Spark SQL cannot complete the evaluation.  For $\Q_{G5}$, all systems cannot finish the evaluation on Wiki dataset due to the large intermediate results created.  \revm{We also observe that both SQLite and MySQL cannot finish all test points for $\Q_{G2}$ and $\Q_{G6}$, and most test points over Wiki dataset. It could be the reason that both systems are not designed for analytical queries. }  Our optimization techniques already achieve a speedup ranging from 2x to 1760x on PostgreSQL, from 1.2x to 270x on Spark SQL, \revm{from 2x to 1848x on DuckDB, from 1.25x to 1095x on SQLite, and from 1.8x to 5.1x on MySQL for graph queries}, even without considering the queries that could not finish within the time limit. \revm{We also observe an unusual test point for $\Q_{G3}$ in MySQL, that our optimized SQL query takes more time than the vanilla SQL query, which may be due to some unknown deficiencies in MySQL internals.\footnote{We review the execution plan in MySQL and find that the predicated run-time of our optimized SQL query is much smaller than the vanilla SQL query, which is also consistent with our observations in other platforms. The actual running time does not match the expected cost because of some unknown deficiencies in MySQL.}}

Figure~\ref{fig:running time} also shows the running time of all query engines on benchmark queries under different scale factors (i.e., parameters used to generate benchmark dataset, which is roughly proportional to the input data size).  \revm{DuckDB and MySQL fail to finish some test points with scale factor 100.} %\xiao{what is scale factor? \qichen{Scale factor is for measuring the size used by all these benchmarks.  You can simply understand that when the scale factor is n, the data size is n GB.}}. 
However, the improvement in benchmark queries achieved by our optimized techniques is minor, as expected. More specifically, the vanilla benchmark query consists of two free-connex sub-queries, hence can be evaluated in $O(N+ \OUT_1+\OUT_2)$ time, and its optimized query can be evaluated in $O(N+\OUT)$ time. Due to the special primary-key foreign-key joins and group-by aggregations, $\OUT_1 \approx \OUT_2 \approx \OUT << N$, such that the input contains a few hundred of million records while the query result only involves thousands of records. \revm{The improvement of our optimized techniques in SQLite, DuckDB, and MySQL is also limited. On some test points, our optimized SQL queries are even more time-consuming than vanilla SQL queries. We find that the vanilla SQL queries can greatly benefit from the indices built for primary-key foreign-key join and outperform our optimized SQL queries, which do not enjoy efficient indices for set difference or anti-join operators. How to build indices to accelerate relational operators in these systems could be interesting future work. Meanwhile, we notice that loading input data and building indices are much more time-consuming than evaluating the query; for example, it takes DuckDB 16 minutes to load a 50G-sized TPC-DS dataset, while only 8 seconds to execute the whole query.}

\smallskip \noindent  {\bf  Impact of $\OUT$, $\OUT_1$ and $\OUT_2$.}  Implied by the theoretical results, the sizes $\OUT_1,\OUT_2$ of sub-queries $\Q_1, \Q_2$ impact the performance of vanilla SQL queries, while only the actual output size $\OUT$ affect the performance of our optimized SQL queries. %It should be noted that $\OUT \ge \OUT_1 - \OUT_2$.  
Below, we study the impact of $\OUT_1$, $\OUT_2$ and $\OUT$ on the performance of both approaches over $\Q_{G4}$.   

\begin{figure}[h]
    \centering
    \includegraphics[width=0.6\linewidth]{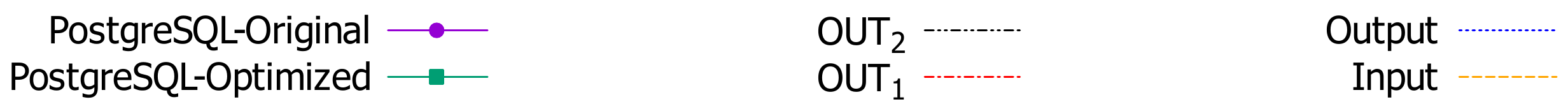}
    \includegraphics[width=0.6\linewidth]{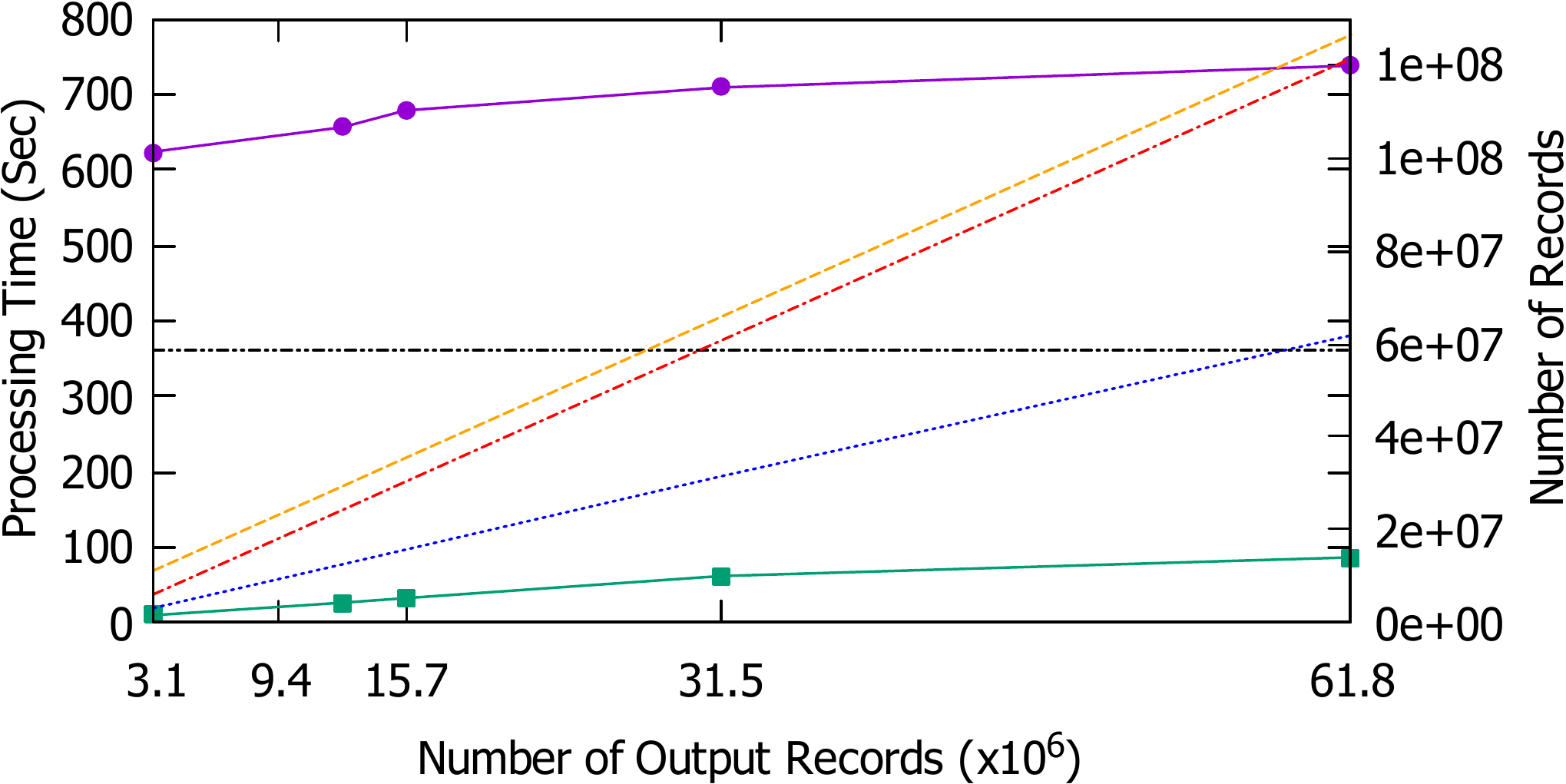}
    \caption{Running time of $\Q_{G4}$ on Google dataset in PostgreSQL with different $\OUT_1$.}
     \label{fig:out1}
\end{figure}

In Figure~\ref{fig:out1}, we investigate the impact of $\OUT_1$ for computing \DCQ. We fix $\Q_2$ (as well as $\OUT_2$) and only vary the size of $\textsf{Triple}$ (as well as $N$ and $\OUT_1$). Note that $\OUT$ also increases as $\OUT_2$ decreases.  The running time of our optimized \revm{SQL query} grows slowly with $\OUT$, while the vanilla \revm{SQL query} incurs a fixed overhead for evaluating $\Q_2$, even when $\OUT_1$ (as well as $\OUT$) decreases to as small as $1$.

\begin{figure}[h]
    \centering
    \includegraphics[width=0.6\linewidth]{figure/F7Title.pdf}
    \includegraphics[width=0.6\linewidth]{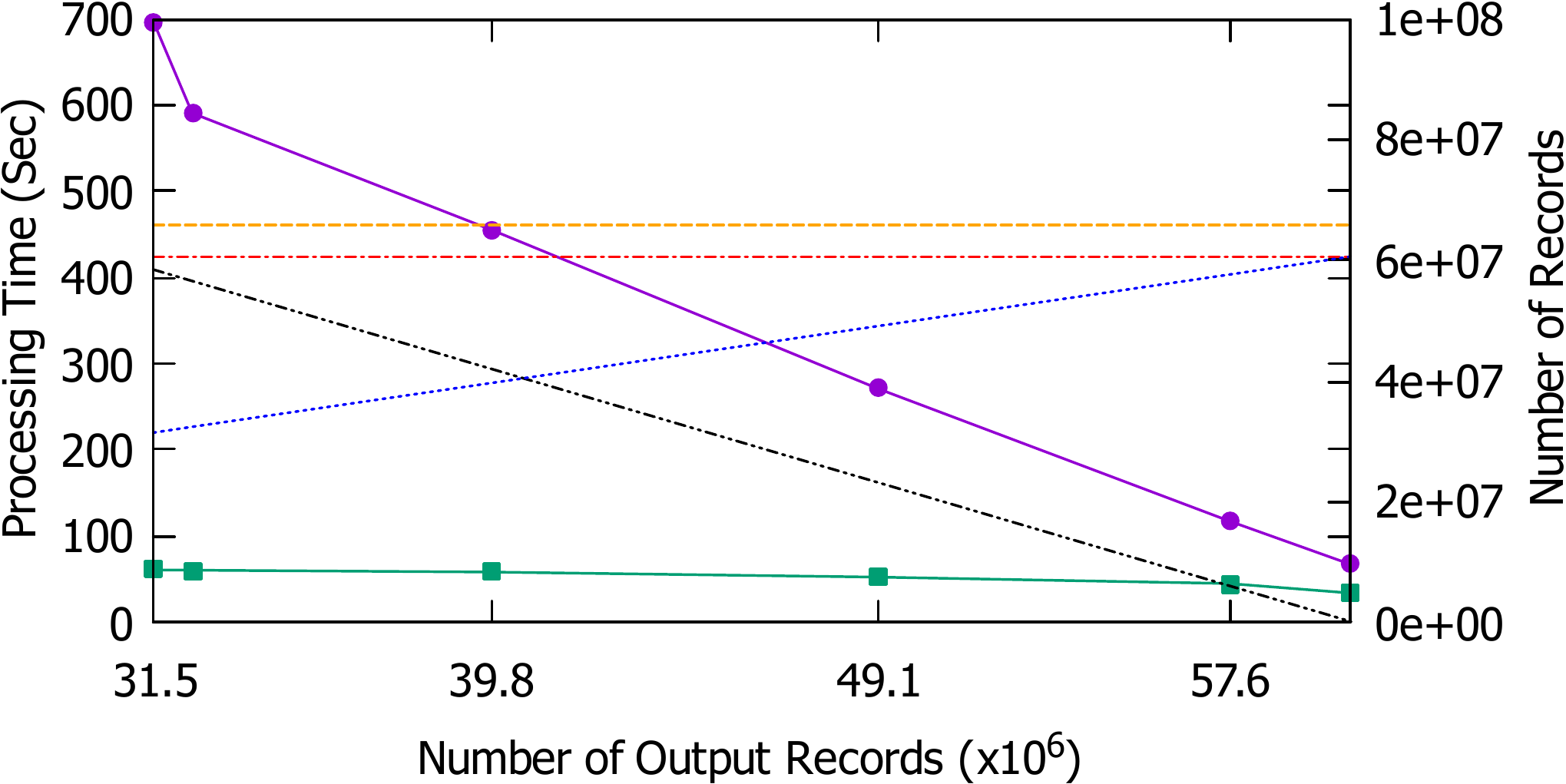}
    \caption{Running time of $\Q_{G4}$ on Google dataset in PostgreSQL with different $\OUT_2$.}
    \label{fig:out2}
\end{figure}

In Figure~\ref{fig:out2}, we investigate the impact of $\OUT_2$ for computing \DCQ. We fix $\Q_1$ (as well as $N$ and $\OUT_1$) and vary a filter predicate applied to relation $\textrm{Graph}$ in $\Q_2$.  When the predicate is more selective, $\OUT_2$ becomes smaller, and $\OUT$ becomes larger.  The running time of vanilla \revm{SQL query} decreases as $\OUT_2$ decreases, and the running time of our optimized \revm{SQL query} does not change, which is only affected by $N$ and $\OUT$. 

\begin{figure}[h]
    \centering
    \includegraphics[width=0.6\linewidth]{figure/F7Title.pdf}
    \includegraphics[width=0.6\linewidth]{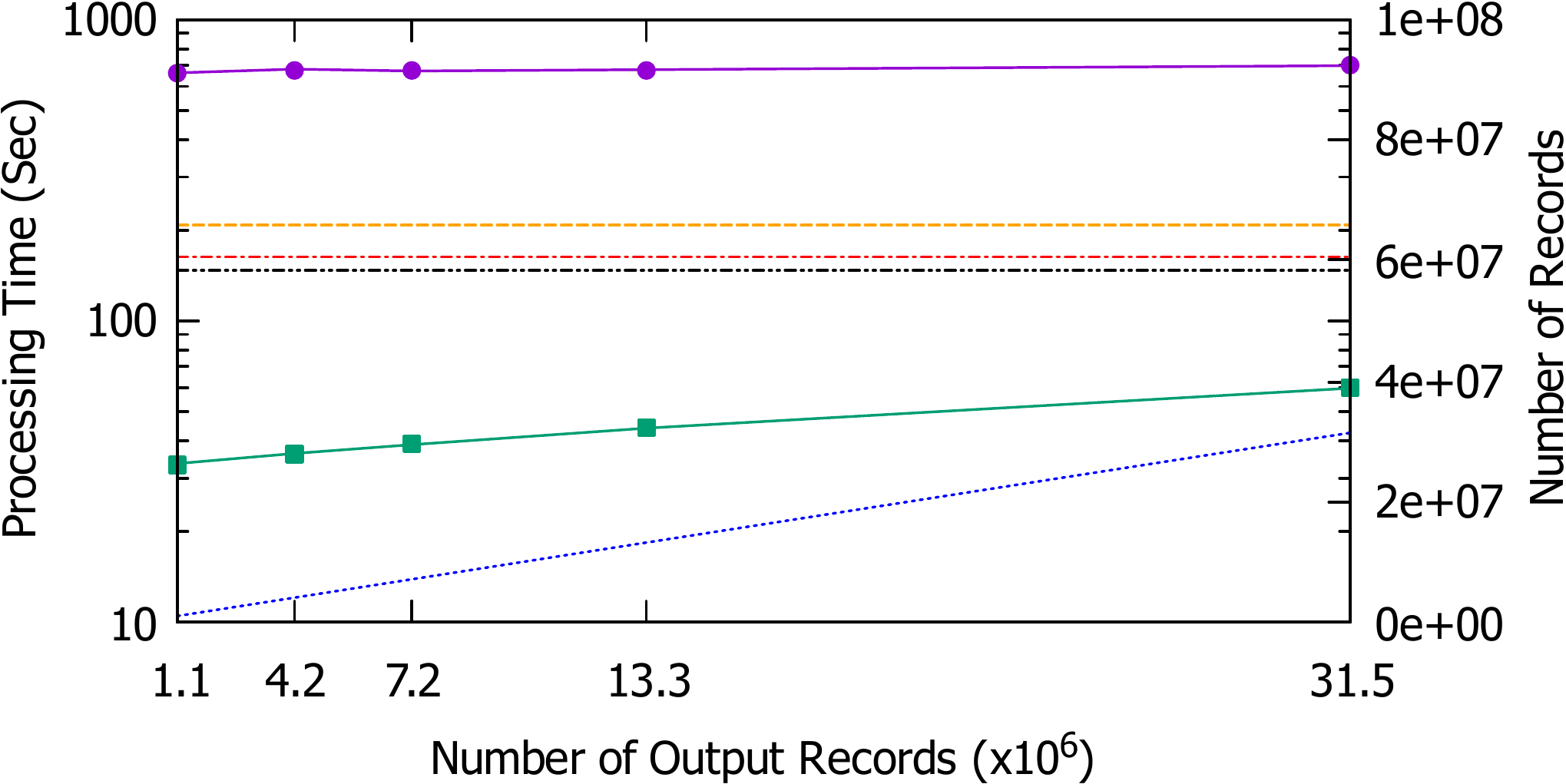}
    \caption{Running time of $\Q_{G4}$ on Google dataset in PostgreSQL with different $\OUT$.}
     \label{fig:out}
\end{figure}

In Figure~\ref{fig:out}, we investigate the impact of $\OUT$ for computing \DCQ. We adjust $\textsf{Triple}$ by changing the proportion of tuples generated by different rules, which will only change $\OUT$, while $\OUT_1$, $\OUT_2$, and $N$ stay the same.  The running time of our optimized \revm{SQL query} increases slowly as $\OUT$ increases. In contrast, the running time of vanilla \revm{SQL query} remains stably high even when $\OUT$ decreases to $1$, since its running time is only impacted by $\OUT_1$ and $\OUT_2$, both of which stay unchanged.

\begin{figure}[h]
    \centering
    \includegraphics[width=0.6\linewidth]{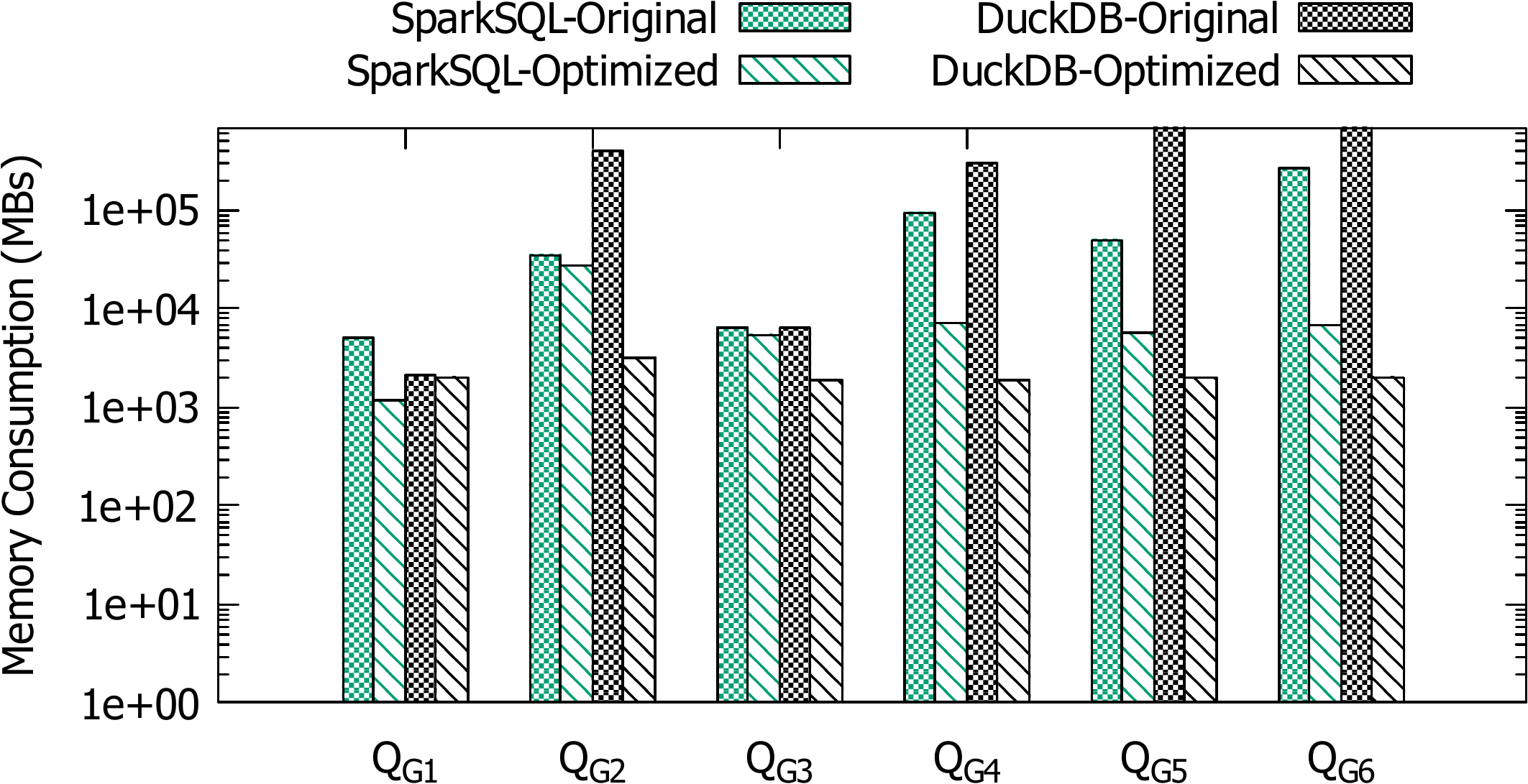}
    \caption{Memory usage by graph queries on Epinions.}
    \label{fig:Memory}
\end{figure}

\smallskip \noindent  {\bf Memory Consumption.}  We also test the memory consumption on both graph and benchmark queries by different engines. \revm{Due to the simplicity of memory consumption measurement, we report the results for PostgreSQL and DuckDB here. For benchmark queries, the optimized and vanilla SQL queries have similar behaviors on memory consumption, since the input size dominates the overall consumption.  Below, we focus on the memory consumption of graph queries. In Figure~\ref{fig:Memory}, our optimized \revm{SQL queries} achieve overall improvements for all graph queries on Epinions dataset in terms of space consumption.} %In particular, when the output size $\OUT_2$ of $\Q_2$ is large in computing $\Q_1 - \Q_2$, our improvement is much more significant. 
For example, our optimized \revm{SQL query} only requires 6.53GB on Spark SQL for evaluating $\Q_{G6}$, while the vanilla \revm{SQL query} fails to finish evaluating $\Q_{G6}$ even using 256G memory.  The improvement of our optimized SQL query is more significant on DuckDB. For $\Q_{G4}$, our optimized SQL query consumes $99.4\%$ less memory than the vanilla SQL query.  For $\Q_{G5}$ and $\Q_{G6}$, our optimized SQL queries consume roughly 2G memory. In contrast, the vanilla SQL queries fail to execute due to out-of-memory errors even after using 738G memory.

%% file: related.tex
\section{Connection with Signed Conjunctive Query}
\label{sec:SCQ}

The class of {\em signed conjunctive queries (SCQ)}~\cite{brault2012negative}, or noted as {\em conjunctive queries with negation}~\cite{lanzinger2021tractability} in the literature, is defined as \[\Q: =  \pi_{\y} \left( \eta_1 R_1(e_1) \Join \cdots \Join \eta_2 R_2(e_2) \Join \cdots \Join \eta_n R_n(e_n) \right),\] where $\eta_i$ is either empty or a negation operator $\neg$. If $\eta_i = \neg$ for all $i \in [n]$, such an SCQ is also known as a {\em negative conjunctive queries} (\NCQ). If $\eta_i = \emptyset$ for all $i \in [n]$, such an \SCQ\ is also known as a \CQ. Recall that $\V = e_1 \cup e_2 \cup \cdots \cup e_n$ and $\E = \{e_1,e_2,\cdots, e_n\}$. The query result of $\Q$ over an instance $D$ denoted as $\Q(D)$ is defined as 
\[\Q(D) = \{t \in \dom(\y):\exists t' \in \dom(\V),  \pi_{e_i} t \in R_i, \forall \eta_i = \emptyset, \pi_{e_j} t \notin R_j, \forall \eta_j = \neg\}.\]
We establish the connection between \SCQ\ and \DCQ\ via Lemma~\ref{lem:DCQ-SCQ} and Lemma~\ref{lem:SCQ-DCQ}.

\noindent {\bf From \DCQ\ to \SCQ.} Intuitively, every \DCQ\ can be rewritten as the union of a set of \SCQ s. Moreover, each resulted \SCQ\ has exactly one negated relation, and each relation of $\Q_2$ participates in one distinct \SCQ\ as the negated relation. For example, $\Q_1 -\Q_2 = R_1(x_1, x_2) \Join R_2(x_2, x_3) - R_3(x_1, x_2) \Join R_4(x_2, x_3)$ can be rewritten as $\left(R_1\Join R_2 \Join \neg R_3\right) \cup \left(R_1 \Join R_2 \Join \neg R_4 \right)$. 
\begin{lemma}
\label{lem:DCQ-SCQ}
    For a \DCQ\ $\Q_1 - \Q_2$, $\Q_1 - \Q_2 = \bigcup_{e \in \E_2} \left(\Q_1 \Join \neg R_e\right)$.
\end{lemma}
\begin{proof}
    \underline{\emph{Direction $\subseteq$}}. For every join result $t \in \Q_1 - \Q_2$, there must exist a relation $e \in \E_2$ such that $\pi_e t \notin R_e$; otherwise, $t \in \Q_2$, coming to a contradiction. Wlog, let $e \in \E_2$ be such a relation for $t$. Together with $t \in \Q_1$, there must be $t \in \Q_1 \Join \neg R_e$. \underline{\emph{Direction $\supseteq$}}. Consider an arbitrary relation $e \in \E_2$, and an arbitrary join result $t \in \Q_1 \Join \neg R_e$. Obviously, $t \notin \Q_2$ since $\pi_e t \notin R_2$. Together with $t \in \Q_1$, there must be $t \in \Q_1 - \Q_2$.
\end{proof}

\noindent {\bf From \SCQ\ to \DCQ.} On the other hand, \SCQ\ can be rewritten as the intersection of a set of \DCQ s. For a \SCQ\ $\Q = (\V,\E)$, let $\E^+, \E^- \subseteq \E$ denote the set of relations with positive, negative sign separately. Let $\V^+, \V^-$ be the set of attributes that appear in positive, negative relations separately. Let $\Q^+ = \left(\Join_{e'\in \E^+} R_{e'} \times_{x \in \V^- - \V^+ } \dom(x)\right)$ denote the {\em positive subquery} defined by positive relation as well as the whole domain of attributes which do not appear in any positive relation.
  
\begin{lemma}
\label{lem:SCQ-DCQ}
    For a \SCQ\ $\Q$, $\Q = \cap_{e \in \E^-} \left(\Q^+ -  \Q^+ \Join R_e\right)$.
\end{lemma}

\begin{proof}
    \underline{\emph{Direction $\subseteq$}}. Consider an arbitrary query result $t \in \Q$. By definition, $\pi_e t \in R_e$ holds for every $e \in \E^+$ and $\pi_e t \notin R_e$ holds for every $e \in \E^-$. This way, for each $e \in \E^-$, we have $t \in \left(\Join_{e'\in \E^+} R_{e'} \times_{x \in \V^- - \V^+ } \dom(x) \right) \Join \neg R_e$. \underline{\emph{Direction $\supseteq$}}. Consider an arbitrary $t$ such that for every $e \in \E^-$, $t \in \left(\Join_{e'\in \E^+} R_{e'} \times_{x \in \V^- - \V^+ } \dom(x) \right) \Join \neg R_e$. Then, $\pi_{e'} t \in R_{e'}$ for every $e' \in \E^+$ but $t \notin R_e$ for every $e \in \E^-$. Thus, $t \in \Q$.
\end{proof}

For example, a \SCQ\ $\Q= R_1(x_2, x_3, x_4) \Join R_2(x_1, x_3,x_4) \Join \neg R_3(x_1, x_2,x_4) \Join \neg R_4(x_1,x_2, x_3)$ can be rewritten as:
$(R_1 \Join R_2 - R_1 \Join R_2 \Join R_3) \cap (R_1 \Join R_2 - R_1 \Join R_2 \Join R_4)$.

\paragraph{Decidability of \SCQ.} Given a \SCQ\ $\Q$, the domain of attributes, and input database $D$, the {\em decidability} problem asks to decide whether there exists a query result in $\Q$. For example, a \NCQ\ $\Q = \neg R_1(x_1, x_2) \Join \neg R_2(x_2, x_3)$ decides if there exists any tuple $(a,b,c) \in \dom(x_1) \times \dom(x_2) \times \dom(x_3)$ such that $(a,b) \notin R_1$ and $(b,c) \notin R_2$, and a \CQ\ $\Q = R_1(x_1, x_2) \Join R_2(x_2, x_3)$ decides if there exists any tuple $(a,b,c)$ such that $(a,b) \in R_1$ and $(b,c) \in R_2$. The decidability problem for \CQ\ , \NCQ\ and \SCQ\ has been well studied separately:

\begin{theorem}[\cite{bagan2007acyclic}]
\label{the:cq}
   A \CQ\ $\Q$ can be decided in linear time if and only if it is $\alpha$-acyclic. 
\end{theorem}

\begin{theorem}[\cite{brault2012negative}]
\label{the:ncq}
   A \NCQ\ $\Q$ can be decided in linear time if and only if it is $\beta$-acyclic.  
\end{theorem}

\begin{theorem}[\cite{brault2013pertinence}]
\label{the:scq}
   A \SCQ\ $\Q$ can be decided in linear time if and only if $(\y, \E^+ \cup S)$ is $\alpha$-acyclic for every $S \subseteq \E^-$.
\end{theorem}

Note that $\beta$-acyclicity is a more restricted notion than $\alpha$-acyclicity, such that $\Q$ is $\beta$-acyclic if all sub-hypergraphs of $\Q$ are $\alpha$-acyclic. Obviously, $\beta$-acyclicity strictly implies $\alpha$-acyclicity. 
In~\cite{lanzinger2021tractability}, this notion of $\beta$-acyclicity has been extended to nest-set width for capturing the tractability of \SCQ\ in terms of both query and data complexity. We won't pursue this direction further.

\paragraph{Decidability of \DCQ.} Implied by Lemma~\ref{lem:DCQ-SCQ} and Theorem~\ref{the:scq}, we come to the following lemma: 

\begin{lemma}
\label{lem:full-if}
    Given two full joins $\Q_1 = (\y, \E_1)$ and $\Q_2 = (\y, \E_2)$, the \DCQ\ $\Q_1 - \Q_2$ can be decided in linear time, if $(\y, \E_1)$ is $\alpha$-acyclic, and $(\y, \E_1 \cup \{e\})$ is $
    \alpha$-acyclic for every $e \in \E_2$.
\end{lemma}

Lemma~\ref{lem:full-if} can be easily proved by a linear-time algorithm.  We can enumerate every tuple in $S_e = \pi_{e} \Join_{e' \in \E} R_{e'}$ within $O(1)$ delay, as $(\y, \E \cup \{e\})$ is $\alpha$-acyclic. For each tuple $t \in S_e$ enumerated, we check whether it belongs to $R_e$. If $t \notin R_e$, a query result of $\Q \Join \neg R_e$ is found; otherwise, we skip it and continue to the next one. It is easy to see that at most $|R_e|$ tuples are checked, so this algorithm runs in $O(N)$ time.

\begin{theorem}
    \label{the:decidability-dcq}
    Given two full joins $\Q_1 = (\y, \E_1)$ and $\Q_2 = (\y, \E_2)$, the \DCQ\ $\Q_1 - \Q_2$ can be decided in linear time, if and only if $(\y, \E_1)$ is $\alpha$-acyclic, as well as $(\y, \E_1 \cup \{e\})$ is $\alpha$-acyclic for every $e \in \E_2$.
\end{theorem}

\begin{proof}
    The if direction follows Lemma~\ref{lem:full-if}. We next distinguish two more cases for the only-if direction. (1) if $(\y, \E_1)$ is cyclic; and (2) if $(\y, \E_1)$ is $\alpha$-acyclic, and there exists some $e \in \E_2$ such that $(\y,\E_1 \cup \{e\})$ is cyclic. (1) follows Theorem~\ref{the:cq} by simply setting $\Q_2 = \emptyset$. (2) follows Lemma~\ref{lem:hardness-3}. 
    \end{proof}

 %\begin{theorem}
 %   \label{the:decidability-dcq}
 %   Given two CQs $\Q_1 = (\y, \V_1, \E_1)$ and $\Q_2 = (\y, \V_2, \E_2)$, %the \DCQ\ $\Q_1 - \Q_2$ can be decided in linear time, if and only if  ... %\end{theorem}

\section{Related Work}
\label{sec:related}

\noindent {\bf Union of CQs.} \cite{carmeli2019enumeration} studied the enumeration complexity of union of conjunctive queries (UCQs), i.e.,  the goal is to find a data structure that after linear preprocessing time, the query answers (without duplication) can be enumerated within a small delay.  Their results implied a linear algorithm in terms of input and output size for the class of union-free-connex UCQs, but whether a linear algorithm can be achieved (and, if possible, how to achieve it) is unknown for the remaining class of UCQ. \cite{christoph2018answering} also investigated the enumeration complexity of UCQs but in the dynamic scenario.

\smallskip \noindent {\bf Selection over CQs.}  Recently, multiple works have studied the complexity of selections over conjunctive queries.  \cite{wang2022conjunctive} investigated the selection in the form of comparisons between two attributes or values.  The work identifies an acyclic condition under which a near-linear-time algorithm can be achieved for conjunctive queries with comparisons.  \cite{abo2022complexity} worked on the selections over intervals, also known as intersection queries, which are special cases for comparison queries since each intersection query can be decomposed into a union of multiple comparison queries.  They show a dichotomy result that an intersection join can be computed in linear time if and only if it is $\iota$-acyclic.   \cite{hu2022computing} studied the complexity of temporal queries, where the intersection condition only exists for one global attribute.  Their result suggested that a temporal query can be solved in linear time if and only if it is r-hierarchical. \cite{tao2022intersection} also investigated the complexity of intersection queries in dynamic settings.  

%% file: appendix.tex
\appendix
\section{SQL Queries}
\label{appendix:query}

\paragraph{Graph Query $\Q_{G1}$}

Original:
\begin{lstlisting}[ language=SQL,
	deletekeywords={IDENTITY},
	deletekeywords={[2]INT},
	morekeywords={clustered},
	mathescape=true,
	xleftmargin=-1pt,
	framexleftmargin=-1pt,
	frame=tb,
	framerule=0pt ]
SELECT g1.src as src, g1.dst as dst 
FROM graph g1
WHERE (g1.src, g1.dst) NOT IN (
    SELECT DISTINCT g1.src, g1.dst 
    FROM graph g1, graph g2, graph g3 
    WHERE g1.dst = g2.src and g2.dst = g3.src);
\end{lstlisting}

Optimized:
\begin{lstlisting}[ language=SQL,
	deletekeywords={IDENTITY},
	deletekeywords={[2]INT},
	morekeywords={clustered},
	mathescape=true,
	xleftmargin=-1pt,
	framexleftmargin=-1pt,
	frame=tb,
	framerule=0pt ]
SELECT g1.src as src, g1.dst as dst 
FROM graph g1 
WHERE NOT EXISTS (
    SELECT * FROM graph g2
    WHERE EXISTS (
        SELECT * FROM graph g3 
        WHERE g3.src = g2.dst 
    ) and g1.dst = g2.src  ); 
\end{lstlisting}

\paragraph{Graph Query $\Q_{G2}$}

Original:
\begin{lstlisting}[ language=SQL,
	deletekeywords={IDENTITY},
	deletekeywords={[2]INT},
	morekeywords={clustered},
	mathescape=true,
	xleftmargin=-1pt,
	framexleftmargin=-1pt,
	frame=tb,
	framerule=0pt ]
SELECT src as A, node1 as B, node2 as C, node3 as D
FROM graph g1, Triple1 T1
WHERE g1.dst = T1.node1
and NOT EXISTS (
  SELECT * FROM Triple2 T2, graph g2
  WHERE T2.node3 = g2.src and T2.node1 = g1.src 
  and T2.node2 = T1.node1 and T2.node3 = T1.node2 
  and g2.dst = T1.node3);
\end{lstlisting}

Optimized:
\begin{lstlisting}[ language=SQL,
	deletekeywords={IDENTITY},
	deletekeywords={[2]INT},
	morekeywords={clustered},
	mathescape=true,
	xleftmargin=-1pt,
	framexleftmargin=-1pt,
	frame=tb,
	framerule=0pt ]
SELECT src as A, node1 as B, node2 as C, node3 as D
FROM graph g1, Triple1 T1
WHERE g1.dst = T1.node1  
and (NOT EXISTS ( SELECT * FROM Triple2 T2
    WHERE T2.node1 = g1.src and T2.node2 = T1.node1 
        and T2.node3 = T1.node2)
    or NOT EXISTS ( SELECT * FROM graph g2
        WHERE g2.src = T1.node2 and g2.dst = T1.node3));
\end{lstlisting}

\paragraph{Graph Query $\Q_{G3}$}

Original:
\begin{lstlisting}[ language=SQL,
	deletekeywords={IDENTITY},
	deletekeywords={[2]INT},
	morekeywords={clustered},
	mathescape=true,
	xleftmargin=-1pt,
	framexleftmargin=-1pt,
	frame=tb,
	framerule=0pt ]
 SELECT node1, node2, node3
 FROM Triple T1
 WHERE NOT EXISTS(
  SELECT *
  FROM graph g1, graph g2, graph g3
  WHERE g1.dst = g2.src and g2.dst = g3.src and g3.dst = g1.src and g1.src = T1.node1 and g2.src = T2.node2 and g3.src = T3.node3);
\end{lstlisting}

Optimized:
\begin{lstlisting}[ language=SQL,
	deletekeywords={IDENTITY},
	deletekeywords={[2]INT},
	morekeywords={clustered},
	mathescape=true,
	xleftmargin=-1pt,
	framexleftmargin=-1pt,
	frame=tb,
	framerule=0pt ]
SELECT node1, node2, node3
FROM Triple T1
WHERE NOT EXISTS (SELECT * FROM graph g1 
    WHERE T1.node1 = g1.src and T1.node2 = g1.dst) 
or NOT EXISTS (SELECT * FROM graph g2 
    WHERE T1.node2 = g2.src and T1.node3 = g2.dst) 
or NOT EXISTS (SELECT * FROM graph g3 
    WHERE T1.node3 = g3.src and T1.node1 = g3.dst); 
\end{lstlisting}

\paragraph{Graph Query $\Q_{G4}$}

Original:
\begin{lstlisting}[ language=SQL,
	deletekeywords={IDENTITY},
	deletekeywords={[2]INT},
	morekeywords={clustered},
	mathescape=true,
	xleftmargin=-1pt,
	framexleftmargin=-1pt,
	frame=tb,
	framerule=0pt ]
SELECT node1, node2, node3
FROM Triple T1
WHERE NOT EXISTS  (
 SELECT *
 FROM graph g1, graph g2, graph g3
 WHERE g1.dst = g2.src and g2.dst = g3.src and g1.src = T1.node1 and g2.src = T1.node2 and g2.dst = T1.node3);
\end{lstlisting}

Optimized:
\begin{lstlisting}[ language=SQL,
	deletekeywords={IDENTITY},
	deletekeywords={[2]INT},
	morekeywords={clustered},
	mathescape=true,
	xleftmargin=-1pt,
	framexleftmargin=-1pt,
	frame=tb,
	framerule=0pt ]
SELECT node1, node2, node3
FROM Triple
WHERE NOT EXISTS
   (SELECT * FROM graph WHERE node1 = src and node2 = dst)
or NOT EXISTS
   (SELECT * FROM graph WHERE node2 = src and node3 = dst)
or NOT EXISTS
   (SELECT * FROM graph WHERE node3 = src);
\end{lstlisting}

\paragraph{Graph Query $\Q_{G5}$}

Original:
\begin{lstlisting}[ language=SQL,
	deletekeywords={IDENTITY},
	deletekeywords={[2]INT},
	morekeywords={clustered},
	mathescape=true,
	xleftmargin=-1pt,
	framexleftmargin=-1pt,
	frame=tb,
	framerule=0pt ]
SELECT g1.src as A, g2.src as B, g3.src as C, g3.dst as D
FROM graph g1, graph g2, graph g3
WHERE g1.dst = g2.src and g2.dst = g3.src and 
NOT EXISTS (SELECT *
    FROM graph g4, graph g5, graph g6
    WHERE g4.dst = g5.src and g5.dst = g6.src and g2.src = g4.src and g5.src = g3.src and g6.src = g3.dst and g6.dst = g1.src);
\end{lstlisting}

Optimized:
\begin{lstlisting}[ language=SQL,
	deletekeywords={IDENTITY},
	deletekeywords={[2]INT},
	morekeywords={clustered},
	mathescape=true,
	xleftmargin=-1pt,
	framexleftmargin=-1pt,
	frame=tb,
	framerule=0pt ]
SELECT g1.src as A, g2.src as B, g3.src as C, g3.dst as D
FROM graph g1, graph g2, graph g3
WHERE g1.dst = g2.src and g2.dst = g3.src and 
NOT EXISTS (
    SELECT * FROM graph g6 WHERE g6.dst = g1.src and g6.src = g3.dst); 
\end{lstlisting}

\paragraph{Graph Query $\Q_{G6}$}

Original:
\begin{lstlisting}[ language=SQL,
	deletekeywords={IDENTITY},
	deletekeywords={[2]INT},
	morekeywords={clustered},
	mathescape=true,
	xleftmargin=-1pt,
	framexleftmargin=-1pt,
	frame=tb,
	framerule=0pt ]
SELECT g1.src as A, g1.dst as B, g2.src as C, g2.dst as D
FROM graph g1, graph g2
WHERE NOT EXISTS (SELECT *  
    FROM graph g3, graph g4, graph g5, graph g6
    WHERE g3.dst = g4.src and g4.dst = g5.dst and g5.src = g3.src and g5.dst = g6.src and g3.src = g1.src and g3.dst = g1.dst and g6.src = g2.src and g6.dst = g2.dst);
\end{lstlisting}

Optimized:
\begin{lstlisting}[ language=SQL,
	deletekeywords={IDENTITY},
	deletekeywords={[2]INT},
	morekeywords={clustered},
	mathescape=true,
	xleftmargin=-1pt,
	framexleftmargin=-1pt,
	frame=tb,
	framerule=0pt ]
SELECT g1.src as A, g1.dst as B, g2.src as C, g2.dst as D
FROM graph g1, graph g2
WHERE NOT EXISTS (SELECT * FROM graph g4 
    WHERE g4.src = g1.dst and g4.dst = g2.src)
or NOT EXISTS ( SELECT * FROM graph g5 
    WHERE g1.src = g5.src and g2.src = g5.dst); 
\end{lstlisting}

\paragraph{TPC-H Query 16}

Original:
\begin{lstlisting}[ language=SQL,
	deletekeywords={IDENTITY},
	deletekeywords={[2]INT},
	morekeywords={clustered},
	mathescape=true,
	xleftmargin=-1pt,
	framexleftmargin=-1pt,
	frame=tb,
	framerule=0pt ]
SELECT p_brand, p_type, p_size, 
count(distinct ps_suppkey) as supplier_cnt
FROM partsupp, part
WHERE p_partkey = ps_partkey
and p_brand <> 'Brand#45'
and p_type NOT LIKE 'MEDIUM POLISHED%'
and p_size IN (49, 14, 23, 45, 19, 3, 36, 9)
and ps_suppkey NOT IN (
  SELECT s_suppkey
  FROM supplier, nation
  WHERE s_nationkey = n_nationkey and n_name = 'CHINA')
GROUP BY p_brand, p_type, p_size;
\end{lstlisting}

Optimized:
\begin{lstlisting}[ language=SQL,
	deletekeywords={IDENTITY},
	deletekeywords={[2]INT},
	morekeywords={clustered},
	mathescape=true,
	xleftmargin=-1pt,
	framexleftmargin=-1pt,
	frame=tb,
	framerule=0pt ]
SELECT p_brand, p_type, p_size, 
count(distinct ps_suppkey) as supplier_cnt
FROM partsupp, part
WHERE p_partkey = ps_partkey
and p_brand <> 'Brand#45'
and p_type NOT LIKE 'MEDIUM POLISHED%'
and p_size IN (49, 14, 23, 45, 19, 3, 36, 9)
and NOT EXISTS ( SELECT * FROM supplier 
    WHERE EXISTS (SELECT * FROM nation 
        WHERE s_nationkey = n_nationkey 
         and n_name = 'CHINA')
  and s_suppkey = ps_suppkey)
GROUP BY p_brand, p_type, p_size;
\end{lstlisting}

\paragraph{TPC-DS Query 35}

Original:
\begin{lstlisting}[ language=SQL,
	deletekeywords={IDENTITY},
	deletekeywords={[2]INT},
	morekeywords={clustered},
	mathescape=true,
	xleftmargin=-1pt,
	framexleftmargin=-1pt,
	frame=tb,
	framerule=0pt ]
SELECT 
  ca_state, cd_gender, cd_marital_status, cd_dep_count,
  count(*) cnt1, stddev_samp(cd_dep_count), 
  sum(cd_dep_count), min(cd_dep_count), 
  cd_dep_employed_count,count(*) cnt2, 
  stddev_samp(cd_dep_employed_count),
  sum(cd_dep_employed_count), min(cd_dep_employed_count),
  cd_dep_college_count, count(*) cnt3,
  stddev_samp(cd_dep_college_count),
  sum(cd_dep_college_count), min(cd_dep_college_count)
FROM
  customer c,customer_address ca,customer_demographics
WHERE
  c.c_current_addr_sk = ca.ca_address_sk and
  cd_demo_sk = c.c_current_cdemo_sk and
  not exists (select *
          from store_sales,date_dim
          where ss_sold_date_sk = d_date_sk and
                c.c_customer_sk = ss_customer_sk and
                d_year = 2001 and
                d_qoy < 4) and
   not exists (select *
            from web_sales,date_dim
            where ws_sold_date_sk = d_date_sk and
                  d_year = 2001 and
                  d_qoy < 4 and
                  ws_bill_customer_sk = c.c_customer_sk) and
   not exists (select *
            from catalog_sales,date_dim
            where cs_sold_date_sk = d_date_sk and
                  d_year = 2001 and
                  d_qoy < 4 and
                  cs_ship_customer_sk = c.c_customer_sk)
group by ca_state,
         cd_gender,
         cd_marital_status,
         cd_dep_count,
         cd_dep_employed_count,
         cd_dep_college_count;
\end{lstlisting}

Optimized:
\begin{lstlisting}[ language=SQL,
	deletekeywords={IDENTITY},
	deletekeywords={[2]INT},
	morekeywords={clustered},
	mathescape=true,
	xleftmargin=-1pt,
	framexleftmargin=-1pt,
	frame=tb,
	framerule=0pt ]
SELECT  
  ca_state, cd_gender, cd_marital_status, cd_dep_count,
  count(*) cnt1, stddev_samp(cd_dep_count), 
  sum(cd_dep_count), min(cd_dep_count), 
  cd_dep_employed_count,count(*) cnt2, 
  stddev_samp(cd_dep_employed_count),
  sum(cd_dep_employed_count), min(cd_dep_employed_count),
  cd_dep_college_count, count(*) cnt3,
  stddev_samp(cd_dep_college_count),
  sum(cd_dep_college_count), min(cd_dep_college_count)
FROM
  customer_address ca,customer_demographics, 
  (select * from customer cu
    where not exists (select * from store_sales 
      where exists (select * from date_dim 
        where d_year = 2001 and d_qoy < 4 and ss_sold_date_sk = d_date_sk)
      and cu.c_customer_sk = ss_customer_sk) 
    and not exists (select * from web_sales 
      where exists (select * from date_dim 
        where d_year = 2001 and d_qoy < 4 and ws_sold_date_sk = d_date_sk)
      and cu.c_customer_sk = ws_bill_customer_sk) 
    and not exists (select * from catalog_sales 
      where exists (select * from date_dim 
        where d_year = 2001 and d_qoy < 4 and cs_sold_date_sk = d_date_sk)
    and cu.c_customer_sk = cs_ship_customer_sk)) as c
WHERE
  c.c_current_addr_sk = ca.ca_address_sk and
  cd_demo_sk = c.c_current_cdemo_sk
group by ca_state,
         cd_gender,
         cd_marital_status,
         cd_dep_count,
         cd_dep_employed_count,
         cd_dep_college_count;
\end{lstlisting}

\paragraph{TPC-DS Query 69}

Original:
\begin{lstlisting}[ language=SQL,
	deletekeywords={IDENTITY},
	deletekeywords={[2]INT},
	morekeywords={clustered},
	mathescape=true,
	xleftmargin=-1pt,
	framexleftmargin=-1pt,
	frame=tb,
	framerule=0pt ]
SELECT
  cd_gender, cd_marital_status, cd_education_status,
  count(*) cnt1, cd_purchase_estimate,
  count(*) cnt2, cd_credit_rating, count(*) cnt3
FROM
  customer c,customer_address ca,customer_demographics
WHERE
  c.c_current_addr_sk = ca.ca_address_sk and
  ca_state in ('IN','ND','PA') and
  cd_demo_sk = c.c_current_cdemo_sk and 
  exists (select *
          from store_sales,date_dim
          where c.c_customer_sk = ss_customer_sk and
                ss_sold_date_sk = d_date_sk and
                d_year = 1999 and
                d_moy between 2 and 2+2) and
   (not exists (select *
            from web_sales,date_dim
            where c.c_customer_sk = ws_bill_customer_sk and
                  ws_sold_date_sk = d_date_sk and
                  d_year = 1999 and
                  d_moy between 2 and 2+2) and
    not exists (select * 
            from catalog_sales,date_dim
            where c.c_customer_sk = cs_ship_customer_sk and
                  cs_sold_date_sk = d_date_sk and
                  d_year = 1999 and
                  d_moy between 2 and 2+2))
group by cd_gender,
         cd_marital_status,
         cd_education_status,
         cd_purchase_estimate,
         cd_credit_rating;
\end{lstlisting}

Optimized:
\begin{lstlisting}[ language=SQL,
	deletekeywords={IDENTITY},
	deletekeywords={[2]INT},
	morekeywords={clustered},
	mathescape=true,
	xleftmargin=-1pt,
	framexleftmargin=-1pt,
	frame=tb,
	framerule=0pt ]
SELECT
  cd_gender, cd_marital_status, cd_education_status,
  count(*) cnt1, cd_purchase_estimate,
  count(*) cnt2, cd_credit_rating, count(*) cnt3
FROM
  customer_address ca,customer_demographics, 
  (select * from customer cu
    where exists (select * from store_sales 
      where exists (select * from date_dim 
        where ss_sold_date_sk = d_date_sk and d_year = 1999 and d_moy between 2 and 2+2) 
      and not exists (select * from web_sales 
        where exists (select * from date_dim 
          where ws_sold_date_sk = d_date_sk and d_year = 1999 and d_moy between 2 and 2+2) 
        and ws_bill_customer_sk = cu.c_customer_sk) 
      and not exists (select * from catalog_sales 
        where exists (select * from date_dim
          where cs_sold_date_sk = d_date_sk and d_year = 1999 and d_moy between 2 and 2+2) 
        and cs_ship_customer_sk = c_customer_sk)
      and ss_customer_sk = cu.c_customer_sk)) as c
WHERE
  c.c_current_addr_sk = ca.ca_address_sk and
  ca_state in ('IN','ND','PA') and
  cd_demo_sk = c.c_current_cdemo_sk
group by cd_gender,
         cd_marital_status,
         cd_education_status,
         cd_purchase_estimate,
         cd_credit_rating;
\end{lstlisting}

    \section{Missing Proofs in Section~\ref{sec:hardness-proof}}
    \label{appendxi:hardness-proof}

    \subsection{Preliminaries on CQs}
        \begin{definition}[GYO Reduction]
            The GYO reduction for a CQ $(\y, \V,\E)$ is an iterative procedure that (1) if an attribute $x \in \V$ only appears in one relation $e\in \E$, then $x$ can be removed from $e$; (2) if there exists a pair of relations $e,e' \in \E$ such that $e \subseteq e'$, then $e$ can be removed.
        \end{definition}
    
        \begin{lemma}[\cite{yannakakis1981algorithms}]
            A query $(\y, \V,\E)$ is $\alpha$-acyclic if the GYO reduction results in an empty query.
        \end{lemma}
 
        \begin{definition}[Path]
            In a CQ $(\y, \V,\E)$, a path between a pair of attributes $x_1, x_k$ is a sequence of attributes $C = \langle x_1, x_2, \cdots, x_k \rangle \subseteq \V$, such that 
            \begin{itemize}[leftmargin=*]
                \item there exists $e \in \E$ with $x_i, x_{i+1} \in e$ for any $1 \le i < k$;
                \item for any $e \in \E$, either $|e \cap C| = 1$, or $e \cap C = \{x_i, x_{i+1}\}$ for some $i \in [k-1]$. 
            \end{itemize}
        \end{definition}
        
        \begin{definition}[Cycle]
        \label{def:cycle}
            In a CQ $(\y, \V,\E)$, a cycle is a sequence of attributes $C = \{x_1, \cdots, x_k\} \subseteq \V$, such that 
            \begin{itemize}[leftmargin=*]
                \item there exists $e \in \E$ with $x_i, x_{i+1} \in e$ for any $1 \le i < k$, and $x_1, x_k \in e$;
                \item for any $e \in \E$, either $|e \cap C| = 1$, or $e \cap C = \{x_i, x_{i+1}\}$ for some $1 \le i < k$, or $e \cap C = \{x_1, x_k\}$.
            \end{itemize}
        \end{definition}
        
        \begin{definition}[Clique]
            In a CQ $(\y, \V,\E)$, a clique is a subset of attributes $C \subseteq \V$, such that for any pair of attributes $x_1, x_2 \in C$, there exists $e \in \E$ with $x_1, x_2 \in e$.
        \end{definition}

        \begin{definition}[Conformal]
            A CQ $(\y, \V,\E)$ is conformal, if every clique $C \subseteq \V$ there exists $e \in \E$ with $C \subseteq e$.
        \end{definition}

        \begin{definition}[Non-conformal Clique]
            Following the definition of conformal of CQ, we define a clique $C \subseteq \V$ as non-conformal in a CQ $(\y, \V,\E)$, if there does not exist $e \in \E$ such that $C \subseteq e$.
        \end{definition}

        \begin{lemma}[\cite{brault2016hypergraph}]
        \label{lem:brault-acyclic}
            A CQ is $\alpha$-acyclic if and only if it is conformal and cycle-free. 
        \end{lemma}
        
        \begin{lemma}[\cite{bagan2007acyclic}]
        \label{lem:acyclic-non-free-connex}
            In an acyclic but non-free-connex CQ $(\y,\V,\E)$, there must exists a sequence of distinct attributes $C = \langle x_1, x_2, \cdots, x_k\rangle$ with $k \ge 3$, such that 
            \begin{itemize}[leftmargin=*]
                \item there exists a relation $e \in \E$ such that $\{x_i, x_{i+1}\} \subseteq e$ for every $i \in \{1,2,\cdots, k-1\}$; 
                \item $x_1, x_k \in \y$ but $x_2, \cdots, x_{k-1} \notin \y$; 
                \item for each $e \in \E$, either $|e \cap C| \le 1$ or $e \cap C = \{x_i, x_{i+1}\}$ for some $i \in \{1,2,\cdots, k-1\}$; 
            \end{itemize}
        \end{lemma}

          \begin{lemma}
            \label{lem:subclique}
                In a CQ $\Q = (\y, \V, \E)$, if there exists a clique $C$, then for any $C' \subset C$, $C'$ is also a clique.
            \end{lemma}

            \begin{proof}
             Since $C$ is a clique, there exists a relation that contains every pair of attributes.  As $C'$ is a subset of $C$, then for any two attributes in $C'$ there also exists a relation that contains both of these two attributes, hence $C'$ is also a clique.
            \end{proof}

            \begin{lemma}
                In a cycle-free CQ $\Q = (\y, \V, \E)$, if there exists a non-conformal clique $C$, then $|C| > 3$.  
            \end{lemma}

            \begin{proof}
                For the clique $C$ of size 1 or 2, it is clear that $C$ is conformal as there is a relation containing the entire clique by definition.  Suppose  there exists a non-conformal clique $C$ with $|C| = 3$, say $C = \{x_1, x_2, x_3\}$.  As the clique is non-conformal, there does not exist a relation that covers all three attributes, but any pair of attributes appears together in one relation.  Then $x_1, x_2, x_3$ will form a triangle, contradicting the fact that $\Q$ is cycle-free. Hence, any non-conformal clique $C$ in a cycle-free CQ must have $|C| \ge 3$.
            \end{proof}
            
            We denote a non-conformal clique as a {\em minimal} if there exists no subset $C' \subseteq C$ such that $C'$ is a non-conformal clique. 

            \begin{lemma} \label{lem:minimal-clique}
                In a CQ $\Q = (\y, \V, \E)$ with a minimal non-conformal clique $C$, for every $\V \subsetneq C$ there exists some $e \in \E$ with $\V \subseteq e$.
            \end{lemma}

            \begin{proof}
                As $C$ is a clique, $\V$ is also a clique for any $\V \subsetneq C$. Meanwhile, as $C$ is the minimal non-conformal clique, $\V$ is a conform clique, which implies a relation $e \in \E$ with $\V \subseteq e$. 
            \end{proof}

    \subsection{Helper Lemmas}
    Now, we are ready to show some helper lemmas, which will be used to prove Lemma~\ref{lem:hardness-2} and Lemma~\ref{lem:hardness-3}.
    
    \begin{definition}[Subquery] 
            For a CQ $\Q= (\y, \V, \E)$, a subquery of $\Q$ induced by a set of attributes $C \subseteq \V$ is denoted as $\Q[C] = (\y \cap C, C, \E[C])$, where $\E[C] = \{e \cap C : e \in \E, e \cap C \neq \emptyset\}$.
        \end{definition}
    
    \begin{lemma}
            \label{lem:subquery}
                Given two CQs $\Q_1 = (\y, \V_1, \E_1)$ and $\Q_2 = (\y, \V_2, \E_2)$, for any $C \subseteq \V_1 \cap \V_2$, if $\Q[C] =\Q_1[C] - \Q_2[C]$ requires $\Omega(N^{1-o(1)})$ time, then $\Q = \Q_1 - \Q_2$ requires $\Omega(N^{1-o(1)})$ time.
            \end{lemma}

            \begin{proof}
                Given any database instance $D$ for $\Q[C]$, we can construct a database instance $D'$ for $\Q$ as follows. For any attribute $x \notin C$, we set its value to be $*$. For any $e \in \E$ with $e \cap C \neq \emptyset$, there exists a corresponding relation $e' = e \cap C$ in the residual query. For each tuple $t'$ in $R_{e'}$, we insert $t$ into $e$ with $\pi_{e \cap C} t' = \pi_{e \cap C} t$.  It is easy to see that there is a one-to-one correspondence between $\Q[C]$ and $\Q$.  Hence, if $\Q$ can be solved in linear time, then $\Q[C]$ can be solved in linear time, coming to a contradiction.
            \end{proof}
            
            \begin{lemma}
    Any algorithm for evaluating the following \DCQ:  \[\Q_1 - \Q_2  = R_1(x_1)- \pi_{x_1} \left( R_2(x_1,x_2) \Join R_3(x_2,x_3) \Join R_4(x_3, x_4) \Join R_5(x_2,x_4) \right),\] requires $\Omega(N^{1-o(1)})$ time, assuming the strong triangle conjecture.
\end{lemma}

\begin{proof}
    Given a graph $G = (V, E)$, we construct $R_3 = R_4 = R_5 = E$, $R_1 = V$ and $R_2 = \{u\} \times V$ for some $u \in V$. Let $N = |E| \ge |V|$.  We note that $\OUT = |V|-1$ if and only if there is a triangle in $G$. Hence, if $\Q$ can be evaluated in $O(N)$ time,  whether there is a triangle in $G$ can be determined in $O(N)$ time, contradicting the detecting triangle conjecture. 
 \end{proof}

\begin{lemma}
\label{lem:hard-4}
	Any algorithm for deciding the following \DCQ $\Q_1 - \Q_2$ requires $\Omega(N^{1-o(1)})$ time, assuming the strong triangle conjecture, where $\Q_1 = R_1(x_1, x_2) \Join R_2(x_2, x_3)$ and $\Q_2 = R_3(x_1, x_3) \Join R_4(x_2)$, or $\Q_2 = R_3(x_1, x_3) \Join R_4(x_2,x_3)$, or $\Q_2 = R_3(x_1, x_3) \Join R_5(x_1, x_2)$, or $\Q_2 = R_3(x_1, x_3) \Join R_4(x_2,x_3) \Join R_5(x_1, x_2)$.
\end{lemma}

\begin{proof}
    In the proof of Lemma~\ref{lem:hard-3}, we have shown the hardness for $\Q_1 - \Q_2 = R_1(x_1, x_2) \Join R_2(x_2, x_3) - R_3(x_1, x_3) \Join R_4(x_2)$. The remaining three queries can be proved similarly. Given an arbitrary graph $G = (V,E)$ with $V$ as the set of vertices and $E$ as the set of edges, we perform an algorithm to detect whether there exists a triangle in $G$. Let $m = |E| = N^{3/4}$ be the number of edges in $G$. Let $\mathcal{N}(u) = \{u \in V: (v,u) \in E\}$ be the neighbor list of vertex $u \in V$. The degree $\textrm{deg}(u)$ of a vertex $u \in V$ is defined as the size of the neighbor list of $u$, i.e., $\textrm{deg}(u) = |\mathcal{N}(u)|$. We partition vertices in $V$ into two subsets: $V^H = \{v \in V: \textrm{deg}(v) > m^{1/3}\}$ and $V^L=V-V^H$. From $G$, we construct following relations: $R=E$, $R_0 =\{(u,v) \in E: u \in V^L \textrm{ or } v \in V^L\}$, $R_1 = \{(u,v) \in E: u \in V^H\}$, $R_2 = \{(u,v) \in E: v \in V^H\}$ and $R_3 = V^H \times V^H - E$. It can be easily checked that each relation contains $O(N)$ tuples.
    
    For $\Q_1 - \Q_2 = R_1(x_1, x_2) \Join R_2(x_2, x_3) - R_3(x_1, x_3) \Join R_4(x_2, x_3)$, we set $R_4 = E$ and consider following queries:
    \begin{align*}
		   & \Q = R(x_1, x_2) \Join R(x_2, x_3) \Join R_0(x_1, x_3) \\
		   & \Q' = R_1(x_1, x_2) \Join R_2(x_2, x_3) - R_3(x_1, x_3) \Join R_4(x_2, x_3)
    \end{align*}
    It can be easily proved that a triangle exists in $G$ if and only if $\Q$ or $\Q'$ is not empty. We point out that $\Q$ can be computed in $O(m^{4/3})$ time, since $|R(x_2, x_3) \Join R_0(x_1, x_3)| \le m^{4/3}$ and $|R(x_1, x_2) \Join R_0(x_1, \\x_3)| \le m^{4/3}$ implied by the definition of $V^L$. This way, if $\Q'$ can be computed in $O(m^{4/3})$ time, then detecting whether there exists a triangle or not takes $O(m^{4/3})$ time, coming to a contradiction to the detecting triangle conjecture.
    
    For $\Q_1 - \Q_2 = R_1(x_1, x_2) \Join R_2(x_2, x_3) - R_3(x_1, x_3) \Join R_5(x_1, x_2)$, We set $R_5 =R$ and consider following queries:
    \begin{align*}
		   & \Q = R(x_1, x_2) \Join R(x_2, x_3) \Join R_0(x_1, x_3) \\
		   & \Q' = R_1(x_1, x_2) \Join R_2(x_2, x_3) - R_3(x_1, x_3) \Join R_5(x_1, x_2)
    \end{align*}
    For $\Q_1 - \Q_2 = R_1(x_1, x_2) \Join R_2(x_2, x_3) - R_3(x_1, x_3) \Join R_4(x_2, x_3) \Join R_5(x_1, x_2)$, we set $R_4 = R_5 = E$ and consider following queries:
        \begin{align*}
		   & \Q = R(x_1, x_2) \Join R(x_2, x_3) \Join R_0(x_1, x_3) \\
		   & \Q' = R_1(x_1, x_2) \Join R_2(x_2, x_3) - R_3(x_1, x_3) \Join R_4(x_2, x_3) \Join R_5(x_1, x_2)
    \end{align*}
    Both cases follow the similar argument as above. Together, we have completed the proof. 
\end{proof}

    \subsection{Proof of Lemma~\ref{lem:hardness-2}}
    We assume $\Q_1$ is reduced. As $\Q_1$ is free-connex, then $\Q_1$ must be an acyclic full join. We consider repeatedly applying the following procedures to $\Q_2 = (\y,\V_2,\E_2)$: (1) if there is an non-output attribute $x \in \V_2 - \y$ only appearing in one relation $e \in \E_2$, remove $x$ from $e$ as well as $\V_2$; (2) if there is a pair of relations $e,e' \in \E_2$ such that $e \subseteq e'$, remove $e$ from $\E_2$. As $\Q_2$ is non-reducible, the residual query must be non-full; otherwise $(\y, \V_2, \E_2 \cup \{\y\})$ is free-connex, contradicting the fact that $\Q_2$ is non-linear-reducible. Hence, we can assume for $\Q_2$ that every non-output attribute must appear in at least two relations, and there exists no relation whose attributes are fully contained in another relation. We distinguish two cases:

    {\bf (Case 1):} $\Q_2$ is acyclic. As $\Q_2$ is non-linear-reducible, $\Q_2$ must be non-free-connex. Implied by Lemma~\ref{lem:acyclic-non-free-connex}, there must exist such a path $\langle x_1, x_2, \cdots, x_k\rangle$ with desired properties. Moreover, for the acyclic full join $\Q_1 = (\y, \E_1)$, we initialize two sets $S_1 = \{x_1\}$ and $S_2 = \{x_k\}$, and repeat the following procedure: if there exists some $e \in \E_1$ such that $e \cap S_1 \neq \emptyset$ and $e \cap S_2 \neq \emptyset$, we just stop; otherwise, we find some $e \in \E_1$ such that $e \cap S_1 \neq \emptyset$, we just add all attributes in $e$ into $S_1$, and remove $e$. Then for each $e \in \E_1$, we have either $e \subseteq S_1$, or $e \subseteq S_2$, or $e \cap S_1 \neq \emptyset$ and $e \cap S_2 \neq \emptyset$.
    
    Given an arbitrary instance of $R_1,R_2,R_3$ in lemma~\ref{lem:hard-2}, we construct an input instance for $\Q_1, \Q_2$ separately as follows.  For $\Q_1$, we set $\dom(x) = \dom(x_1)$ for every $x \in S_1$, $\dom(x) = \dom(x_2)$ for every $x \in S_2$, and $\dom(x) = \{*\}$ for every $x \in \V_1 - S_1 \cup S_2$. Then, the result of $\Q_1$ degenerates to $R_1(x_1,x_k)$.
    %\begin{itemize}[leftmargin=*]
    %    \item If $e \subseteq S_1$, we add a tuple $(a,a,\cdots,a)$ for every $a \in \dom(x_1)$.
    %    \item If $e \subseteq S_2$, we add a tuple $(b,b,\cdots,b)$ for every $b \in \dom(x_2)$.
    %    \item If $e \cap S_1 \neq \emptyset$ and $e \cap S_2 \neq \emptyset$, we add a tuple $t$ with $\pi_x t =a$ if $x \in e \cap S_1$ and $\pi_x t =b$ if $x \in e\cap S_2$, for every $(a,b) \in R_1$.
    %\end{itemize}
    For $\Q_2$, we simply set $\dom(x) = \{*\}$, $x_2 = x_3 = \cdots= x_{k-1}$, and $\dom(x_1), \dom(x_k)$ as the same as that in $\Q_1$. Implied by the properties of the path found, every relation in $\E_2$ must either contains a single attribute from $\{x_1, x_2, \cdots, x_k\}$, or degenerates to one edge of the path. Hence, the result of $\Q_2$ degenerates to $\pi_{x_1, x_k}R_2(x_1, x_2) \Join R_3(x_2, x_k)$, which is exactly captured by Lemma~\ref{lem:hardness-2}.
    
    {\bf (Case 2):} $\Q_2$ is cyclic. Then, there exists a cycle or a non-conformal clique in $\Q_2$. We further distinguish the following cases.
    
    {\bf (Case 2.1):} there is a cycle $C$ such that $C \subseteq \V -\y$. We can reduce $\Q_1 -\Q_2$ to $1- \pi_{\emptyset} R_1(x_1, x_2) \Join R_2(x_2,x_3) \Join R_3(x_1,x_3)$.
    
    {\bf (Case 2.2):} there is a cycle $C$ such that $C -\y \neq \emptyset$ and $C \cap \y \neq \emptyset$. We can reduce $\Q_1 -\Q_2$ to $R_1(x_1) - \pi_{x_1} (R_2(x_1,x_2) \Join R_3(x_2, x_3) \Join R_4(x_1,x_3))$ if $|C \cap \y|=1$, and $R_1(x_1, x_2) - \pi_{x_1, x_2} (R_2(x_1, x_2) \Join R_3(x_2, x_3) \Join R_4(x_1,x_3))$ otherwise.
    
    {\bf (Case 2.3):} there exists no cycle but a non-conformal clique $C$ such that $C -\y \neq \emptyset$. %If there exists some $e \in \E_2$ such that $C \cap \y \subseteq e$, we will proceed with the next step. Otherwise, $C' = C \cap \y$ is also a non-conformal clique. we   In this case, we can always find a minimal non-conformal clique $C$ such , i.e., there exists no subset $C' \subseteq C$ such that $C'$ is also a non-conformal clique. 
    In this case, we will show the hardness of  \DCQ\ $\Q[C] = \Q_1[C] - \Q_2[C]$, based on the hardness of $\Q_2[C]$. Consider an arbitrary instance $D$ for $\Q_2[C]$. For simplicity, assume the domain of each attribute in $\y$ is $[N]$. We construct the following instance $D'$ for $\Q_1[C]$. There is a one-to-one mapping between any pair of attributes $x,x' \in C \cap \y$. We also set $\dom(x) = \{*\}$ for every $x \in C - \y$. It can be easily checked that $\Q_1[C]$ contains exactly $N$ results, and therefore $\Q[C]$ contains at most $N$ results. Moreover, $\Q_2[C]$ is empty if and only if $|\Q[C]| = N$.  Suppose we have an algorithm that can compute $\Q[C]$ in linear time, then we can determine whether $\Q_2[C]$ is empty or not in linear time, contradicting the fact that $\Q_2[C]$ cannot be determined in $O(N)$ time.  As $\Q[C]$ cannot be computed in linear time, combining with Lemma~\ref{lem:subquery}, $\Q$ cannot be computed in linear time.

    {\bf (Case 2.4):} $C\subseteq \y$ holds for every cycle $C$, as well as every non-conformal clique $C$. Recall that there exists no non-output attribute only appearing in one relation, and there exists no relation whose attributes are contained by another relation. %For any cycle $C$ witnessed by a sequence of relations $e_0, e_1, e_2, \cdots, e_k$ as in Definition~\ref{def:cycle}, we note that no pair of relations sharing a non-output attribute; otherwise, a new cyclic has been found with non-output attributes, coming to a contradiction. 
    Let $\Q_2' =(\y,\V_2, \E_2 \cup \{\y\})$.
    Every non-conformal clique $C$ in $\Q_2$ becomes conformal in $\Q_2'$ due to the existence of $\{\y\}$. Similarly, every cycle will disappear in $\Q_2'$ due to the existence of $\{\y\}$. 
    In this case, $\Q'_2$ must be acyclic, implied by Lemma~\ref{lem:brault-acyclic}. Meanwhile, as $\Q_2$ is non-linear-reducible, $\Q'_2$ must be non-free-connex. As $\Q_2'$ is acyclic and non-free-connex, there must exist a path in $\Q_2$ as characterized by Lemma~\ref{lem:acyclic-non-free-connex}.
    Following the similar argument as {\bf (Case 1)}, we can reduce $\Q_1 -\Q_2$ to $R_1(x_1, x_3) - \pi_{x_1, x_3} R_2(x_1, x_2) \Join R_3(x_2, x_3)$.

\subsection{Proof of Lemma~\ref{lem:hardness-3}}
    Given a free-connex CQ $\Q_1 = (\y,\V_1, \E_1)$ and a linear-reducible CQ $\Q_2 = (\y,\V_2, \E_2)$, we denote $(\y, \E'_1)$ and $(\y, \E'_2)$ as the reduced queries of $\Q_1,\Q_2$ respectively. Let $e' \in \E_2$ be the relation such that $(\y,\E'_1 \cup \{e'\})$ is cyclic. As $\Q_1$ is free-connex, $(\y, \E'_1)$ is acyclic. Our proof proceeds with the following steps: 
        \begin{itemize}[leftmargin=*]
            \item {\bf Step 1:} In $(\y, \E'_1 \cup \{e'\})$, there exists no $e \in \E'_1$ such that $e' \subseteq e$;
            \item {\bf Step 2:} There exists a pair of attributes $x_1, x_n \in e'$, such that there exists no $e \in \E'_1$ with $x_1, x_n \in e$;
            \item {\bf Step 3:} There is a cycle $C = \langle x_1,x_2, \cdots, x_n\rangle$ in $(\y, \E'_1 \cup \{e'\})$ with $e' \cap C = \{x_1, x_n\}$, and $\langle x_1, x_2, \cdots,x_n \rangle$ is a path in $(\y, \E'_1)$; 
            \item {\bf Step 4:} There is a reduction from Lemma~\ref{lem:hard-4} to $\Q_1 - \Q_2$; 
        \end{itemize}

        For {\bf Step 1}, if there exists some $e \in \E'_1$ such that $e' \subseteq e$, then $(\y, \E'_1 \cup \{e'\})$ is acyclic if and only if $(\y, \E'_1 \cup \{e'\})$, contradicting the fact that $(\y, \E'_1 \cup \{e'\})$ is cyclic but $(\y, \E'_1)$ is acyclic. 
        
        For {\bf Step 2}, we first show that $|e'| \ge 2$. Suppose $|e'| =1$, say $e  = \{x\}$. There must exist $e \in \E'_1$ such that $x \in e$, hence $e' \subseteq e$, coming to a contradiction of {\bf Step 1}. Hence, $|e'| \ge 2$. Moreover, if for every pair of attributes $x_1, x_n \in e'$, there exists some $e \in \E'_1$ such that $x_1, x_n \in e$, then we find a clique of attributes in $e$ in $(\y, \E'_1)$. As $(\y, \E'_1)$ is acyclic, there must exist some $e'' \in \E'_1$ such that $e' \subseteq e''$, coming to a contradiction of {\bf Step 1}. Hence, we can always find a pair of attributes $x_1, x_n \in e'$ as desired.

        For {\bf Step 3}, since $(\y, \E'_1)$  is acyclic but $(\y, \E'_1 \cup \{e'\})$ is cyclic, either a cycle or a non-conformal clique is formed by the addition of $e'$. Let's consider the case where a new non-conformal clique $S\subseteq \y$ is formed. By definition, there exists no relation $e \in \E'_1 \cup \{e'\}$ such that $S \subseteq e$. We partition $S$ into two subsets $S_1, S_2$ such that $S_1 = S \cap e'$ and $S_2 = S - e'$. It is clear that $S_2 \neq \emptyset$; otherwise $S \subseteq e'$, contradicting the fact that $S$ is non-conformal. Moreover, for any $x \in S_1$, $\{x\} \cup S_2$ is also a clique. As $(\y, \E'_1)$ is acyclic, $\{x\} \cup S_2$ must be a conformal clique, i.e., there exists some $e \in \E'_1$ such that $\{x\} \cup S_2 \subseteq e$. Meanwhile, $|S_1| \ge 2$; otherwise $S$ is a non-conformal clique in $(\y, \E'_1)$, contradicting the fact that $(\y,\E'_1)$ is acyclic.
        We can also identify two different attributes $x_1, x_2 \in S_1$ such that there exists no relation $e \in \E'_1$ with $\{x_1,x_2\} \cup S_2 \subseteq e$; otherwise, $S$ is a non-conformal clique in $(\y,\E'_1)$, contradicting the fact that $(\y,\E'_1)$ is acyclic. Let $e_1 \in\E'_1$ be the relation that $\{x_1\} \cup S_2 \subseteq e_1$, and $e_2 \in \E'_2$ be the relation that $\{x_2\} \cup S_2 \subseteq e_2$. From above, we note that $x_2 \notin e_1$ and $x_1 \notin e_2$. Let $x \in S_2$ be an attribute such that there exists no relation $e \in \E'_1$, such that $x_1, x_2, x \in e$. It is always feasible to find such an attribute $x$, since there exist no relation $e \in \E'_1$ such that $\{x_1, x_2\} \cup S \subseteq e$. 
        
        In either way, a cycle of $\langle x_1, x_2, x\rangle$ forms after the addition of $e'$. Hence, a new cycle $C$ must be formed by the addition of $e'$, say $C =\langle x_1, x_2, \cdots, x_n\rangle$. Let $e' \cap C = \{x_1, x_n\}$. The existence of $C$ also implies a path of $\langle x_1, x_2, \cdots, x_n\rangle$ in $(\y, \E'_1)$. 

        For {\bf Step 4}, we show the following reduction. For simplicity, we set $x_2 = x_3 = \cdots = x_{n-1}$. For any attribute $x \notin \{x_1, x_2, \cdots, x_n\}$, we set $\dom(x) =\{*\}$. If ignoring all attributes with domain as $\{*\}$, each relation in $\E_1$ falls into $R_1(x_1, x_2)$ or $R_2(x_2, x_n)$. As $x_2 \notin e'$, there exists at least one relation in $e \in \E'_2$ such that $x_2 \in e$. We distinguish four more cases on such $e$:
        \begin{itemize}[leftmargin=*]
            \item $R_1(x_1, x_2) \Join R_2(x_2, x_n) - R_{e'}(x_1, x_n) \Join R_e(x_2)$;
            \item $R_1(x_1, x_2) \Join R_2(x_2, x_n) - R_{e'}(x_1, x_n) \Join R_e(x_2, x_n)$;
             \item $R_1(x_1, x_2) \Join R_2(x_2, x_n) - R_{e'}(x_1, x_n) \Join R_{e}(x_1, x_2)$;
            \item $R_1(x_1, x_2) \Join R_2(x_2, x_n) - R_{e'}(x_1, x_n) \Join R_e(x_2, x_n) \Join R_{e''}(x_1, x_2)$;
        \end{itemize}
        which follows the proof of Lemma~\ref{lem:hard-4}. 
        
        So far, we have shown the hardness of computing $\Q_1[\y] - \Q_2[\y]$, and the hardness of computing $\Q_1 - \Q_2$ follows by Lemma~\ref{lem:subquery}.

\section{Missing Materials in Section~\ref{sec:extension}}
\label{appendix:extension}

\subsection{Difference of Multiple CQs}
\begin{algorithm}[h]
\caption{{\sc DMCQ}$(\Q_1, \Q_2, \cdots, \Q_k, D_1, D_2, \cdots, D_k)$}
\label{alg:multiple}

 \lIf{$k=2$}{\Return {\sc EasyDCQ}$(\Q_1, \Q_2, D_1, D_2)$}
 \lIf{$\y \neq \V_1$}{$((\y, \E_1), D_1) \gets \textsc{Reduce}(\Q_1, D_1)$}
 \lIf{$\y \neq \V_2$}{$((\y, \E_2), D_2) \gets \textsc{Reduce}(\Q_2, D_2)$}
 $\mathcal{S} \gets \emptyset$\;
 \ForEach{$e \in \E_2$}{
    $S_e \gets \textsc{Yannakakis}((e,\y,\E_1), D)$\;
    $\Q'_1 \gets (\y, \y, \E_1 \cup \{e\})$\;
    $D'_1 \gets \textsc{Yannakakis}(\Q'_1, D_1 \cup \{S_e - R_e\})$\;
    $\mathcal{S} \gets \mathcal{S} \cup \textsc{DMCQ}(\Q'_1, \Q_3, \cdots, \Q_k, D'_1, D_3, \cdots, D_k)$\;
    }
\Return $\mathcal{S}$\;
\end{algorithm}

\subsection{Proof of Theorem~\ref{the:bag}}

    In the bag semantics, for any free-connex CQ $\Q = (\y,\V,\E)$ and an instance $D$, it is still possible to reduce the query and instance in linear time, while preserving the correctness of the query results. We invoke Algorithm~\ref{alg:reduce}, but incorporate the semi-join and projection operators in the bag semantics. For a semi-join result $t$ of $R_e \ltimes R_{e'}$, we define:
    %\begin{itemize}
    %    \item For a semi-join result $t$ of $R_e \ltimes R_{e'}$, 
        \[\displaystyle{w(t) = \sum_{t' \in R_{e'}: \pi_{e \cap e'} t = \pi_{e \cap e'} t'} w(t').}\]
    %\end{itemize}
    Then, we are left with two full joins that share the same query structure after applying the reduce procedure to both $\Q_1$ and $\Q_2$. 
    
    Suppose we are given two instances $D_1, D_2$ for the full join $\Q = (\V,\E)$. Let $R_e, R'_e$ be the corresponding relations to $e \in \E$ in $D_1, D_2$. Again, assume that each tuple $t$ is associated with a positive count $w(t)>0$. Let $w_1, w_2$ be the count functions of $\Q_1,\Q_2$ respectively. Generalizing the algorithm in Example~\ref{exp:bag}, the high-level idea is to find all join results $t \in \Q_1$ such that $\prod_{e \in \E} \frac{w_1(\pi_e t)}{w_2(\pi_e t)} > 1$. For each $e \in \E$, we distinguish tuples in $R_{e}$ into three case: $R_{e\emptyset} = \{t \in R_e: t \notin R'_e\}$, $R_{e<} = \{t \in R_e: t \in R'_e, w_1(t) \le w_2(t)\}$ and $R_{e>} = \{t \in R_e: t \in R'_e: w_1(t) > w_2(t)\}$. We can rewrite it as:
    
    \begin{lemma}
        Given two full CQs $\Q_1 = \Q_2 = (\V,\E)$, 
        \begin{align*}
            \Q_1 - \Q_2 =&  \cup_{\bar{E} \in \E} \left(\Join_{e \in \bar{E}} R_{e\emptyset} \right) \Join \left(\Join_{e \in \E - \bar{E}} (R_{e<} + R_{e>}) \right) +\cup_{\bar{E} \subseteq \E} \left(\Join_{e \in \bar{\E}} R_{e<}\right) \Join_\theta \left(\Join_{e \in \E - \bar{\E}} R_{e>}\right),
        \end{align*}
        where a pair of tuples $(t_1, t_2)$ can be $\theta$-joined if and only if $w_1(t_1) \cdot w_1(t_2) > w_2(t_1) \cdot w_2(t_2)$.
    \end{lemma}
    
    The first part of $\bigcup_{\bar{E} \in \E} \left(\Join_{e \in \bar{E}} R_{e\emptyset} \right) \Join \left(\Join_{e \in \E - \bar{E}} (R_{e<} + R_{e>}) \right)$ can be computed similarly as we have done in the set semantics. We next focus on the second part. 
    Each $\bar{\E} \subseteq \E$ derives a $\theta$-joins, which will be computed by the following procedure \textsc{BagDCQ}. For simplicity, let $S_e =R_{e<}$ if $e \in \bar{E}$ and $S_e = R_{e>}$ otherwise. We maintain additional variable $\zeta_t$ for every tuple $t \in S_e$ for every $e \in \E$. Initially, $\zeta_t = \frac{w_1(t)}{w_2(t)}$ if $t \in R_e$ and $t \in R'_e$,  $\zeta_t = +\infty$ if $t \in R_e$ and $t \notin R'_e$, and $\zeta_t = 0$ if $t \notin R_e$. Algorithm~\ref{alg:bag-DCQ} consists of two phases. In the first phase, it updates the value of $\zeta_t$ for every tuple $t$ over a join tree $\T$. More specifically, suppose $t \in S_e$ for some $e \in \E$. Let $\T_e$ be the subtree of $\T$ rooted at $e$. Then,
    \[\zeta_t = \max_{t' \in \Join_{e' \in \T_e} S_{e'}: \pi_{e}t' = t} \prod_{e' \in \T_e} \frac{w_1(\pi_e t')}{w_2(\pi_e t')},\]
    i.e., the maximum product of $\frac{w_1(\cdot)}{w_2(\cdot)}$ over all join results in the subtree rooted at $e$, participated by $t$. As a result, a tuple $t$ in the root node participates in any query result if and only if $\zeta_t > 1$. In the second phase, we invoke $\textsc{Enumerate}$ procedure for every $t \in S_r$ with $\zeta_t > 1$, and enumerate all the query results participated by $t$. 

    The procedure $\textsc{Enumerate}(\T, t, \tau)$ takes three parameters, which returns all join results over the join tree $\T$ participated by $t$ (from the root relation of $\T$), whose product of ratios over participated tuples is at least $\tau$. Let $r$ be the root node of $\T$. In the base case when $\T$ is a single node, we just return $t$. As we prove later, there must be $\zeta_t  = \frac{w_1(t)}{w_2(t)} > \tau$ in this case. In general, we distinguish two more cases. If $r$ contains a single child, say $u$, it suffices to find all tuples $t' \in S_u$ such that $\zeta_{t'} \cdot \frac{w_1(t)}{w_2(t)} > \tau$, i.e., participate in at least one join result. For each such a tuple $t'$, we recursively enumerate the query results in $\T_u$ participated by $t$, whose product of $\frac{w_1(\cdot)}{w_2(\cdot)}$ is at least $\tau \cdot \frac{w_2(t)}{w_1(t)}$, which can be done by $\textsc{Enumerate}\left(\T_u, t', \tau \cdot \frac{w_2(t)}{w_1(t)}\right)$ (line 6). Otherwise, $r$ contains at least two child nodes. We also play with recursion and shrink the join tree $\T$ by removing a subtree rooted at one child node. W.l.o.g., assume $\{u_1, u_2, \cdots, u_k\}$ is the set of child nodes of $r$. We first find out tuples in $R_{u_k}$ that will participate in any query result with $t$. This can be done by first finding the maximum $\zeta$-value of tuples in another child node that can be joined with $t$, and then finding the minimum $\zeta$-value that tuples in $u_k$ should satisfy (line 10). Then, we enumerate all query results in the subtree $\T_{u_k}$ whose product of $\frac{w_1(\cdot)}{w_2(\cdot)}$ is at least $\tau'$, which are exactly those will participate in the final query results (line 11). For each such a tuple enumerated (at line 12-14), we in turn find out the query results in the remaining subtree of $\T - \T_{u_k}$ whose product of $\frac{w_1(\cdot)}{w_2(\cdot)}$ is at least (updated) $\tau/\tau'$, where $\tau'$ is the product of $\frac{w_1(\cdot)}{w_2(\cdot)}$ for $t$. At last, we just output their combination as a Cartesian product.

 \begin{algorithm}
    \caption{\textsc{BagDCQ}$(\Q=(\V, \E), \bar{\E}, D_1, D_2)$}
    \label{alg:bag-DCQ}
    \SetKwInOut{Input}{Input}
    \SetKwInOut{Output}{Output}
 
    \Input{A full join $(\V,\E)$ and two instances $D_1, D_2$}
    \Output{Results of $\Q(D_1) - \Q(D_2)$;}
 
    $\T \gets$ the join tree of $(\V, \E)$ with root $r$\;
    \ForEach{$u \in \T$ in a bottom-up way (excluding leaf)}{
        \ForEach{$t \in S_u$}{
            \lIf{$\exists v \textrm{is a child of u}$, s.t. $\not \exists t' \in S_v$ with $\pi_{u\cap v} t' = \pi_{u \cap v} t$}{
                $\displaystyle{\zeta_t} \gets 0$}
            \lElse{
                $\displaystyle{\zeta_t \gets \zeta_t \cdot \prod_{v: v \textrm{is a child of u}}\max_{t' \in S_v: \pi_{u\cap v} t' = \pi_{u \cap v} t} \zeta_{t'}}$}
        }
        \If{$u \neq r$}{
            Sort $S_u$ by the join attribute(s) between $u$ and its parent first, and then in decreasing ordering of $\zeta$\; 
        }
    }
    \ForEach{$t \in S_r$ with $\zeta_t > 1$}{
        {\sc Enumerate}($\T$, $t$, $1$)\;
    }
    \end{algorithm}